\documentclass[12pt]{article}

\usepackage{scicite}

\usepackage{times}
\usepackage{amsfonts} 

\usepackage[]{graphicx}
\usepackage{mathtools}

\newcommand{\tkey}[3]{\protect\raisebox{#3}{\protect\includegraphics[width=#2]{#1.pdf}}}

\graphicspath{{./}}

\newenvironment{sciabstract}{%
\begin{quote} \bf}
{\end{quote}}

\title{
Accelerated Simulations of Molecular Systems through Learning of their Effective Dynamics
} 

\author
{
Pantelis R.~Vlachas$^{1}$,
Julija Zavadlav$^{2}$, \\
Matej Praprotnik$^{3,4}$,
Petros Koumoutsakos$^{1,5,\dagger}$ 
\\
\normalsize{
$^{1}$
Computational Science and Engineering Laboratory, 
}\\
\normalsize{
ETH Zurich, CH-8092, Switzerland
}\\
\normalsize{
$^{2}$
Professorship of Multiscale Modeling of Fluid Materials,
}\\
\normalsize{
Department of Mechanical Engineering, Technical University of Munich
}\\
\normalsize{
80333, Munich, Germany
}\\
\normalsize{
$^{3}$
Laboratory for Molecular Modeling, National Institute of Chemistry
}\\
\normalsize{
SI-1001 Ljubljana, Slovenia
}\\
\normalsize{
$^{4}$
Department of Physics, Faculty of Mathematics and Physics, University of Ljubljana
}\\
\normalsize{
SI-1000 Ljubljana, Slovenia
}\\
\normalsize{
$^{5}$
John A. Paulson School of Engineering and Applied Sciences
}\\
\normalsize{
Harvard University, Cambridge, MA 02138, USA
}\\
\normalsize{$^\dagger$ Corresponding author. E-mail:  petros@seas.harvard.edu}
}

\date{}

\begin{document} 

\baselineskip24pt

\maketitle 

\begin{sciabstract}
Simulations are vital for understanding and predicting the evolution of complex molecular systems.
However, despite advances in algorithms and special purpose hardware, accessing the timescales necessary to capture the structural evolution of bio-molecules remains a daunting task. In this work we present a novel framework to advance simulation timescales by up to three orders of magnitude, by learning the effective dynamics (LED) of molecular systems.
LED augments the equation-free methodology by employing  a probabilistic mapping between coarse and fine scales using mixture density network (MDN) autoencoders and evolves the non-Markovian latent dynamics using long short-term memory MDNs.
We demonstrate the effectiveness of LED in the M\"ueller-Brown potential, the Trp Cage protein, and the alanine dipeptide. LED  identifies explainable reduced-order representations and can generate, at any instant, the respective all-atom molecular trajectories.
We believe that the proposed framework provides a dramatic increase to simulation capabilities and opens new horizons for the effective modeling of complex molecular systems.
\end{sciabstract}



\section*{Introduction}
\label{sec:intro}
Over the last 30 years molecular dynamics (MD) simulations of biological macro-molecules have advanced our  understanding of their structure and function\cite{Karplus2002}. Today MD simulations have become an essential tool for scientific discovery in the fields of biology, chemistry, and medicine.  However, they remain hampered by their  limited access to timescales of biological relevance for protein folding pathways, conformational dynamics, and rare-event kinetics. 

In order to resolve this bottleneck two complementary approaches have been pursued.
First efforts centered around innovative hardware solutions started with crowd sourcing for compute cycles~\cite{Shirts2000} and have more recently received a boost with the Anton machine~\cite{Shaw2009} enabling remarkable, second-long, simulations for small bio-molecules. Complementary algorithmic efforts aim to advance time scales by systematic coarse graining of the system dynamics.
One of the first such studies used  the principal component or normal mode analysis to simulate the conformational changes in proteins~\cite{balsera1996principal,brooks1983harmonic,skjaerven2011principal,praprotnik2005molecular}.
Several  coarse-graining (CG) methods  reduce the complexity of molecular systems by modeling several atoms as a single particle~\cite{noid2013perspective,zavadlav2019bayesian,Voth:2016}.
Backmapping techniques~\cite{pezeshkian2020backmapping, stieffenhofer2020adversarial,Kremer:2006} can be subsequently utilized to recover the atomistic degrees of freedom from a CG representation.
Multiscale approaches combine the atomistic and coarse-grained/continuum models~\cite{werder2005hybrid,ayton2007multiscale,annurev2008} to augment the accessible timescales while significant efforts have focused on enhanced sampling techniques~\cite{huber1994local,voudouris1998guided, dellago1998efficient, laio2002escaping,van2003novel, maragliano2006string, jaffrelot2020high}.
Several of these methods exploit the fact that coarse kinetic dynamics on the molecular level are often governed by a few, slow collective variables (CVs) (also termed reaction coordinates)~\cite{peters2006obtaining, stamati2010application,bittracher2018data,bonati2020data}, or by transitions between a few long-lived metastable states~\cite{schutte2011markov,bittracher2018transition}.

The CVs are typically specified a priori and their choice crucially impacts the performance and success of the respective sampling methods.
Similar to to the CG models, the CVs provide a low order representation of the molecular system, albeit without a  particle representation.
CVs are of much lower dimensionality than CG models, and retrieving atomistic configurations from CVs is a more challenging problem.
While many research efforts have addressed the coarse to fine mapping in CG models, the literature is still scarce on methods to retrieve atomistic configurations from CVs.

Machine learning (ML) methods~\cite{Michie1968,Bishop2006}, exploiting the expressive power of deep networks and their scalability to large datasets, have been used to alleviate the computational burden associated with the simulation of proteins, leading to profound scientific discoveries~\cite{noe2020machine, butler2018machine, noe2020machine2}. 

The pioneering work in Ref.~\cite{behler2007generalized}, utilized neural networks to learn an approximate potential energy surface of density functional theory (DFT) in bulk silicon from quantum mechanical calculations, performing MD simulations with this approximate potential and accelerating the DFT simulations.
The field of data-driven learning of potential energy surfaces and force fields is rapidly attracting attention with important recent extensions and applications~\cite{rupp2012fast, chmiela2018towards, schutt2018schnet, rowe2020accurate, bartok2017machine, imbalzano2018automatic, hansen2013assessment, faber2017prediction, cheng2019ab}.
ML is employed to identify CG models for MD in Refs.~\cite{zhang2018deepcg, wang2019machine, durumeric2019adversarial}.
Boltzmann generators are proposed in Ref.~\cite{noe2019boltzmann} to sample from the equilibrium distribution of a molecular system directly surpassing the need to perform MD.

Early ML methods for the identification of CVs are building on the variational approach~\cite{nuske2014variational} leading to the time-lagged independent analysis (TICA)~\cite{perez2016hierarchical}.
TICA is based on the Koopman operator theory, suggesting the existence of a latent transformation to an infinite-dimensional space that linearizes the dynamics on average.
As a consequence, slow CVs are modeled as linear combinations of feature functions of the state of the protein (atom coordinates, or internal structural coordinates).
Coarse-graining of the molecular dynamics is achieved by discretizing the state space and employing indicator vector functions as features~\cite{buchete2008coarse, noe2011dynamical, nuske2014variational, ribeiro2018reweighted}.
Consequently, the feature state dynamics reduce to the propagation law of a Markov State Model (MSM).
More recently the need for expert knowledge to construct the latent feature functions has been alleviated by learning the latent space using neural networks ~\cite{mardt2018vampnets,wehmeyer2018time}.
The dynamics on the latent space are assumed to be linear and Markovian.
For example, VAMPnets~\cite{mardt2018vampnets,chen2019nonlinear} learn nonlinear features of the molecular state with autoencoder (AE) networks.
However, they are not generative and cannot recover the detailed configuration of the protein (decoding part).
Moreover, the method requires the construction of an MSM to sample the latent dynamics and approximate the time-scales of the dynamics.
Time-lagged AE have been utilized to identify a reaction coordinate embedding and propagate the dynamics in Ref.~\cite{wehmeyer2018time} but they are not generative, as the learned mappings are deterministic, while the effective dynamics are assumed to be Markovian.

Extensions to generative approaches include Refs.~\cite{wu2018deep,hernandez2018variational,sidky2020molecular}.
In Ref.~\cite{wu2018deep}, a deep generative MSM is utilized to capture the long-timescale dynamics and sample realistic alanine dipeptide configurations.
Even though Mixture Density Networks (MDNs) are employed in Ref.~\cite{sidky2020molecular} to propagate the dynamics in the latent space, memory effects are not taken into account.
The proposed method is based on the autocorrelation loss, which suffers from the dependency on the batch size~\cite{hernandez2018variational}.
In~\cite{ribeiro2018reweighted},the Reweighted autoencoded variational Bayes for enhanced sampling (RAVE) method is proposed that alternates between iterations of MD and a Variational AE (VAEs) model. 
RAVE is encoding each time-step independently without  taking into account the temporal aspect of the latent dynamics.
RAVE requires the transition to the high-dimensional configuration space to progress the simulation in time, which can be computationally expensive.

The works mentioned above  imply  memory-less (Markovian) latent space dynamics by selecting an appropriate time-lag in the master equations~\cite{buchete2008coarse, noe2011dynamical}.
The time-lag is usually estimated heuristically, balancing the requirements to be large enough so that the Markovian assumption holds, and at the same time small enough to ensure that the method samples the configuration space efficiently. We remark that 
in cases where a protein is interacting with a solvent, only the configuration of the protein is taken into account and not the solvent.
This renders the Markovian assumption in the latent dynamics rather unrealistic. This issue is addressed in this work by employing Long Short-Term Memory (LSTM)~\cite{hochreiter1997long} Recurrent Neural Networks (RNNs) that capture memory effects of the latent dynamics.
 
Here we propose a novel data-driven generative framework that relies on  Learning the Effective Dynamics (LED) of the molecular systems \cite{vlachas2020learning}.
LED is founded on the  equation-free framework (EFF)~\cite{kevrekidis2003equation} and it enriches it by employing ML methodologies to evolve the latent space dynamics with the Mixture Density Network - Long Short-Term Memory RNN (MDN-LSTM) and the two-way mapping between coarse and fine scales with Mixture Density Network Autoencoders (MDN-AEs)~\cite{bishop1994mixture}.
These enrichments are essenetial in extending the applicability of EFF to non-Markovian settings and problems with strong non-linearities.
We demonstrate the effectiveness of the LED framework in simulations of the M\"ueller-Brown potential (MBP), the Trp Cage miniprotein, and the alanine dipeptide in water.
LED can accurately capture the statistics, and reproduce the free energy landscape from data.
Moreover, LED uncovers low-energy metastable states in the free energy projected to the latent space and recovers the transition time-scales between them.
We find that in simulations of the alanine dipeptide and the Trp Cage miniprotein, LED is three orders of magnitude faster than the classical MD solver.
As a data-driven generative method, LED has the ability to sample novel unseen configurations interpolating the training data and accelerating the exploration of the state space.


\section*{Materials and Methods}
\label{sec:method}

\begin{figure*}[t!]
\centering
\includegraphics[width=1.0\textwidth,clip]{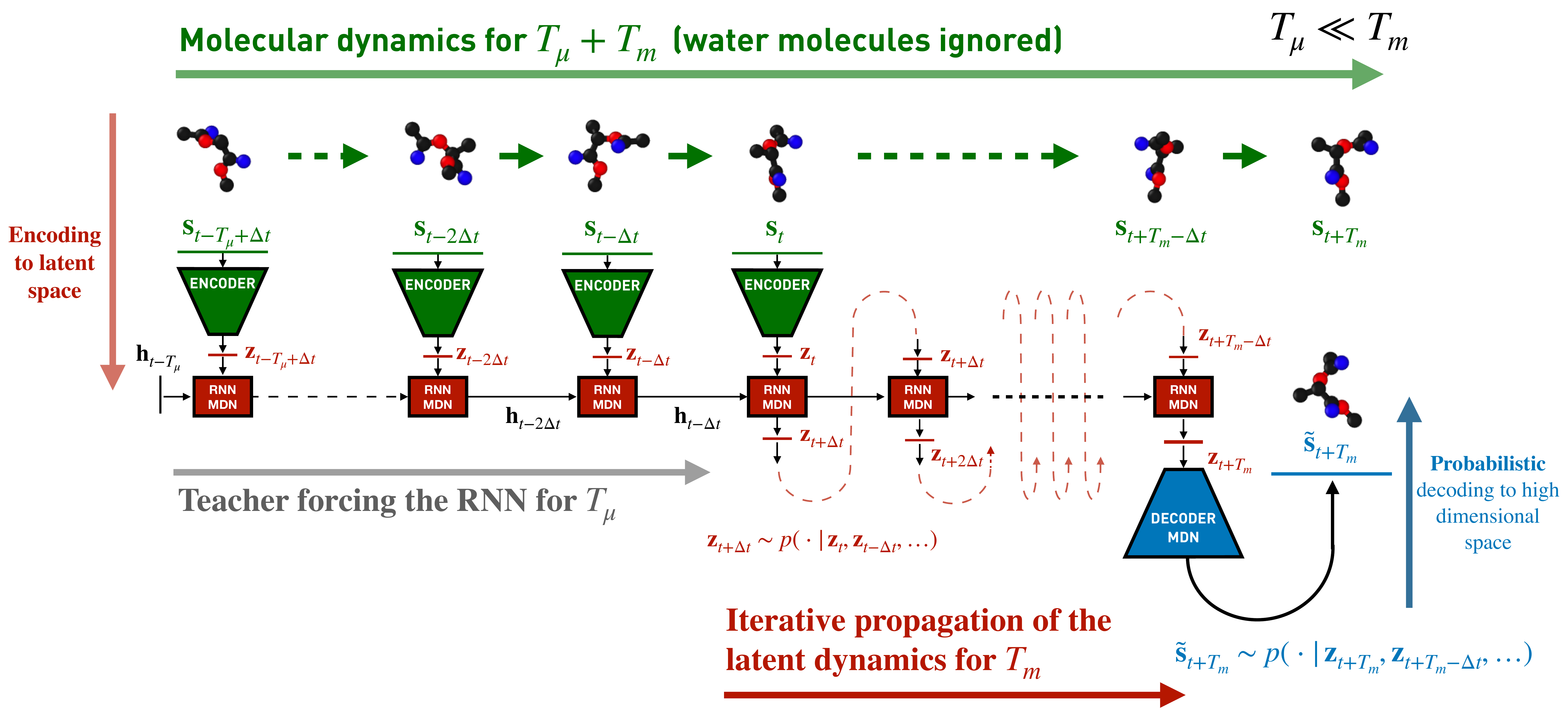}
\centering
\caption{
High dimensional (fine scale) dynamics $\mathbf{s}_t$ are simulated for a short period ($T_{\mu}$).
During this warm-up period, the state $\mathbf{s}_t$ is passed through the encoder network.
The outputs of the encoder $\mathbf{z}_t$ provide the time-series input to the LSTM, allowing for the update of its hidden state $\mathbf{h}_t$, thus capturing non-Markovian effects.
The output of the LSTM is a parametrization of the probabilistic non-Markovian latent dynamics $p(\mathbf{z}_t | \mathbf{h}_{t})$.
Starting from the last latent state $\mathbf{z}_t$, the LSTM iteratively samples $p(\mathbf{z}_t | \mathbf{h}_{t})$ and propagates the low order latent dynamics up to a total horizon of $T_m$ time units, with $T_m>T_{\mu}$.
The LED decoder may be utilized at any desired time-scale to map the latent state $\mathbf{z}_t$ back to a high-dimensional representation $\mathbf{s}_t \sim p(\cdot | \mathbf{z}_t, \mathbf{z}_{t-\Delta t}, \dots)$.
Propagation in the low order space unraveled by LED is orders of magnitude cheaper than evolving the high dimensional system based on first principles (molecular dynamics/density functional theory, etc.).
}
\label{fig:figures-led:led}
\end{figure*}

The LED framework \cite{vlachas2020learning}  for molecular systems is founded on the equation-free framework (EFF)~\cite{kevrekidis2003equation}.
It addresses the key bottlenecks of EFF namely, the coarse to fine mapping and the evolution of the latent space using an MDN-AE and an MDN-LSTM respectively. 
An illustration of the LED framework is given in Figure~\ref{fig:figures-led:led}.

In the following, the state of a molecule at  time $t$ is described by a high dimensional vector $\boldsymbol{s}_t \in \Omega  \subseteq \mathbb{R}^{d_{\boldsymbol{s}}}$, where $d_{\boldsymbol{s}} \in \mathbb{N}$ denotes its dimension.
The state vector can include the atom positions or their rotation/translation invariant features obtained using for example the Kabsch transform~\cite{kabsch1976solution}.
A trajectory of this system is obtained by an MD integrator and the 
state of the molecule after a timestep $\Delta t$ is described by the probability distribution function (PDF):

\begin{equation}
p( \boldsymbol{s}_{t + \Delta t}| \boldsymbol{s}_{t} ).
\label{eq:transitionalpdf}
\end{equation}
The transition distribution in Equation~\ref{eq:transitionalpdf} depends on the choice of $\Delta t$.

\paragraph*{Mixture Density Network (MDN) Autoencoder (AE):} Here the MDN-AE is utilized to identify the latent (coarse) representation and upscale it probabilistically to the high dimensional state space. MDNs\cite{Bishop2006} are neural architectures that can represent arbitrary conditional distributions. The MDN output is a parametrization of the distribution of a multivariate random variable conditioned on the input of the network. 

The latent state is computed by $\boldsymbol{z}_t = \mathcal{E} (\boldsymbol{s}_t ; \boldsymbol{w}_{ \mathcal{E}})$, where $\mathcal{E}$ is the encoder (a deep neural network) with trainable parameters $\boldsymbol{w}_{ \mathcal{E}}$ and $\boldsymbol{z}_t \in \mathbb{R}^{d_{\boldsymbol{z}}}$ with $d_{\boldsymbol{z}} \ll d_{\boldsymbol{s}}$.
Since $\boldsymbol{z}_t$ is a coarse approximation, many states can be mapped to the same $\boldsymbol{z}_t$.
As a consequence, a deterministic mapping $\boldsymbol{z}_t \to \boldsymbol{s}_t$ like the one used in Refs.~\cite{mardt2018vampnets,wehmeyer2018time} is not suitable.
Here, an MDN is employed to model the upscaling conditional PDF $p( \boldsymbol{s}_{t} | \boldsymbol{z}_{t} )$ described by the parameters $\boldsymbol{w}_{\boldsymbol{s} | \boldsymbol{z} }$.
These parameters are the outputs of the decoder with weights $\boldsymbol{w}_{ \mathcal{D}}$ and are a function of the latent representation $\boldsymbol{z}_t$, i.e.

\begin{equation}
\boldsymbol{w}_{\boldsymbol{s} | \boldsymbol{z} } (\boldsymbol{z}_t) = \mathcal{D} (\boldsymbol{z}_t ; \boldsymbol{w}_{ \mathcal{D}}).
\label{eq:mdndecoder}
\end{equation}
The state of the molecule can then be sampled from $
p( \boldsymbol{s}_t | \boldsymbol{z}_t )  \vcentcolon= p ( \boldsymbol{s}_t ; \boldsymbol{w}_{\boldsymbol{s} | \boldsymbol{z} } )
$.

Including in the state $\boldsymbol{s}_t$ the rotation/translation invariant features of the molecule under study~\cite{kabsch1976solution}, ensures that the MDN samples physically meaningful molecular configurations.
The state $\boldsymbol{s}_t$ is composed of states representing bond lengths $\boldsymbol{s}_t^{b}\in \mathbf{R}^{d_{\boldsymbol{s}}^b}$, and angles $\boldsymbol{s}_t^{a}\in \mathbf{R}^{d_{\boldsymbol{s}}^a}$.
Initially, the MD data of the bonds are scaled to $[0,1]$.
An auxiliary variable vector $\boldsymbol{v}_t \in \mathbf{R}^{d_{\boldsymbol{s}}^b}$ is defined to model the distribution of bonds.
In particular, $p( \boldsymbol{v}_{t} | \boldsymbol{z}_{t} )$ is modeled as a Gaussian mixture model with $K_{\boldsymbol{s}}$ mixture kernels as

\begin{equation}
p( \boldsymbol{v}_t | \boldsymbol{z}_t ) =
\sum_{k=1}^{K_{\boldsymbol{s}}} \pi^{k}_{ \boldsymbol{v} }(\boldsymbol{z}_t) \, \mathcal{N}  \bigg( \, \boldsymbol{\mu}_{ \boldsymbol{v} }^k(\boldsymbol{z}_t), \boldsymbol{\sigma}_{\boldsymbol{v}}^k(\boldsymbol{z}_t) \, \bigg),
\label{eq:mdnbonds}
\end{equation}
and the mapping $\boldsymbol{s}_t^b= \ln(1+\exp(\boldsymbol{v}_t))$ is used to recover the distribution of the scaled bond lengths at the output.
The functional form of the mixing coefficients $\pi^{k}_{ \boldsymbol{v} }(\boldsymbol{z}_t)$, the means $\boldsymbol{\mu}_{ \boldsymbol{v} }^k(\boldsymbol{z}_t)$, and the variances $\boldsymbol{\sigma}_{\boldsymbol{v}}^k(\boldsymbol{z}_t)$ is a deep neural network (decoder $\mathcal{D}$).
The distribution of the angles is modeled with the circular normal (von Mises) distribution, i.e.

\begin{equation}
p( \boldsymbol{s}_t^a | \boldsymbol{z}_t ) =
\sum_{k=1}^{K_{\boldsymbol{s}}} \pi^{k}_{ \boldsymbol{s}^a }(\boldsymbol{z}_t) \, \frac{ 
\exp \bigg(\boldsymbol{\nu}^{k}_{ \boldsymbol{s}^a }(\boldsymbol{z}_t) \, \cos \Big( \boldsymbol{s}_t^a - \boldsymbol{\mu}_{ \boldsymbol{s}^a }^k(\boldsymbol{z}_t) \Big) \bigg) 
}{2 \pi I_{0} \big(\boldsymbol{\nu}^{k}_{ \boldsymbol{s}^a }(\boldsymbol{z}_t) \big)},
\label{eq:mdnangles}
\end{equation}
where $I_{0}(\boldsymbol{\nu}^{k}_{ \boldsymbol{s}^a })$ is the modified Bessel function of order $0$.
Here, again the functional form of $\pi^{k}_{ \boldsymbol{s}^a }(\boldsymbol{z}_t)$, $\boldsymbol{\mu}_{ \boldsymbol{s}^a }^k(\boldsymbol{z}_t)$ and $\boldsymbol{\nu}^{k}_{ \boldsymbol{s}^a }(\boldsymbol{z}_t)$ is a deep neural network (decoder $\mathcal{D}$).

In total, the outputs of the decoder $\mathcal{D}$ that parametrize $p( \boldsymbol{s}_t | \boldsymbol{z}_t )$ are

\begin{equation}
 \boldsymbol{w}_{\boldsymbol{s} | \boldsymbol{z} }  = \{ 
 \pi^{k}_{ \boldsymbol{v} }, \boldsymbol{\mu}_{ \boldsymbol{v} }^k, \boldsymbol{\sigma}_{\boldsymbol{v}}^k,
 \pi^{k}_{ \boldsymbol{s}^a }, \boldsymbol{\mu}_{ \boldsymbol{s}^a }^k, \boldsymbol{\nu}^{k}_{ \boldsymbol{s}^a }
 \}_{k \in \{1,\dots, K_{\boldsymbol{s}} \}},
\end{equation}
which are all functions of the latent state $\boldsymbol{z}_t$, which is the decoder input.
The MDN-AE is trained to predict the mixing coefficients minimizing the data likelihood

\begin{equation}
\begin{aligned}
\boldsymbol{w}_{ \mathcal{E}}, \boldsymbol{w}_{ \mathcal{D}}
=&
\underset{ \boldsymbol{w}_{ \mathcal{E}}, \boldsymbol{w}_{ \mathcal{D}} }{\operatorname{argmax}}
\, 
p( \boldsymbol{s}_t | \boldsymbol{z}_t )
\\
=& 
\underset{ \boldsymbol{w}_{ \mathcal{E}}, \boldsymbol{w}_{ \mathcal{D}} }{\operatorname{argmax}}
\, 
p \big( \boldsymbol{s}_t ;  \boldsymbol{w}_{\boldsymbol{s} | \boldsymbol{z} }  \big),
\end{aligned}
\end{equation}
where $\boldsymbol{w}_{\boldsymbol{s} | \boldsymbol{z} } = \mathcal{D} \big( \mathcal{E} (\boldsymbol{s}_t ; \boldsymbol{w}_{ \mathcal{E}} ) ; \boldsymbol{w}_{ \mathcal{D}} \big)$ is the output of the MDN-AE and $\boldsymbol{s}_t$ are the MD data.
The details of the training procedure can be found in~\cite{vlachas2020backpropagation}.

\paragraph*{Long Short-Term Memory recurrent neural network (LSTM)}
The latent dynamics may be characterized by non-Markovian effects, i.e.
$$p(\boldsymbol{z}_{t+\Delta t} | \boldsymbol{z}_{t}, \boldsymbol{z}_{t-\Delta t}, \dots),$$ due to the neglected degrees of freedom (solvent) or the selection of a relatively small time-lag $\Delta t$.

\begin{figure}[t!]
\centering
\includegraphics[width=0.4\textwidth,clip]{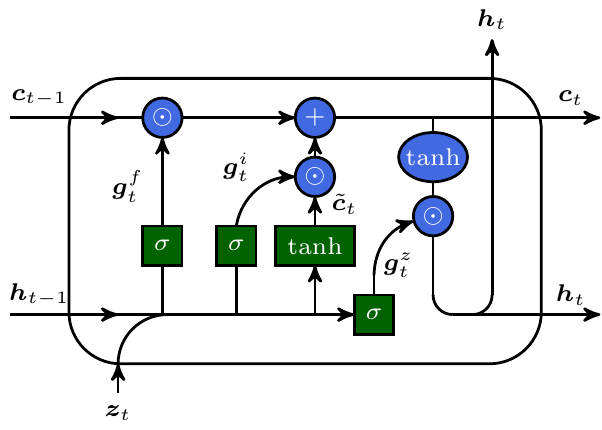}
\caption{
Information flow in an LSTM cell.
}
\label{fig:lstm}
\end{figure}

Here the LSTM cell architecture~\cite{hochreiter1997long} is utilized to evolve the nonlinear and non-Markovian latent dynamics.
The propagation in the LSTM is given by:
\begin{equation}
\boldsymbol{h}_{t}, \boldsymbol{c}_{t} =
\mathcal{R}
\big(
\boldsymbol{z}_t, \boldsymbol{h}_{t-\Delta t}, \boldsymbol{c}_{t-\Delta t}
; 
\boldsymbol{w}_{ \mathcal{R}}
\big)
,
\label{eq:lstmrec}
\end{equation}
where the hidden-to-hidden recurrent mapping $\mathcal{R}$ takes the form

\begin{equation}
\begin{aligned}
\boldsymbol{g}^f_t &= \sigma_f \big(W_f [\boldsymbol{h}_{t-\Delta t}, \boldsymbol{z}_t ] + \boldsymbol{b}_f\big)  \\
\boldsymbol{g}^{i}_t &= \sigma_i \big( W_i [\boldsymbol{h}_{t-\Delta t}, \boldsymbol{z}_t ] +\boldsymbol{b}_i \big) \\
\tilde{\boldsymbol{c}}_t &=\tanh \big( W_c [\boldsymbol{h}_{t-\Delta t}, \boldsymbol{z}_t ] +\boldsymbol{b}_c \big)  \\
\boldsymbol{c}_t &=\boldsymbol{g}^f_t \odot \boldsymbol{c}_{t-\Delta t} + \boldsymbol{g}^{i}_t \odot \tilde{\boldsymbol{c}}_t   \\
\boldsymbol{g}^{\boldsymbol{z}}_t &= \sigma_h \big( W_h [\boldsymbol{h}_{t-\Delta t}, \boldsymbol{z}_t ] + \boldsymbol{b}_h \big)  \\
\boldsymbol{h}_t &=  \boldsymbol{g}^{\boldsymbol{z}}_t \odot  \tanh(\boldsymbol{c}_t),
\end{aligned}
\label{eq:lstmequations}
\end{equation}
where $\boldsymbol{g}^f_t, \boldsymbol{g}^{i}_t, \boldsymbol{g}^{\boldsymbol{z}}_t \in \mathbb{R}^{d_{\boldsymbol{h}}}$,
are the gate vector signals (forget, input and output gates),
$\boldsymbol{z}_{t} \in \mathbb{R}^{d_{\boldsymbol{z}}}$ is the latent input at time $t$,
$\boldsymbol{h}_{t} \in \mathbb{R}^{d_{\boldsymbol{h}}}$ is the hidden state,
$\boldsymbol{c}_{t}\in \mathbb{R}^{d_{\boldsymbol{h}}}$ is the cell state,
while $W_f$, $W_i$, $W_c, W_h$ $\in \mathbb{R}^{d_{\boldsymbol{h}} \times (d_{\boldsymbol{h}}+d_{\boldsymbol{z}})}$,
are weight matrices and $\boldsymbol{b}_f, \boldsymbol{b}_i, \boldsymbol{b}_c, \boldsymbol{b}_h \in \mathbb{R}^{d_{\boldsymbol{h}}}$  biases.
The symbol $\odot$ denotes the element-wise product.
The activation functions $\sigma_f$, $\sigma_i$ and $\sigma_h$ are sigmoids.
The dimension of the hidden state $d_{\boldsymbol{h}}$ (number of hidden units) controls the capacity of the cell to encode history information.
The set of trainable parameters of the LSTM are

\begin{equation}
\boldsymbol{w}_{ \mathcal{R}}
= 
\{
\boldsymbol{b}_f, \boldsymbol{b}_i, \boldsymbol{b}_c, \boldsymbol{b}_h,
W_f, W_i, W_c, W_h
\}
.
\end{equation}
An illustration of the information flow in a LSTM cell is given in Figure~\ref{fig:lstm}.
The cell state can encode the history of the latent state evolution and capture non-Markovian effects.

\paragraph*{Mixture Density LSTM Network (MDN-LSTM)}
The LSTM captures the history of the latent state and the non-Markovian latent transition dynamics are expressed as:

\begin{equation}
p(\boldsymbol{z}_{t+\Delta t} | \boldsymbol{z}_{t}, \boldsymbol{z}_{t-\Delta t}, \dots) 
= p ( \boldsymbol{z}_{t+\Delta t} | \boldsymbol{h}_{t}),
\label{eq:latentsampling}
\end{equation}
where $\boldsymbol{h}_{t}$ given in Equation~\ref{eq:lstmrec}.
A second MDN is used to model the conditional distribution $p ( \boldsymbol{z}_{t+\Delta t} | \boldsymbol{h}_{t})$ of the latent transition dynamics.
This MDN is conditioned on the hidden state of the LSTM $\boldsymbol{h}_{t}$ and implicitly conditioned on the history, i.e., $p(\boldsymbol{z}_{t+\Delta t} | \boldsymbol{z}_{t}, \boldsymbol{z}_{t-\Delta t}, \dots) 
\vcentcolon= p ( \boldsymbol{z}_{t+\Delta t} ; \boldsymbol{w}_{\boldsymbol{z} | \boldsymbol{h}} )$, so it can capture non-Markovian dynamics.
The distribution $p ( \boldsymbol{z}_{t+\Delta t} | \boldsymbol{h}_{t})$ is modeled as a Gaussian mixture with $K_{\boldsymbol{z}}$ mixture kernels

\begin{equation}
p( \boldsymbol{z}_{t + \Delta t} | \boldsymbol{h}_{t} ) =
\sum_{k=1}^{K_{\boldsymbol{z}}} \pi^{k}_{ \boldsymbol{z} }(\boldsymbol{h}_{t}) \, \mathcal{N}  \bigg( \, \boldsymbol{\mu}_{ \boldsymbol{z} }^k(\boldsymbol{h}_{t}), \boldsymbol{\sigma}_{\boldsymbol{z}}^k(\boldsymbol{h}_{t}) \, \bigg),
\label{eq:mdnlatent}
\end{equation}
with parameters $\boldsymbol{w}_{\boldsymbol{z} | \boldsymbol{h}}$ given by

\begin{equation}
\boldsymbol{w}_{\boldsymbol{z} | \boldsymbol{h}} (\boldsymbol{h}_t)  = \{ 
 \pi^{k}_{ \boldsymbol{z} } (\boldsymbol{h}_t), \boldsymbol{\mu}_{ \boldsymbol{z} }^k(\boldsymbol{h}_t), \boldsymbol{\sigma}_{\boldsymbol{z}}^k(\boldsymbol{h}_t)
 \},
\end{equation}
that are a function of $\boldsymbol{h}_{t}$.
These parameters are the outputs of the neural network $\mathcal{Z}(\boldsymbol{h}_{t} ; \boldsymbol{w}_{ \mathcal{Z}})$, with trainable weights $\boldsymbol{w}_{ \mathcal{Z}}$, and are a function of the hidden state, i.e.

\begin{equation}
\begin{aligned}
& p ( \boldsymbol{z}_{t+\Delta t} | \boldsymbol{h}_{t}) \vcentcolon= p ( \boldsymbol{z}_{t+\Delta t} ; \boldsymbol{w}_{\boldsymbol{z} | \boldsymbol{h}} ), \\
&  \boldsymbol{w}_{\boldsymbol{z} | \boldsymbol{h}} (\boldsymbol{h}_{t}) = \mathcal{Z}(\boldsymbol{h}_{t} ; \boldsymbol{w}_{ \mathcal{Z}})
.
\label{eq:wzch}
\end{aligned}
\end{equation}

The weights of the LSTM $\boldsymbol{w}_{ \mathcal{R}}$ and the latent MDN $\boldsymbol{w}_{ \mathcal{Z}}$ are trained to output the parameters $\boldsymbol{w}_{\boldsymbol{z} | \boldsymbol{h}}$ that maximize the likelihood of the latent evolution

\begin{equation}
\begin{aligned}
 \boldsymbol{w}_{ \mathcal{R}}, \boldsymbol{w}_{ \mathcal{Z}}
=
& \, \underset{ \boldsymbol{w}_{ \mathcal{R}}, \boldsymbol{w}_{ \mathcal{Z}} }{\operatorname{argmax}}
\, p \big( \boldsymbol{z}_{t + \Delta t} ;  \boldsymbol{w}_{\boldsymbol{z} | \boldsymbol{h}} 
\big),
\label{eq:wrwz}
\end{aligned}
\end{equation}
where $\boldsymbol{w}_{\boldsymbol{z} | \boldsymbol{h}}$ is defined in Equation~\ref{eq:wzch}, and $\boldsymbol{h}_t$ appearing in Equation~\ref{eq:wzch} is defined in Equation~\ref{eq:lstmrec}.
During the training phase, the MD trajectory data $\boldsymbol{s}_t$ are provided at the input of the trained MDN-AE $\boldsymbol{z}_{t} = \mathcal{E}(\boldsymbol{s}_t ; \boldsymbol{w}_{ \mathcal{E}})$.
The encoder outputs the latent dynamics $\boldsymbol{z}_t$ that are used to update the hidden state of the LSTM and optimize its weights according to Equation~\ref{eq:wrwz}.
In contrast to the linear operator utilized in MSMs, the recurrent functional form in Equation~\ref{eq:lstmrec} can be nonlinear and incorporate memory effects, via the hidden state of the LSTM.

\paragraph*{Learned Effective Dynamics}

The LED framework can be employed to accelerate MD simulations and enable more efficient exploration of the state space and uncovering of novel protein configurations (shown in SM Section~\ref{sec:appendix:alanine:novelconf}).
The networks in LED are trained on trajectories from MD simulations in two phases.
First, the MDN-AE provides a reduced-order representation, maximizing the data likelihood (Ref.~\cite{vlachas2020learning}).
The MDN-AE is trained with Backpropagation~\cite{rumelhart1985learning} using the adaptive stochastic optimization method Adam~\cite{kingma2014adam}.
Adding a pre-training phase fitting the kernels $\boldsymbol{\mu}^k, \boldsymbol{\sigma}^k$ of the MDN-AE to the data, and fixing them during MDN-AE training led to better results.
Next, the MDN-LSTM is trained to forecast the latent space dynamics (the MDN-AE weights are considered fixed) to maximize the latent data likelihood.
MDN-LSTM is trained with Backpropagation through time (BPTT)~\cite{werbos1988generalization} with Adam optimizer.

The LED propagates the computationally inexpensive dynamics on its latent space.
Starting from an initial state from a test dataset (unseen during training), a short time history $T_{\mu}$ of the state evolution is utilized to warm up the hidden state of the LED.
The MDN-LSTM is used to propagate the latent dynamics for a time horizon $T_m\gg T_{\mu}$.
High-dimensional state configurations can be recovered at any time instant by using the probabilistic decoder part of MDN-AE. We find that the LED framework can accelerate MD simulations by three orders of magnitude.


\section*{Results}
\label{sec:results}

The LED framework is tested in three systems, single-particle Langevin dynamics using the two-dimensional MBP, the Trp Cage miniprotein, and the alanine dipeptide, widely adopted as benchmarks for molecular dynamics modeling~\cite{muller1979location,nuske2014variational,mardt2018vampnets,wehmeyer2018time,sidky2020molecular}.


\paragraph*{M\"uller-Brown potential (MBP)}
\label{sec:mbp}
The Langevin dynamics of a particle in the MBP are characterized by the stochastic differential equation

\begin{equation}
m \ddot{\boldsymbol{x}}(t) = -\nabla V \big(\boldsymbol{x}(t) \big)  - \gamma  \dot{\boldsymbol{x}}(t)+ \sqrt{2 k_B T} R(t),
\end{equation}
where $\boldsymbol{x} \in \mathbb{R}^2$ is the position, $\dot{\boldsymbol{x}} $ is the velocity, $\ddot{\boldsymbol{x}} $ is the acceleration, $V(\boldsymbol{x})$ is the MBP (defined in SM Section~\ref{sec:appendix:mbp}), $k_B$ is the Boltzmann's constant, $T$ is the temperature, $\gamma$ is the damping coefficient, and $R(t)$
a delta-correlated stationary Gaussian process with zero-mean.
The nature of the dynamics is affected by the damping coefficient $\gamma$.
Low damping coefficients lead to an inertial regime.
High damping factors lead to a diffusive regime (Brownian motion) with less prominent memory effects.
Here, a low damping $\gamma=1$ is considered, along with $k_B T=15$.

The equations are integrated with the Velocity Verlet algorithm with timestep $\delta t = 10^{-2}$, starting from $96$ initial conditions randomly sampled uniformly from $\boldsymbol{x} \in [-1.5, 1.2] \times [-0.2, 2]$ till $T=10^4$, after truncating an initial transient period of $\tilde{T}=10^3$.
The data are sub-sampled keeping every 50\textsuperscript{th} data point to create the training and testing  datasets for LED.
The coarse time-step of LED is $\Delta t = 0.5$. We use 
$32$ initial conditions for training, $32$ for validation and all $96$ for testing.
LED is trained with a one-dimensional reduced order latent representation $\boldsymbol{z}_t \in \mathbb{R}$.
The reader is referred to the SM Section~\ref{sec:appendix:bmp:hyp} for further information regarding the MBP parameterization of Ref.~\cite{muller1979location} and hyperparameters of LED.

\begin{figure*}[t!]
\centering
\includegraphics[width=0.9\textwidth,clip]{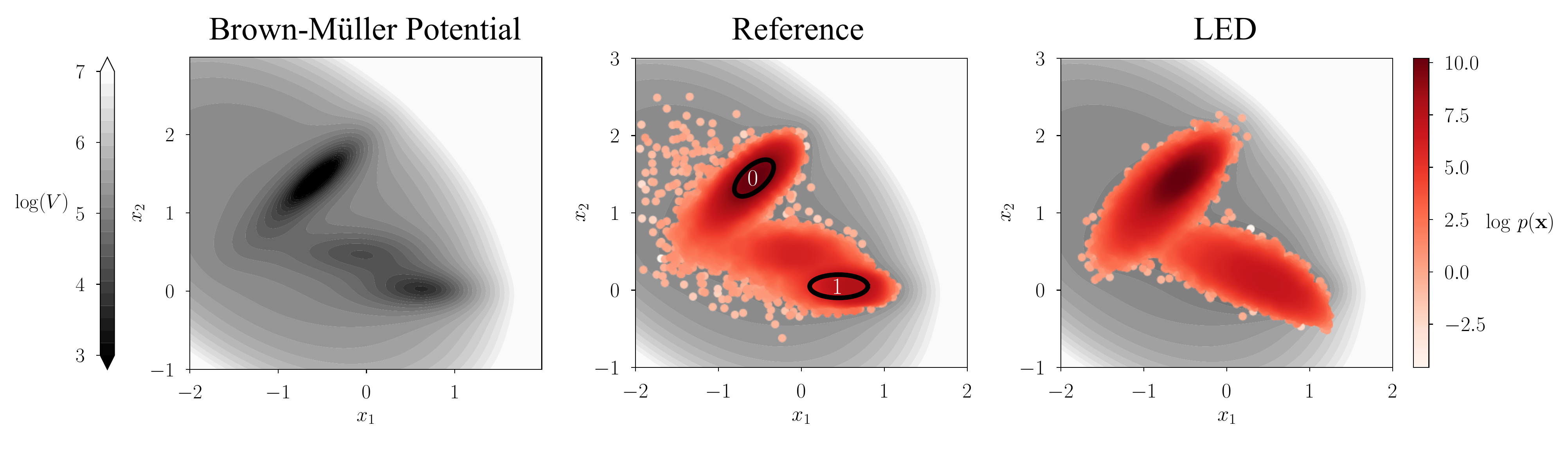}
\caption{
From left to right: the M\"uller-Brown potential, a scatter plot of the joint state distribution computed from reference data (with annotation of two long-lived metastable states), and the same scatter plot obtained by LED sampled trajectories.
}
\label{fig:bmp:bmp_density_clusters}
\end{figure*}

The MBP is shown in Figure~\ref{fig:bmp:bmp_density_clusters}, along with a density scatter plot of the joint distribution of the MBP states computed from the testing data and LED. 
The joint distribution reveals two long-lived metastable states that correspond to the low-energy regions.
The LED learns to transition probabilistically between the metastable states, mimicking the dynamics of the system and reproducing the state statistics. 

\begin{figure*}[h!]
\centering
\includegraphics[width=0.8\textwidth,clip]{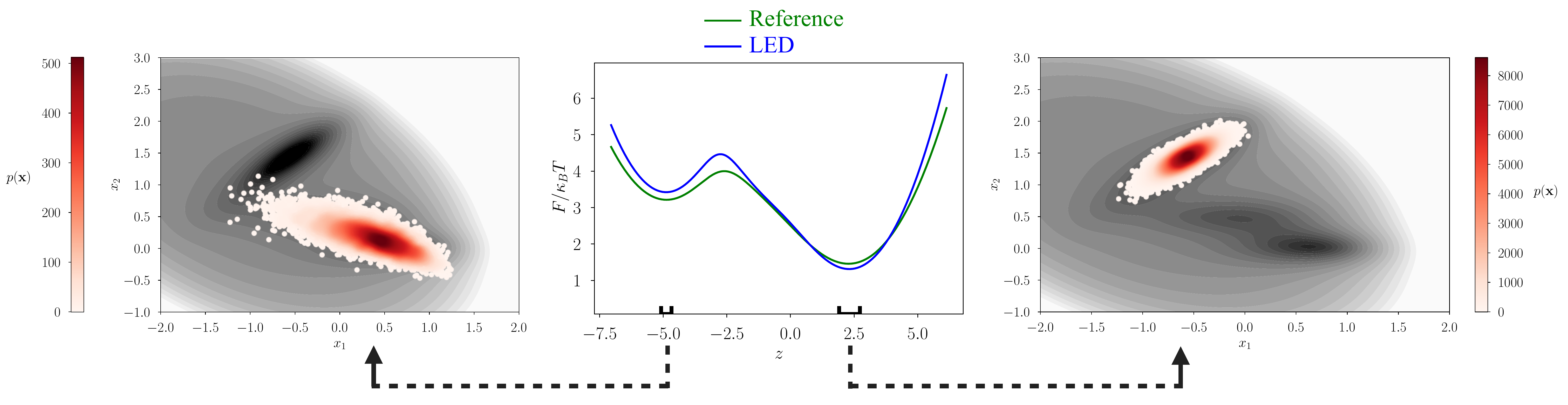}
\centering
\caption{
Middle: free energy profile projected on the latent space learned by the LED encoder, i.e., $F=-\kappa_B T \ln p(\boldsymbol z_t)$.
The free energy profile computed by LED (propagation of the latent dynamics with LED) matches closely the one from the reference data.
Quantitatively, the root mean square error is $0.74 \kappa_B T$.
LED recovers two low-energy regions that are mapped to the two long-lived metastable states (left and right) in the two-dimensional state space $\boldsymbol{s}_t \in \mathbb{R}^2$.
}
\label{fig:bmp:bmp_free_energy_clusters}
\end{figure*}

The free energy projected on the latent space, i.e., $F=-\kappa_B T \log \, p(z_t)$ is plotted in Figure~\ref{fig:bmp:bmp_free_energy_clusters}.
The free energy profile of the trajectories sampled from LED matches closely the one from the reference data with a root mean square error between the two free energy profiles of $\approx 0.74 \kappa_B T$.
LED reveals two minima in the free energy profile.
Utilizing the LED decoder, the latent states in these regions are mapped to their image in the two-dimensional state representation $\boldsymbol{s}_t \in \mathbb{R}^2$ (here corresponding to $\boldsymbol{x}_t \in \mathbb{R}^2$) in Figure~\ref{fig:bmp:bmp_free_energy_clusters}.
LED is mapping the low-energetic regions in the free energy profile to the long-lived metastable states in the two dimensional space of the MBP.

\begin{figure}[t!]
\centering
\includegraphics[width=0.45\textwidth,clip]{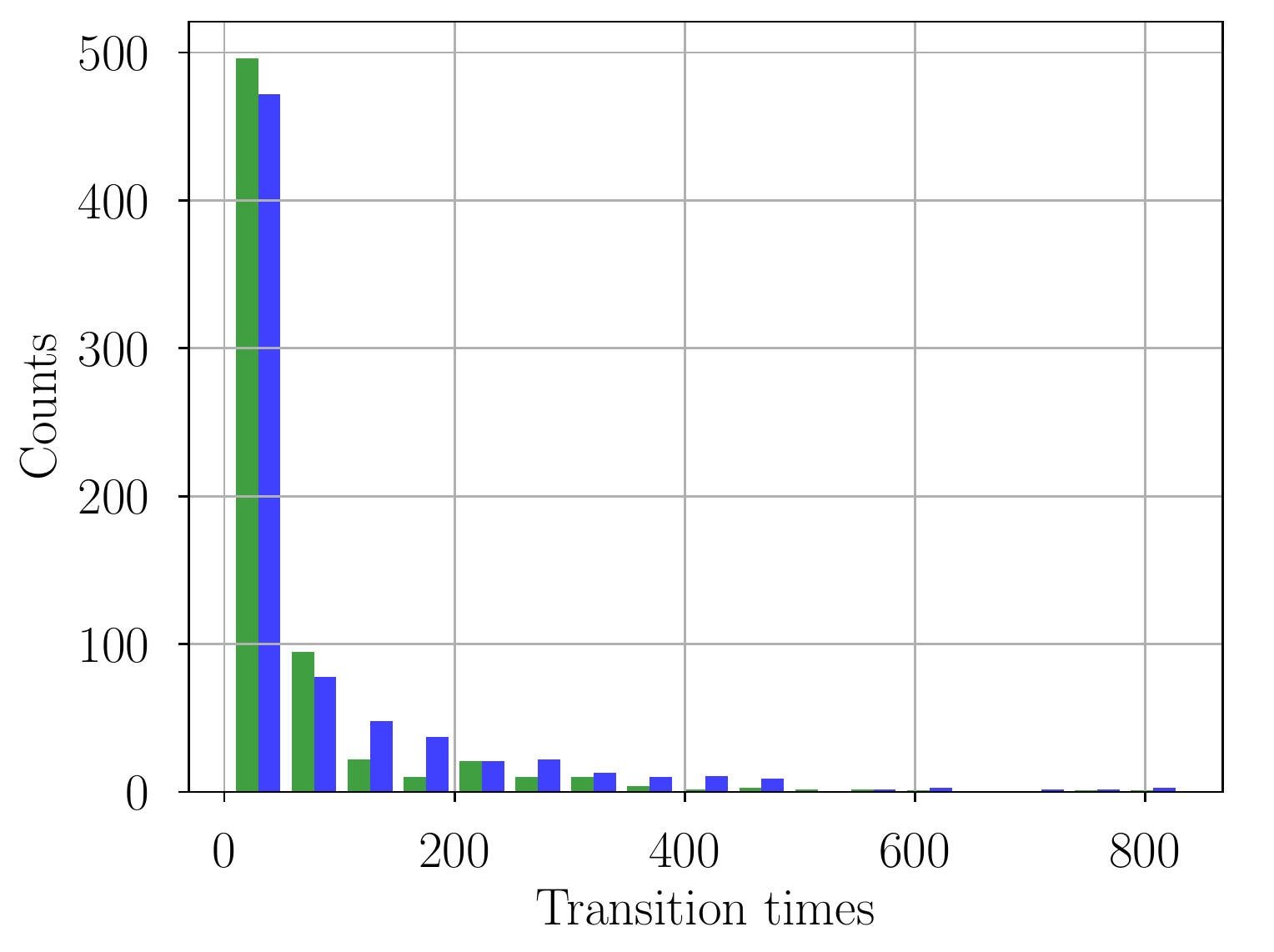}
\hfill 
\includegraphics[width=0.45\textwidth,clip]{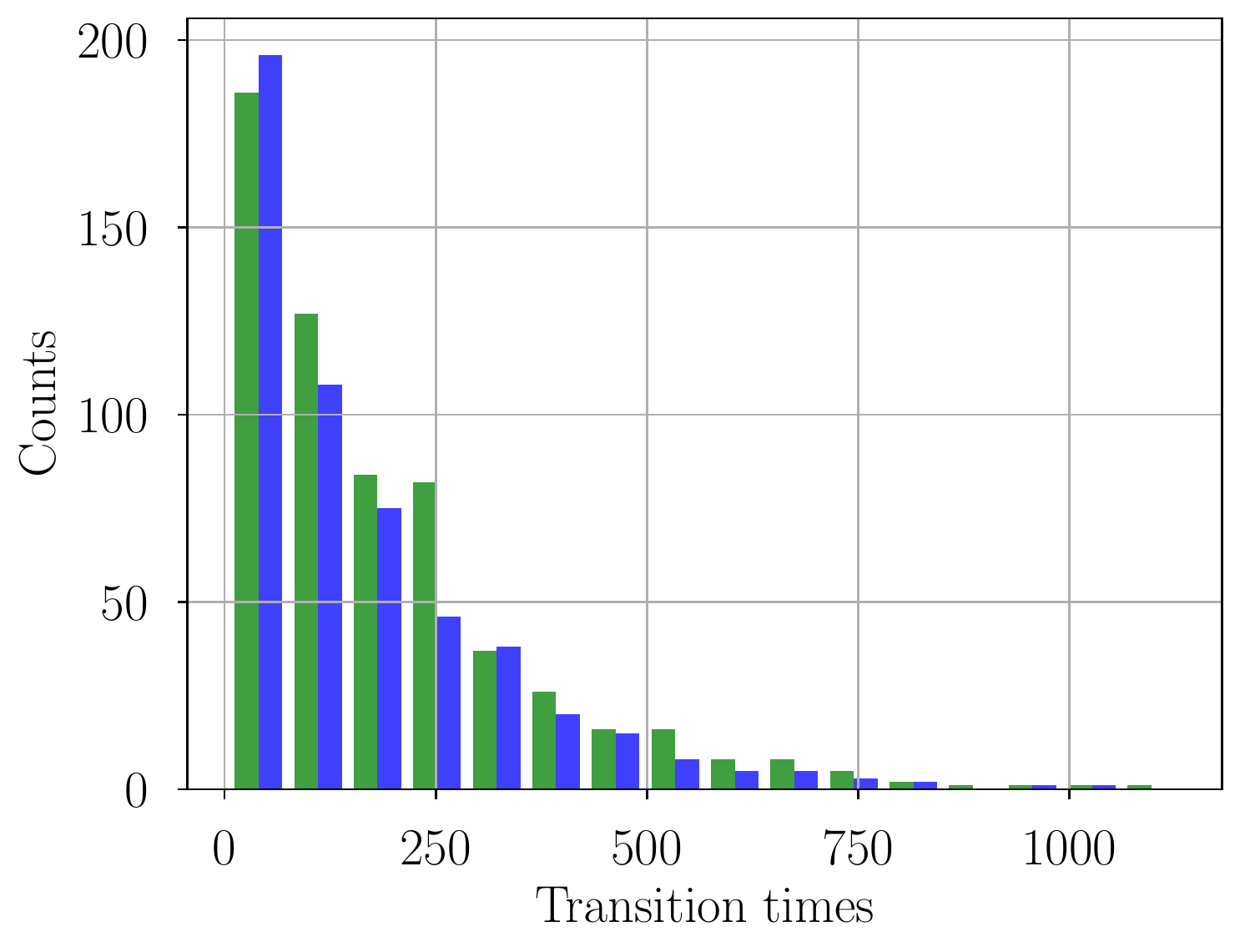}
\caption{
Distribution of the transition times learned by LED \tkey{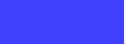}{20pt}{0.2pt}, computed from sampled trajectories, matches the original fine scale transition times of the MBP dynamics \tkey{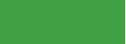}{20pt}{0.2pt}.
Left: Histogram of $T_{0 \to 1}$.
Mean $T_{0\to1}$ of MD trajectories is $61$, mean $T_{0\to1}=91$ for LED.
Right:  Histogram of $T_{1 \to 0}$.
Mean $T_{1 \to 0}$ of MD trajectories is $188$, mean $T_{1 \to 0}=164$ for LED.
LED has learned to propagate the effective dynamics (a one dimensional latent state $\boldsymbol{z}$) and capture the non-Markovian effects.
}
\label{fig:bmp:iterative_latent_forecasting_test_MBP_trans_time}
\end{figure}

Next, we evaluate the LED framework in reproducing the transition times between the long-lived states.
In LED, metastable states can be defined either on the reduced order latent space $\boldsymbol{z}_t\in \mathbb{R}$ or the state space $\boldsymbol{s}_t \in \mathbb{R}^2$ (as the decoder can map any latent state to a state space).
In the following, two metastable states are defined as ellipses on the state space depicted in Figure~\ref{fig:bmp:bmp_density_clusters} (defined in the SM Section~\ref{sec:appendix:bmp:metastablestatescenters}).
The time-scales will vary depending on the definition of the metastable states in the phase space.
The distribution of transition times computed from LED trajectories is compared with the transition time distribution from the test data in Figure~\ref{fig:bmp:iterative_latent_forecasting_test_MBP_trans_time}.
LED captures qualitatively the transition time distributions and the mean values are close to each other.
In SM Section~\ref{sec:appendix:bmp:ledlatentspacetimescale}, we also report the transition times obtained with metastable states definition on the latent space.
This approach has the benefit of not requiring the prior knowledge about the metastable states in the state space.
In conclusion, LED is capturing the joint state distribution on the MBP, and matching the timescales of the system.

\paragraph*{Trp Cage}
\label{sec:trp}

The Trp-cage is  considered a prototypical miniprotein for the  study of protein folding~\cite{sidky2020molecular}.
The protein is simulated with MD~\cite{guzman2019espressopp} with a time-step $\delta t=1\text{fs}$, up to a total time of $T=100\text{ns}$.
The data is sub-sampled at $\Delta t=0.1\text{ps}$, creating a trajectory with $N=10^6$ samples.
The data is divided into $248$ sequences of $4000$ samples ($T=400\text{ps}$ each).
The first $96$ sequences are used for training (corresponding to $38.4\text{ns}$), the next $96$ sequences for validation, while all the data is used for testing. 

The protein positions are transformed into rototranslational invariant features (internal coordinates), composed of bonds, angles, and dihedral angles, leading to a state with dimension $d_{\boldsymbol{s} }=456$.
LED is trained with a latent space $\boldsymbol{z}_t \in \mathbb{R}^2$, i.e., $d_{\boldsymbol{z}}=2$.
LED is tested by starting from the initial condition in each of the $248$ test sequences, iteratively propagating the latent space to forecast $T=400\text{ps}$.
For more information on the hyperparameters of LED, refer to the SM Section~\ref{sec:appendix:trp:hyp}.

\begin{figure}[h!]
\centering
\includegraphics[width=0.8\textwidth]{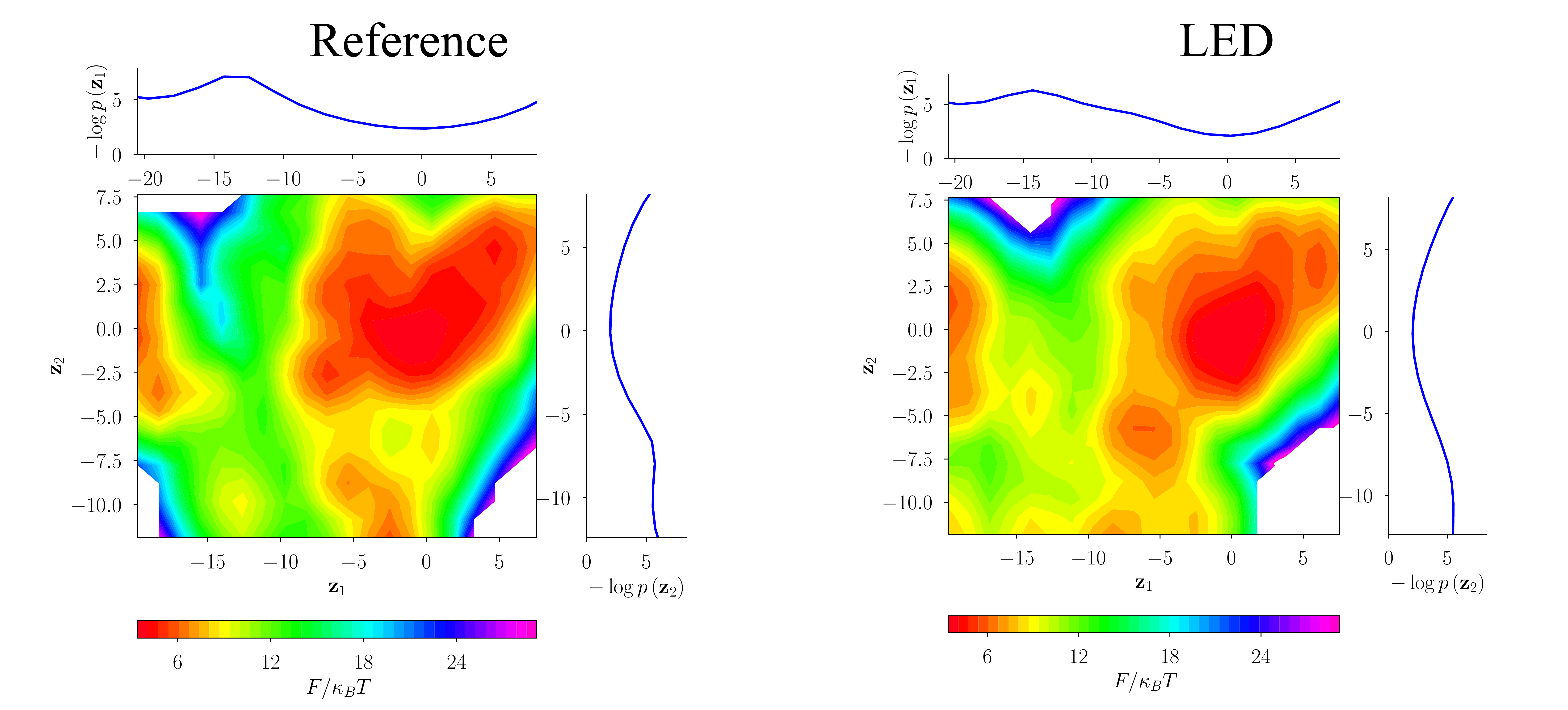}
\caption{
Free energy projection on the latent space $F=-\kappa_B T \log p(\boldsymbol{z}_t)$, with $\boldsymbol{z}_t \in \mathbb{R}^2$.
Left: MD data projected to the LED latent space.
Right: the free energy of trajectories sampled from LED.
LED is capturing the free energy profile.
}
\label{fig:trp:latent_dynamics_free_energy_test}
\end{figure}

The projection of MD trajectory data to LED latent space is illustrated in Figure~\ref{fig:trp:latent_dynamics_free_energy_test} left, in the form of the free energy, i.e., $F=-\kappa_B T \log p(\boldsymbol{z}_t)$, with $\boldsymbol{z}_t = (\mathbf{z}_1,\mathbf{z}_2)^T  \in \mathbb{R}^2$.
The free energy on the latent space computed from trajectories sampled from LED is given in Figure~\ref{fig:trp:latent_dynamics_free_energy_test} on the right.
LED successfully captures the three metastable states of the Trp Cage miniprotein, while being three orders of magnitude faster compared to the MD solver.
Quantitatively, the two profiles agree up to an error margin of approximately $22.5 \kappa_B T$.
The SM Section~\ref{sec:appendix:trp} provides additional results on the agreement of the marginal state distributions, and realistic samples of the protein configuration sampled from LED.


\paragraph*{Alanine dipeptide}
\label{sec:alanine}

The alanine dipeptide is often used as the testing ground for enhanced sampling methods~\cite{MacCarthy:2017}.
LED is evaluated in learning and propagating the dynamics of alanine dipeptide in water.
The molecule is simulated with MD~\cite{guzman2019espressopp} with a time-step $\delta t=1\text{fs}$, up to $T=100\text{ns}$.
We subsample the data, keeping every $100$\textsuperscript{th} datapoint, creating a trajectory with $N=10^6$ samples.
LED is thus operating on a timescale $\Delta t=0.1\text{ps}$.
The data is divided into $248$ sequences of $4000$ samples ($T=400\text{ps}$ each).
The first $96$ sequences are used for training (corresponding to $38.4\text{ns}$), the next $96$ sequences for validation, while all the data is used for testing. 
LED is tested by starting from the initial condition in each of the $248$ test sequences, iteratively propagating the latent space to forecast $T=400\text{ps}$.

The dipeptide positions are transformed into rototranslational invariant features (internal coordinates), composed of bonds, angles, and dihedral angles, leading to a state with dimension $d_{\boldsymbol{s} }=24$.
In order to demonstrate that LED can uncover the dynamics in a drastically reduced order latent space, the dimension of the later is set to one $d_{\boldsymbol{z}}=1$, i.e. $\boldsymbol{z}_t \in \mathbb{R}$.
For more information on the hyperparameters of LED, refer to the SM Section~\ref{sec:appendix:alanine:hyp}.

\begin{figure}[h!]
\centering
\includegraphics[width=0.9\textwidth]{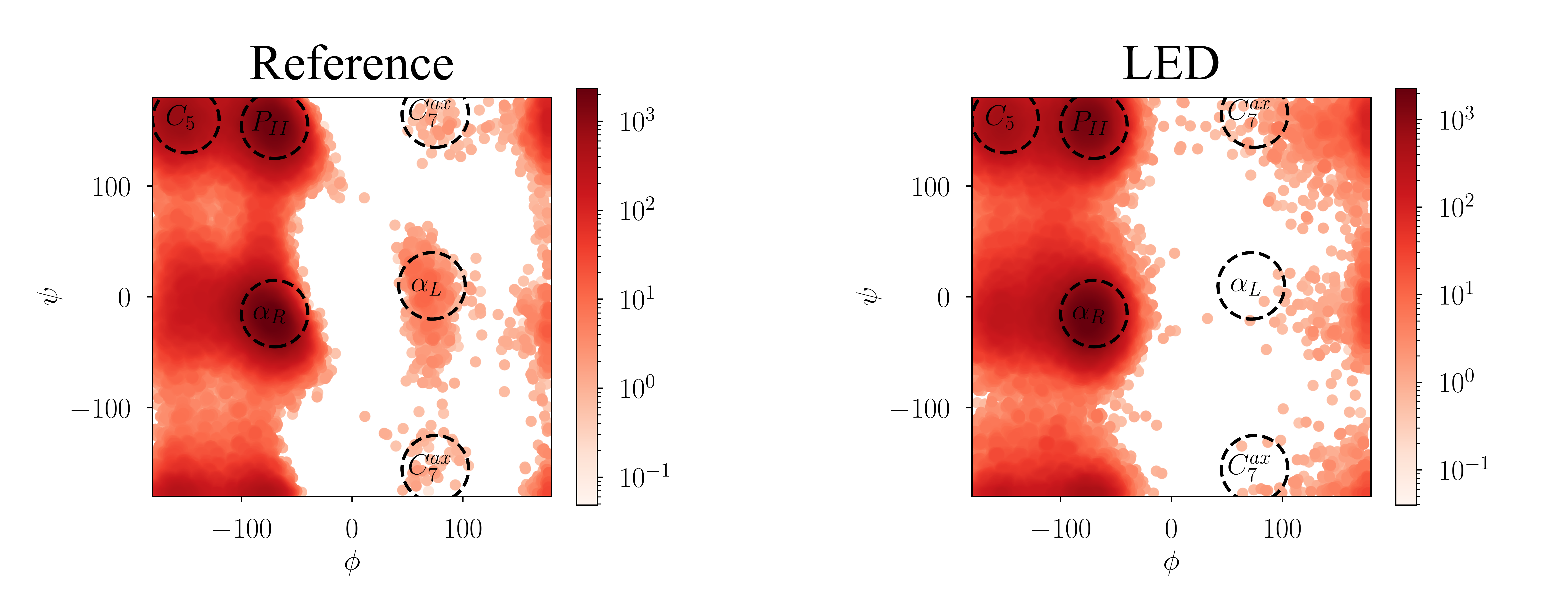}
\centering
\caption{
Ramachandran plot of the alanine dipeptide, i.e., space spanned by two backbone dihedral angles $(\phi, \psi)$.
Scatter plots are colored based on the joint density of $(\phi, \psi)$.
Left: test data.
Right: LED trajectories.
We observe five energetically favorable metastable states denoted with $\{C5, P_{II},\alpha_R, \alpha_L, C_7^{ax} \}$.
LED captures the three dominant metastable states $\{C5, P_{II},\alpha_R,  \}$.
The states $\{ \alpha_L, C_7^{ax} \}$ are rarely observed in the training data.
}
\label{fig:alanine:RamachandranPlot_badNOrotTr_waterNVT}
\end{figure}

The metastable states of the dynamics are represented in terms of the energetically favored regions in the state space of two backbone dihedral angles, $\phi$ and $\psi$, i.e., the Ramachandran space~\cite{ramachandran1963stereochemistry} plotted in Figure~\ref{fig:alanine:RamachandranPlot_badNOrotTr_waterNVT}.
Specifically, previous works consider five low-energy clusters, i.e., $\{C5, P_{II},\alpha_R, \alpha_L, C_7^{ax} \}$.
The trained LED is qualitatively reproducing the density in the Ramachandran plot in Figure~\ref{fig:alanine:RamachandranPlot_badNOrotTr_waterNVT} qualitatively, identifying the three dominant low-energy metastable states $\{C5, P_{II},\alpha_R \}$.
LED, however, fails to capture the state density on the less frequently observed states in the training data $\{\alpha_L, C_7^{ax}\}$.
The marginal distributions of the trajectories generated by LED match the ground-truth ones (MD data) closely, as depicted in Figure~\ref{fig:alanine:iterative_latent_forecasting_state_dist_bar} in the SM Section~\ref{sec:appendix:alanine}.
Even though LED is propagating a one-dimensional latent state, it can reproduce the statistics while being three orders of magnitude faster than the MD solver.
%

\begin{figure*}[h]
\centering
\includegraphics[width=1.0\textwidth]{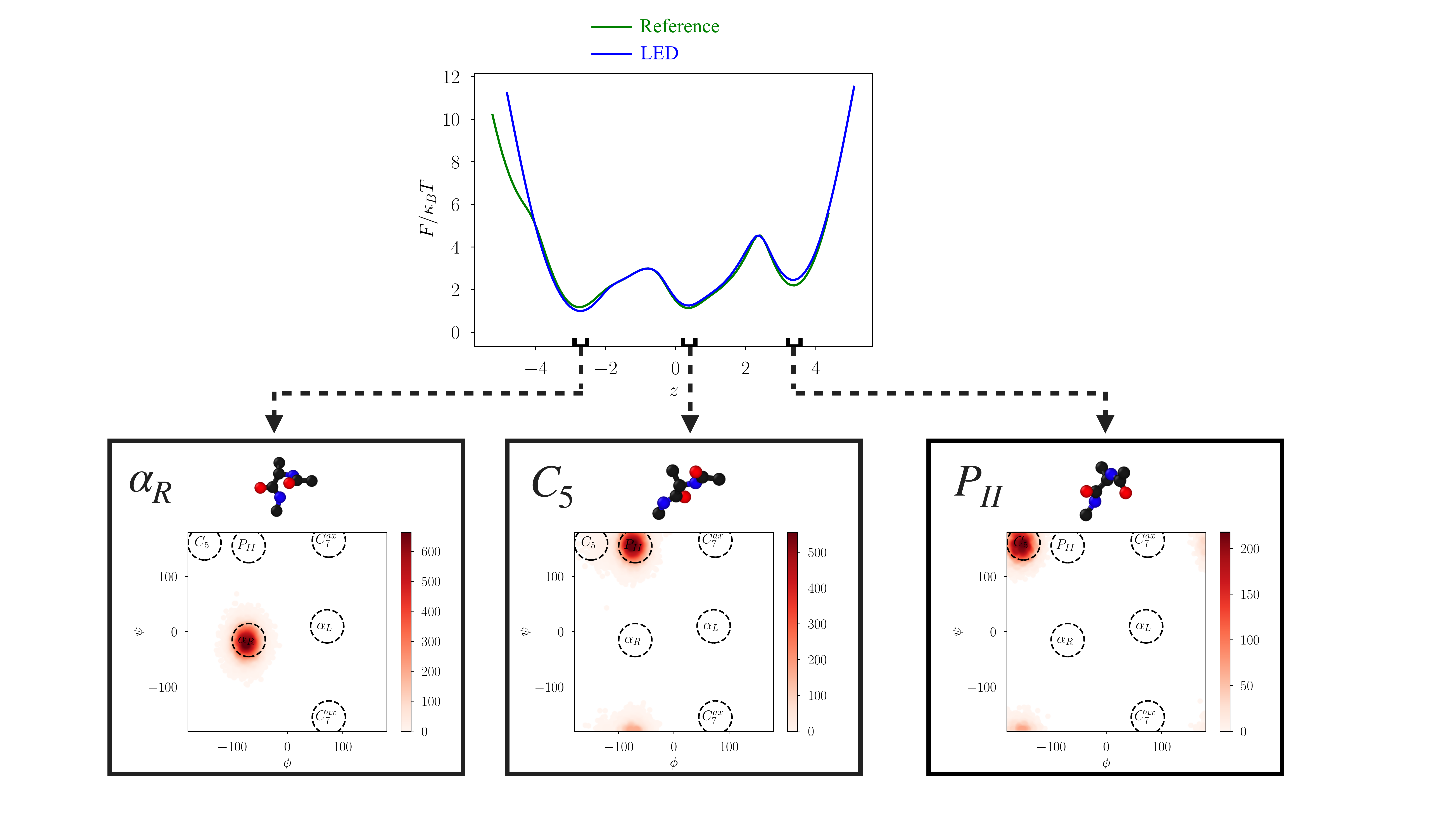}
\centering
\caption{
Plot of the free energy profile projected on the latent state learned by the LED, i.e., $F=-\kappa_B T \ln \, p(\boldsymbol{z}_t)$.
The latent free energy profile of MD trajectories is compared with the latent free energy profile of trajectories sampled from LED.
The two profiles agree up to a root mean square error of $2 \kappa_B T$.
Utilizing the LED decoder, the low-energy regions in the latent space (close to the minima) can be mapped to the corresponding protein configurations and metastable states in the Ramachandran plot.
The LED uncovers the three dominant metastable states $\{C_5, P_{II}, \alpha_R \}$ in the free energy surface (minima).
The LED captures the free energy profile and the dominant metastable states while being computationally three orders of magnitude cheaper than MD.
}
\label{fig:alanine:alanine-clusters}
\end{figure*}

The free energy is projected to the latent space, i.e., $F=-\kappa_B T \ln(p(\boldsymbol{z}_t))$, and plotted in Figure~\ref{fig:alanine:alanine-clusters}.
The free energy projection computed from MD trajectories (test data) is compared with the one computed from trajectories sampled from LED.
The two free energy profiles agree up to a root mean square error of $2 \kappa_B T$. 
Note that LED unravels three dominant minima in the latent space.
These low-energy regions correspond to metastable states of the dynamics.

The Ramachandran space $(\phi,\psi)$ is frequently used to describe the long-term behavior and metastable states of the system~\cite{wehmeyer2018time,trendelkamp2016efficient}.
The latent encoding of the LED is evaluated based on the mapping between the latent space and the Ramachandran space.
Utilizing the MDN decoder, the LED can map the latent state $\boldsymbol{z}$ to the respective rototranslational invariant features (bonds and angles) and regions in the Ramachandran plot.
As illustrated in Figure~\ref{fig:alanine:alanine-clusters}, the LED is mapping the three low-energy regions in the latent space to the three dominant metastable states in the Ramachandran plot $\{C_5, P_{II}, \alpha_R \}$.
Even though LED is propagating a reduced-order one-dimensional latent state, it captures the stochastic dynamics of the system.

\begin{figure*}[t]
\includegraphics[width=0.48\textwidth]{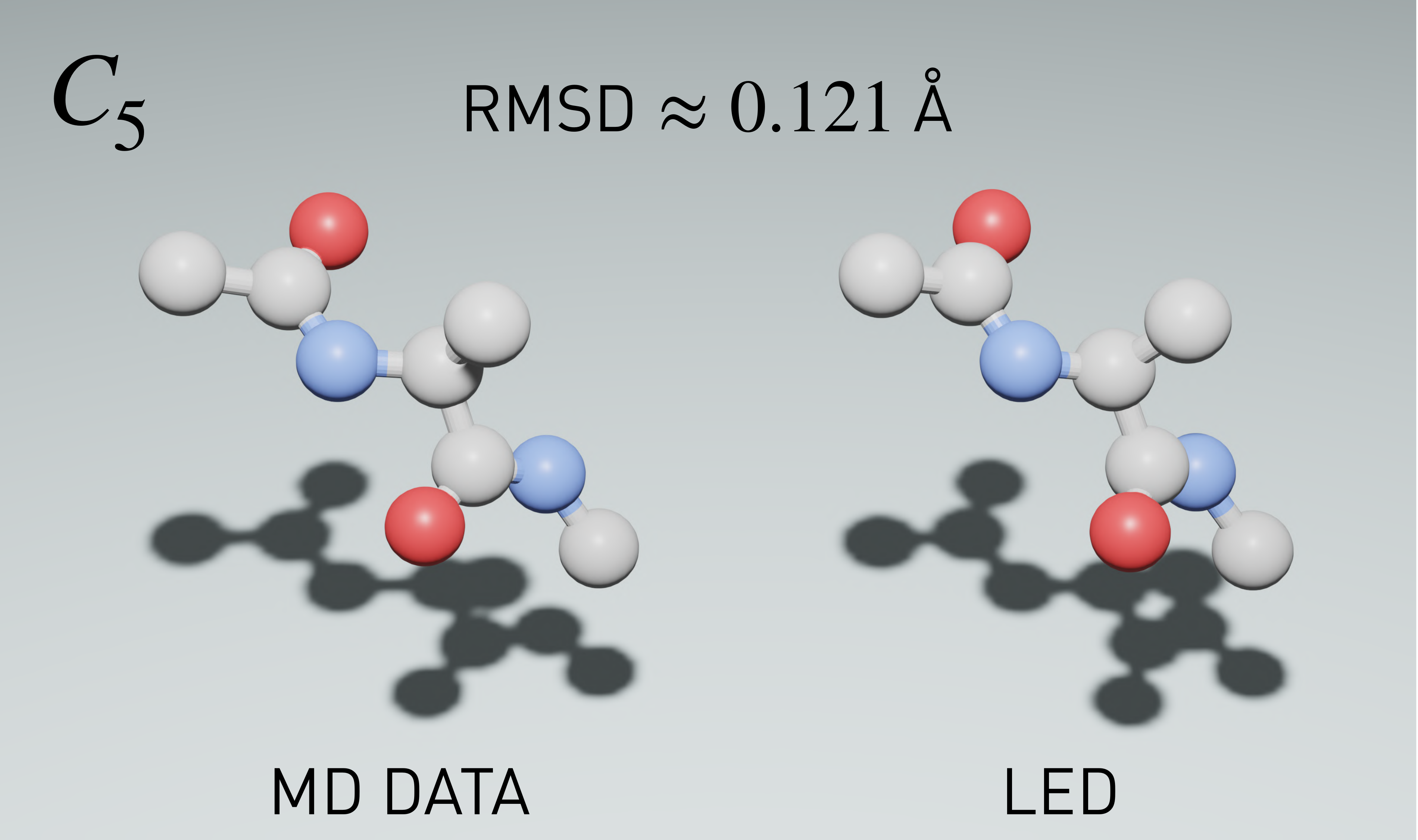}
\hfill
\includegraphics[width=0.48\textwidth]{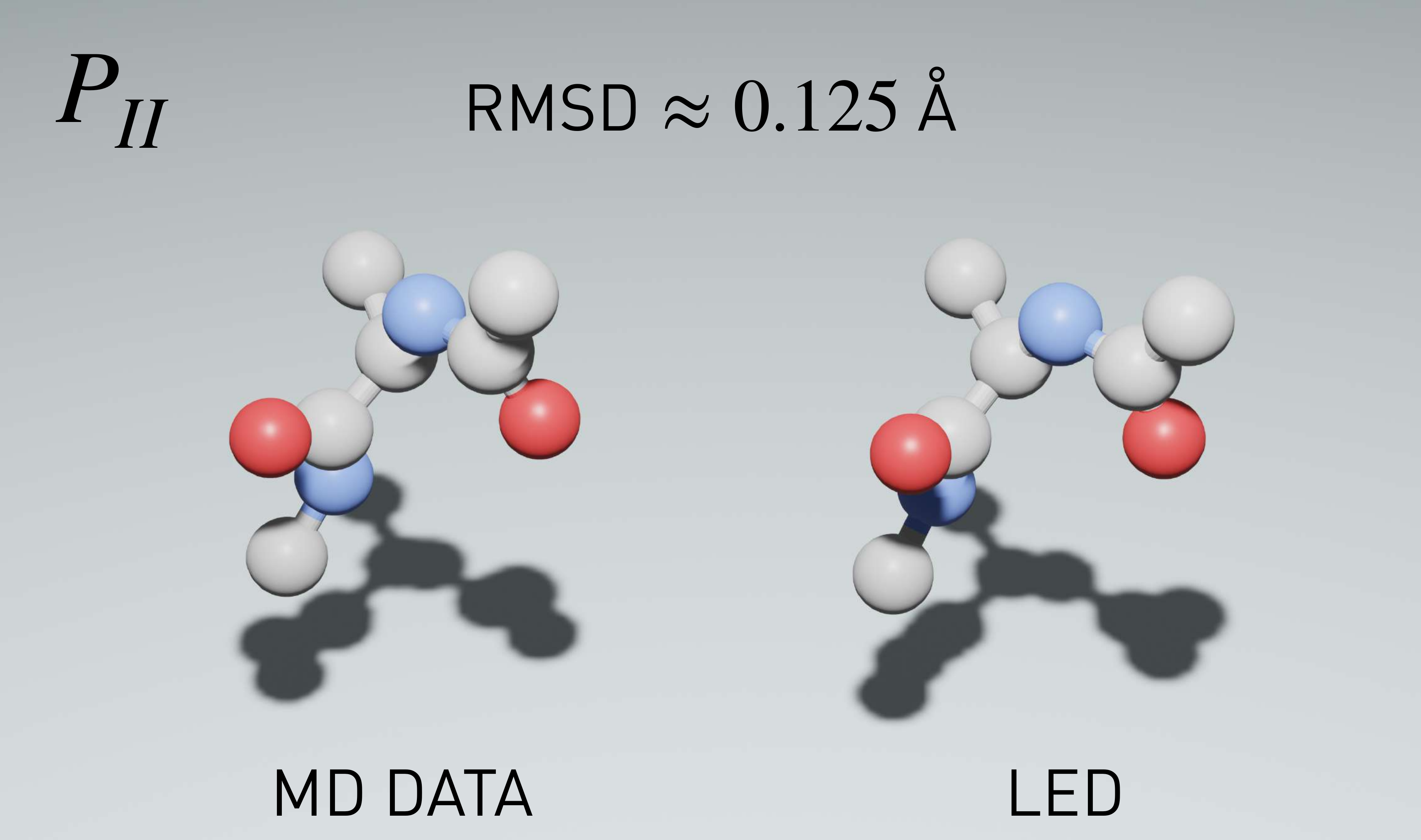}
\includegraphics[width=0.48\textwidth]{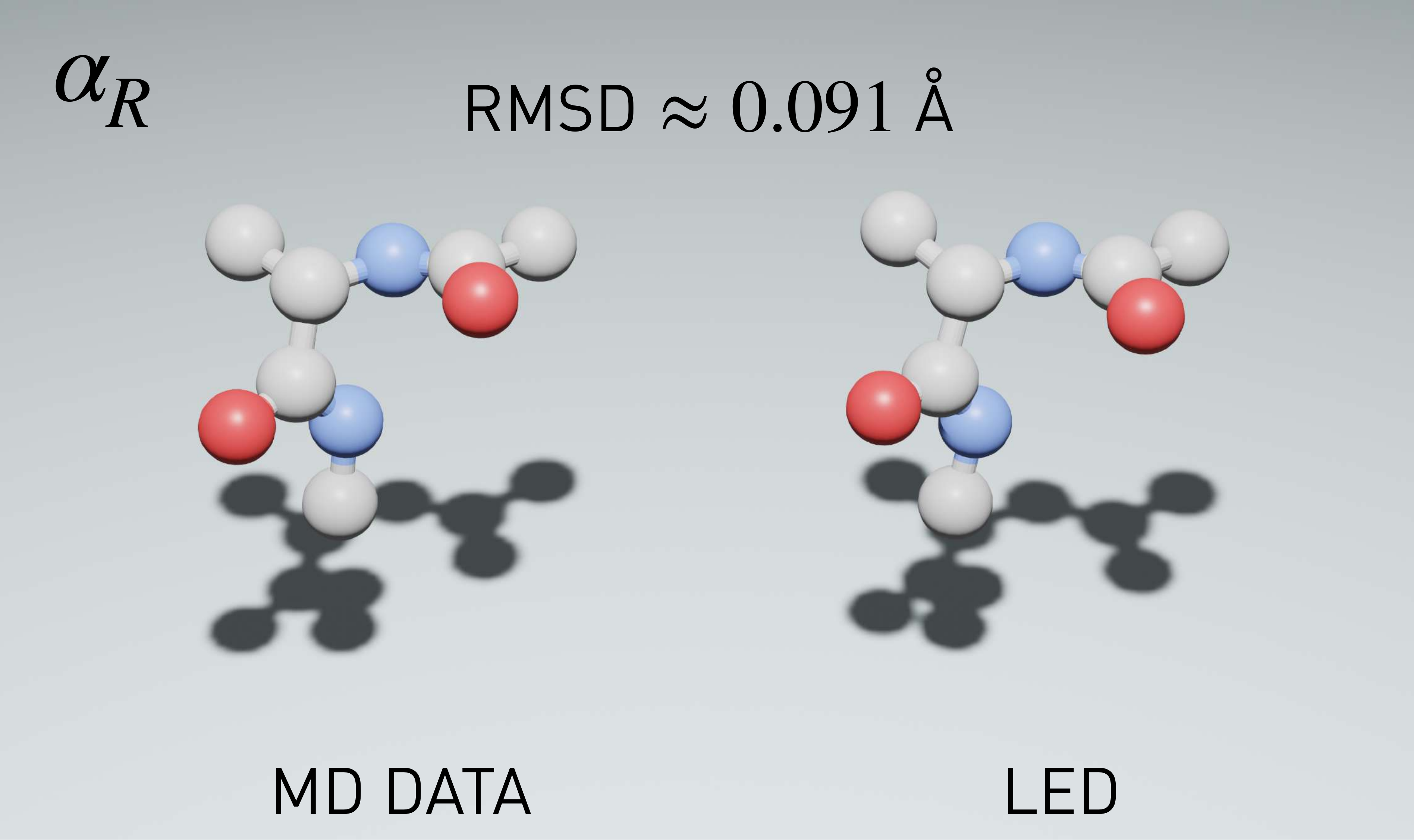}
\hfill
\includegraphics[width=0.48\textwidth]{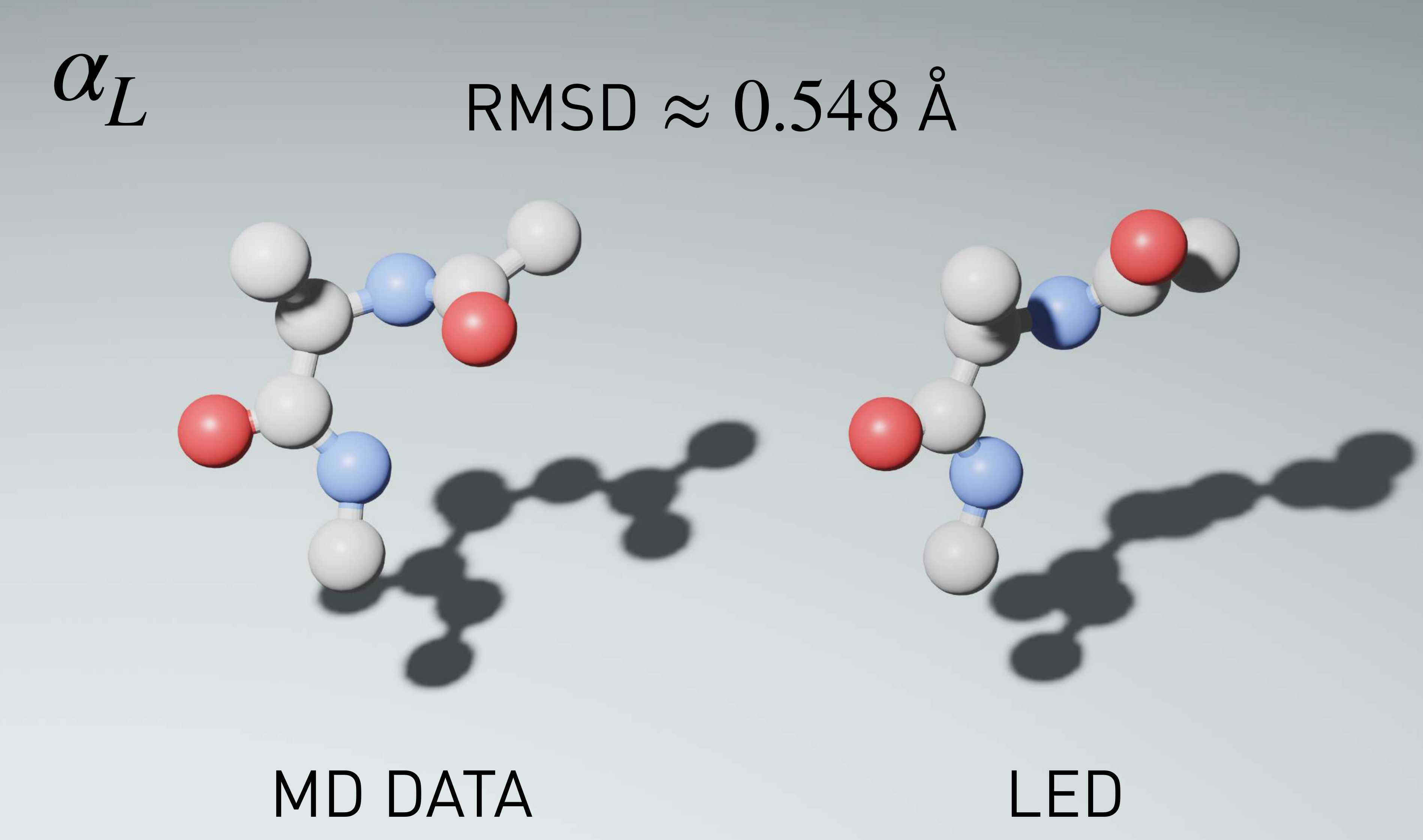}
\caption{
For each metastable state, a random alanine dipeptide configuration sampled from MD data is compared against the closest configuration sampled from the LED with $d_{\mathbf{z}}=1$.
The Root Mean Square Deviation (RMSD) in $\mathring{A}$ between the two is plotted for reference.
}
\label{fig:alanine:blender}
\end{figure*}

In Figure~\ref{fig:alanine:blender}, a configuration randomly sampled from MD data is given for each metastable state.
The closest configuration sampled from LED is compared with the MD data sample in terms of the Root Mean Square Deviation (RMSD) score. 
The LED samples realistic configurations with low RMSD errors for all metastable states.
The mean and standard deviation of the RMSD scores of the $10$ closest neighbors sampled from LED are $\mu \pm \sigma=  0.148  \pm 0.021 \mathring{A}$ for the $C_5$ MD sample configuration (Figure~\ref{fig:alanine:blender} top left).
This score for the rest of the metastable states is $0.340  \pm 0.463 \mathring{A}$ for $P_{II}$, $0.101 \pm 0.019 \mathring{A}$ for $\alpha_R$, $0.885 \pm 0.162 \mathring{A}$ for $\alpha_L$, and $0.383 \pm 0.125 \mathring{A}$ for $C_{7}^{ax}$.
The LED samples similar configurations with low RMSD scores for the most frequently observed metastable states $\{C_5, P_{II}, \alpha_R \}$.
The average RMSD error is slightly higher and fluctuates more for the less frequently observed $\{ \alpha_R, C_{7}^{ax} \}$.

The dynamics learned by LED are evaluated according to the mean first-passage times (MFPTs) between the dominant metastable states.
The MFPT is the average time-scale to reach a final metastable state, starting from any initial state.
The MFPTs are computed a posteriori from trajectories sampled from the LED and the MD test trajectories, using the PyEMMA software~\cite{scherer_pyemma_2015}.
The metastable states considered here are given in the SM Section~\ref{sec:appendix:alanine:metastablestates}.

As a reference for the MFPTs, we consider an MSM fitted to the MD data (test dataset).
The reference MFPTs agree with previous literature~\cite{trendelkamp2016efficient,jang2006multiple,chekmarev2004long,wang2014exploring}.
The time-lag of the MSM is set to $\Delta t_{MSM}=10\text{ps}$ to ensure the necessary Markovianity of the dynamics.
This time-lag is two orders of magnitude larger than the timestep of LED.
Fitting an MSM with a time-lag of $\Delta t_{MSM}=1\text{ps}$ on the MD data results in very high errors ($\approx 85\%$ on average) in the computation of MFPTs.
This emphasizes the need for non-Markovian models that can reproduce the system's dynamics and statistics independent of the selection of the time-lag.

The MFPTs of trajectories sampled from LED are estimated with an MSM with a time-lag $\Delta t_{MSM}=10\text{ps}$.
Note that the LED is operating on a time-step $\Delta t = 0.1 \text{ps}$.
The MFPTs are identified with a low average error of $10.51\%$.
The results on the MFPT are summarized in Table~\ref{tbl:alanine:mfpt}.
LED captures very well the transitions that are dominant in the data e.g. $T_{C_5 \to P_{II} }$, $T_{P_{II} \to C_5 }$ or $T_{\alpha_{R} \to C_5 }$.
In contrast, LED exhibits high MFPT errors in transitions that are rarely observed in the training data.

LED identifies the dominant MFPT successfully by utilizing a very small amount of training data ($38.4\text{ns}$ for training and $38.4\text{ns}$ validation) and propagating the latent dynamics on a reduced order space ($d_{\boldsymbol{z}}=1$).
LED trajectories are three orders of magnitude cheaper to obtain compared to MD data.
At the same time, MSM fitting is a relatively fast procedure once the clustering based on the metastable states is obtained.
In contrast, a careless selection of the time-lag in the MSM that fails to render the dynamics Markovian, (e.g. $\Delta t =1 \text{ps}$) leads to a surrogate model that fails to capture the system time-scales. 
This emphasizes the need to model non-Markovian effects with LED in case of limited data sampled at a high frequency (small time-steps $\Delta t$).
A more informative selection of the time-lag may alleviate this problem, rendering the dynamics Markovian as in the reference MSM.
Still, the consequent sub-sampling of the data can lead to omissions of effects whose time-scales are smaller than the time-lag.
As a consequence, the heuristic selection of the time-lag is rendering the modeling process error-prone.

\begin{table}
\caption{
Mean first-passage times (MFPT) between the metastable states of alanine dipeptide in water in [ns].
MFPTs are estimated by fitting MSMs with different time-lags ($10 \text{ps}$ and $1 \text{ps}$) on trajectories generated by MD, or the LED framework.
The average relative error is given for reference.
}
\label{tbl:alanine:mfpt}
\centering
\includegraphics[width=0.8\textwidth]{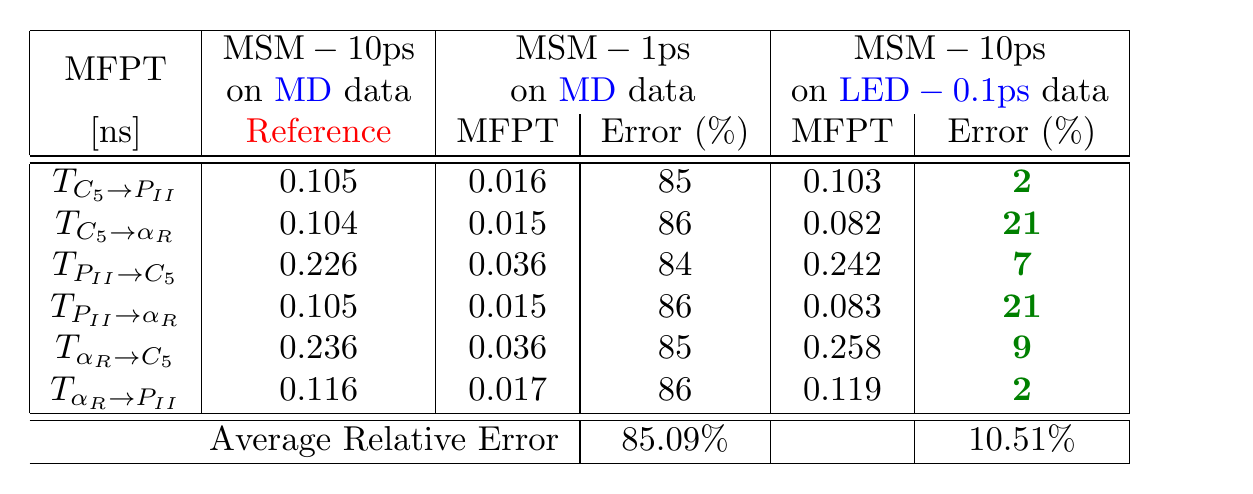}
\end{table}

The SM Section~\ref{sec:appendix:alanine:metastablestateslatent} provides additional results on the MFPTs estimated based on metastable state definition in the latent space of LED (without prior knowledge).
Furthermore, the effectiveness of LED to interpolate the training data and unravel novel configurations of the protein (state-space) is also illustrated.


\section*{Discussion}
\label{sec:conclusion}

This work proposes a data-driven framework, LED, to learn and propagate the effective dynamics of molecular systems accelerating MD simulations by orders of magnitude.
Previous state-of-the-art methods are based on the Markovian assumption on the latent state, or minimize the autocorrelation or the variational loss on the data.
The latter take into account the error on the long-term equilibrium statistics explicitly to capture the system time-scales but suffer from a dependency on the batch size~\cite{hernandez2018variational}.
In contrast, the LED is trained to maximize the data likelihood and identify a continuous reduced-order latent representation.
The nonlinear dynamics are propagated in the latent space and the memory effects are captured through the hidden state of the LSTM.
Moreover, the method is generative and the decoder part of the MDN-AE can be employed to sample high dimensional configurations on any desired time-scales.

The encoder of LED is analogous to the coarse graining model design, while the decoder is implicitly learning a backmapping to atomistic configurations. The LED automates the dimensionality reduction often associated with the empirical a-priori selection of Collective Variables in molecular simulations~\cite{maragliano2006string, wehmeyer2018time}. 
At the same time the MDN-LSTM propagates the dynamics on the latent space in a form that is comparable to  nonlinear, non-Markovian metadynamics~\cite{MacCarthy:2017}.

The effectiveness of LED is demonstrated for three systems.
In the case of the Langevin dynamics using BMP, LED can recover the free energy landscape in the latent space, identify two low-energetic states corresponding to the long-lived metastable states of the potential, and capture the transition times between the metastable states.
For the Trp Cage miniprotein, LED captures the free energy projection on the latent space and unravels three metastable states.
Lastly, for the system of alanine dipeptide in water, LED captures the configuration statistics of the system accurately while being three orders of magnitude faster than MD solvers.
It identifies three low-energetic regions in the free energy profile projected to the one-dimensional latent state that corresponds to the three dominant metastable states $\{\alpha_R, C_5, P_{II} \}$.
LED is also able to capture the dominant mean first-passage times in contrast to the MSM operating on the same time-scale, owing to the non-Markovian latent propagation in the latent state with the MDN-LSTM.
Furthermore, we showcase how our framework is capable of unraveling novel protein configurations interpolating on the training data.

The speed-up achieved by LED depends on the MD solver used, the dimensionality, and the complexity of the protein under study.
Still, it is expected that the computationally cheap propagation in the latent space of the LED is orders of magnitude faster than the MD solver.

The success of LED paves the way for faster exploration of the conformational space of molecular systems.
Future research efforts will target the application of LED to larger proteins and the  investigation of LED's capabilities and limitations in uncovering the metastable states as minima in the free energy profile.
An alternative research direction is the automatic extraction of features directly from the raw position and velocity data (not using rototranslational invariant features).
Moreover, further studies will concentrate on coupling LED with an MD solver in an alternating fashion for faster exploration of the state space.

\section*{Acknowledgments}
The authors thank Ioannis Mandralis, Pascal Weber, and Fabian Wermelinger for fruitful discussions and providing feedback.
The authors also acknowledge the Swiss National Supercomputing Centre (CSCS) support providing the necessary computational resources under Project s930.
The authors declare no competing interests.

\section*{Data and materials availability}
\label{sec:code}

Code and data to reproduce the findings of this study will be made openly available in a public repository upon publication.
The PyEMMA software package~\cite{scherer_pyemma_2015} is employed in the current work for MSM fitting and MFPT estimation.

\section*{Supplementary Materials}
Section S1, Definition of the MBP, the metastable states and time-scale analysis on the LED latent space\\
Section S1, Figure 10. Marginal state statistics of LED in the MBP\\
Section S1, Table 3. AE hyperparameter tuning in the MBP\\
Section S1, Table 4. LSTM hyperparameter tuning in the MBP\\
Section S1, Table 5. LED hyperparameters in the MBP\\
Section S2, Figure 11. Marginal state statistics of LED in the Trp Cage\\
Section S2, Figure 12. Configuration sampled from LED in the Trp Cage\\
Section S2, Table 6. AE hyperparameter tuning in the Trp Cage\\
Section S2, Table 7. LSTM hyperparameter tuning in the Trp Cage\\
Section S2, Table 8. LED hyperparameters in the Trp Cage\\
Section S3, Information on the simulation of alanine, definition of the metastable states, time-scale analysis on the LED latent space, and study on unraveling novel configurations with LED\\
Section S3, Table 9. Metastable state defintions in alanine\\
Section S3, Table 10. Mean First Passage Time analysis in alanine\\
Section S3, Figure 13. Marginal state statistics of LED in alanine\\
Section S3, Figure 14. Unraveling novel configurations with LED in alanine\\
Section S3, Table 11. AE hyperparameter tuning in alanine\\
Section S3, Table 12. LSTM hyperparameter tuning in alanine\\
Section S3, Table 13. LED hyperparameters in alanine\\

\bibliographystyle{Science}

\newpage
\clearpage
\section{SI: M\"uller-Brown Potential}
\label{sec:appendix:mbp}

The MBP has the form
\begin{equation}
\begin{aligned}
V(\boldsymbol{x}) = \sum_{k=1}^4
A_k
\exp
\big(
& \alpha_k (x_1 - \hat{X}_{1,k} ) +\\
& b_k (x_1 - \hat{X}_{1,k} ) (x_2 - \hat{X}_{2,k}  ) +\\
&  c_k (x_2 - \hat{X}_{2,k} )
\big),
\end{aligned}
\end{equation}
where $\boldsymbol{x}=[x_1, x_2]^T$ is the position.
The parametrization
\begin{equation}
\begin{aligned}
\alpha &=[-1, -1, -6.5, 0.7]^T, \\
b &= [0,0,11,0.6]^T, \\
c &=[-10, -10, -6.5, 0.7]^T, \\
A &=[-200, -100, -170, 15]^T, \\
\hat{X} &=
 \begin{bmatrix}
 1&0&-0.5&-1\\
 0&0.5&1.5&1
 \end{bmatrix},
\end{aligned}
\end{equation}
is followed according to Ref.~\cite{muller1979location}.

\begin{figure}[h!]
\centering
\includegraphics[width=0.45\textwidth,clip]{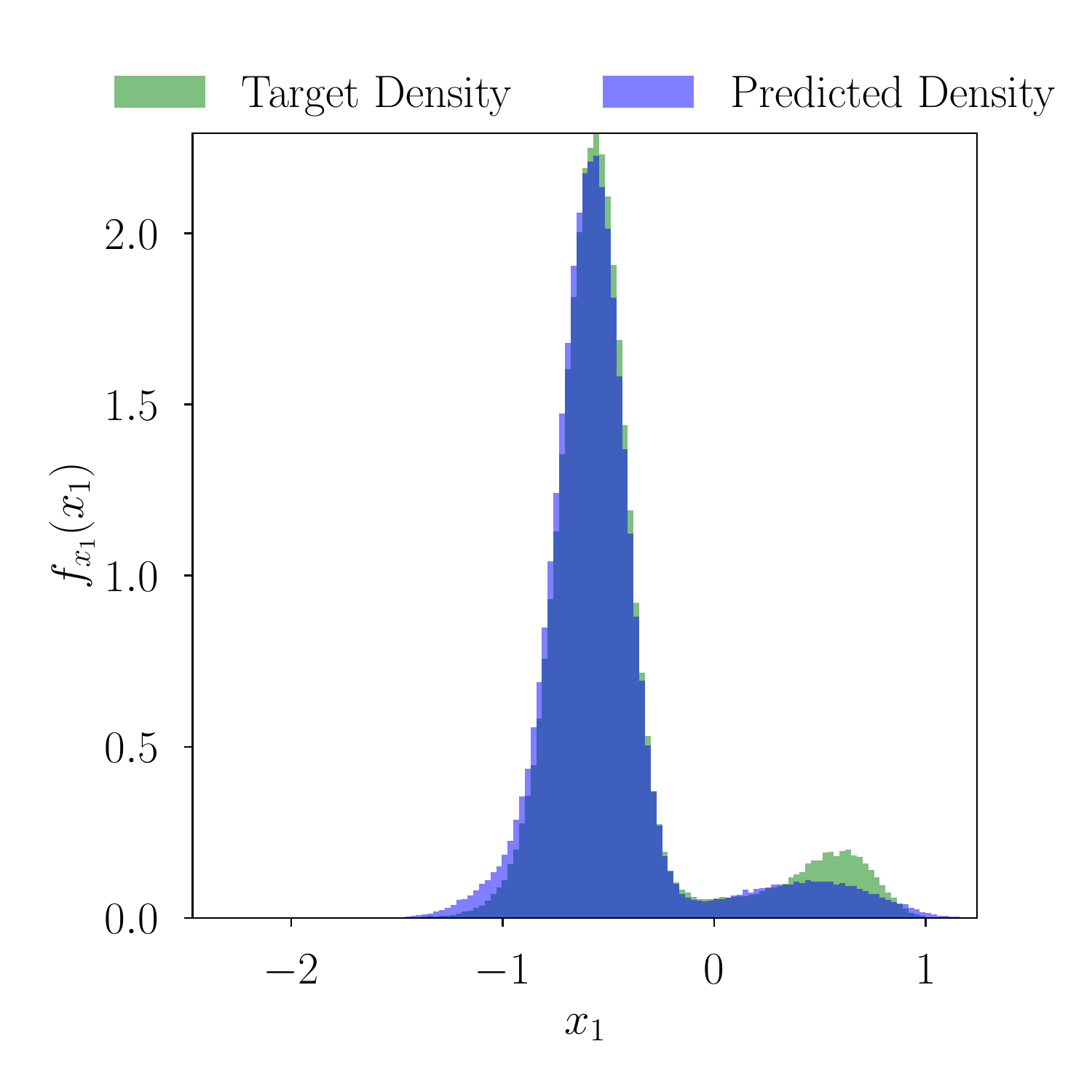}
\hfill 
\includegraphics[width=0.45\textwidth,clip]{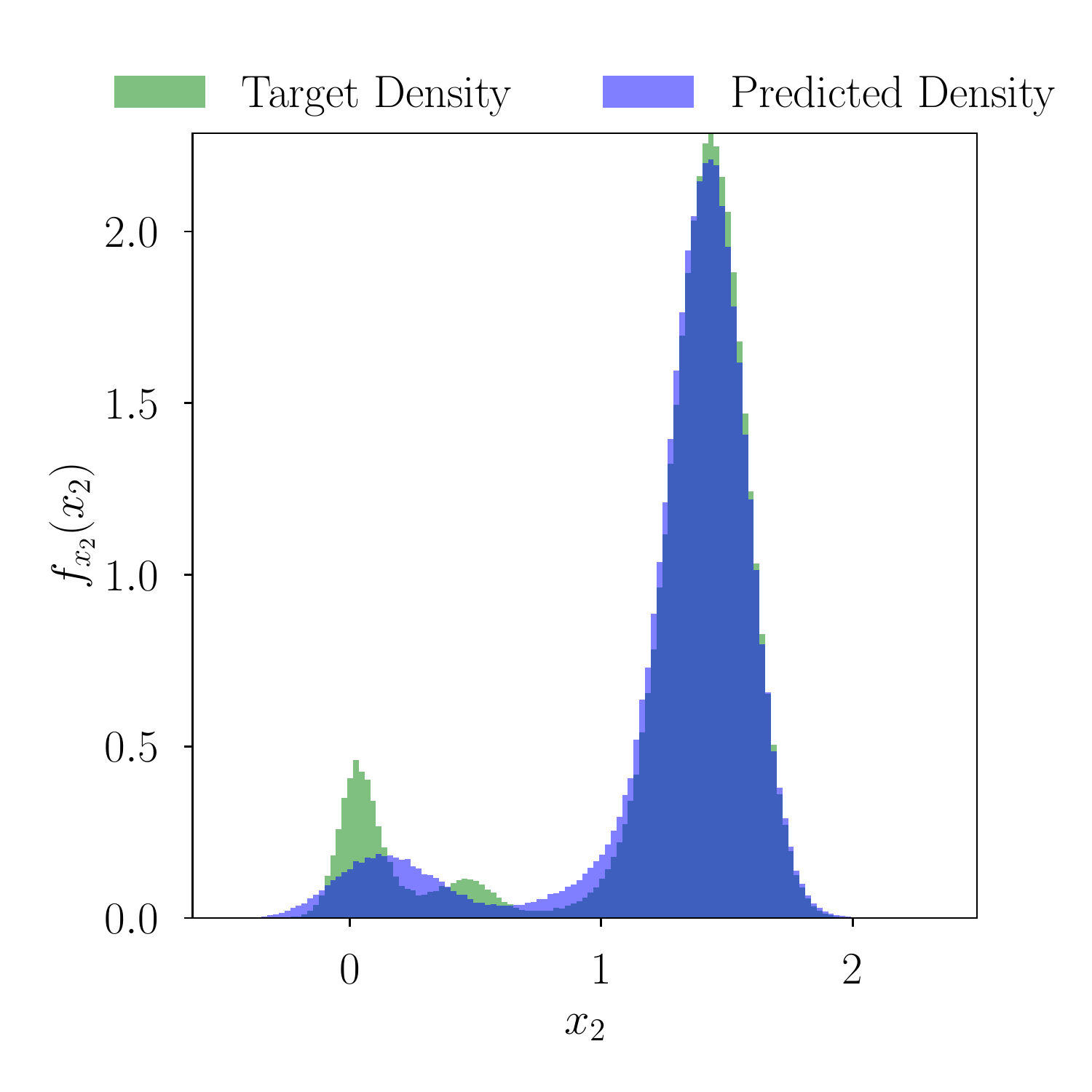}
\caption{
Comparison of the marginal distributions of the MBP states $x_1$ and  $x_2$ between the test data and trajectories of LED.
The LED is propagating the dynamics on a one dimensional reduced order latent state, i.e., $d_{\boldsymbol{z}}=1$.
}
\label{fig:bmp:iterative_latent_forecasting_state_dist_bar}
\end{figure}

The marginal distributions of the MBP states from trajectories sampled from the LED is compared with the groundtruth (test data) in Figure~\ref{fig:bmp:iterative_latent_forecasting_state_dist_bar}.

\paragraph*{Definition of Metastable States}
\label{sec:appendix:bmp:metastablestatescenters}

The metastable states of the MB potential are defined as ellipses in the $\boldsymbol{x} \in \mathbb{R}^2$ space.
The centers and axes given in Table~\ref{tbl:bmp:metastablestatescenters}.
\begin{table}
\caption{
Metastable states in the MBP modeled as ellipses
$x_1^2 / \alpha^2 + 
x_2^2 / \beta^2
\leq 1
$.
The ellipses are rotated by $\theta$.
}
\label{tbl:bmp:metastablestatescenters}
\centering
\begin{tabular}{|c|c|c|c|}
\hline
State & Center $(x_1, x_2)$ & Axes ${\alpha}, \beta)$ & $\theta$ \\
\hline \hline
0 & $(-0.57, 1.45)$ & $(0.15, 0.3)$ &  $\pi /4 $\\ \hline
1 & $(0.45, 0.05)$ & $(0.35, 0.15)$ & $ 0$ \\ \hline
\end{tabular}
\end{table}

\paragraph*{Time-scales in the LED Latent Space}
\label{sec:appendix:bmp:ledlatentspacetimescale}

The latent space learned by LED can be utilized to identify low-energy metastable states without the need for prior knowledge.
The definition of the metastable states in the rotationally and translationally invariant space constitutes such prior knowledge.
Minima in the free energy projection on the LED latent space, constitute probable metastable states.

The trajectories sampled with LED are clustered based on these latent metastable clusters depicted in Figure~\ref{fig:bmp:bmp_free_energy_clusters}.
An MSM is fitted on the clustered trajectories.
The time-lag of the MSM is set to $100$ time units to ensure Markovianity.
The timescales computed by MSM are $\overline{T}_{0 \to 1}=49$ and $\overline{T}_{1 \to 0}=321$.
LED is overestimating $\overline{T}_{1 \to 0}$ and underestimating $\overline{T}_{0 \to 1}$.
The order of the timescales, however, is captured.
In contrast, an MSM with a time-lag of $\Delta_t = 0.5$, which is the timestep of the LED, fails to capture the order of the timescales due to the violated Markovianity assumption ($\overline{T}_{0 \to 1}=3$ and $\overline{T}_{0 \to 1}=21$).

\clearpage
\paragraph*{LED Hyperparameters}
\label{sec:appendix:bmp:hyp}

In order to prepare the dataset for training, validation, and testing of the LED in the MBP, $96$ initial conditions are sampled from $\boldsymbol{x} \in [-1.5, 1.2] \times [-0.2, 2]$.
The dynamics are solved with the Velocity Verlet algorithm, with time-step $\delta t = 10^{-2}$ up to $T=5000$, after an initial transient period of $\tilde{T}=10^3$ discarded from the data.
The data are sub-sampled to $\Delta t = 0.5$, keeping every  50\textsuperscript{th} data point.
In this way, $96$ trajectories of $N=10^4$ samples, each corresponding to $T=5000$ time units are created.
LED is trained on $32$ of these trajectories. $32$ trajectories are used for validation, while all $96$ trajectories are used for testing.

The number and size of hidden layers are the same for the encoder $\mathcal{E}$, the decoder $\mathcal{D}$, and the latent MDN $\mathcal{Z}$.
In the first phase, the MDN-AE is trained, tuning its hyperparameters based on a grid search reported in Table~\ref{app:tab:bmp:hyp:auto}.
The autoencoder with the smallest error on the state statistics on the validation dataset is picked.
Next, the MDN-LSTM is trained, tuning its hyperparameters based on a grid search reported in Table~\ref{app:tab:bmp:hyp:lstm}.
The LED model with the smallest error on the state statistics on the validation dataset is picked.
Both networks are trained with validation based early stopping.
The LED is tested on the total $96$ initial conditions.
For more information of the training technicalities the interested reader is referred to Ref.~\cite{vlachas2020learning}.
\begin{table}[tbhp]
\caption{Hyperparameter tuning of AE for MBP}
\label{app:tab:bmp:hyp:auto}
\centering
\begin{tabular}{ |c|c|c| } 
\hline
\text{Hyperparameter} & 
\text{Values} \\  \hline \hline
Batch size& $32$ \\
Initial learning rate & $0.001$ \\
Weight decay rate & $\{0, 10^{-5} \}$ \\
Number of AE layers & $\{2,3 \}$ \\
Size of AE layers & $\{10,20,40\}$ \\
Activation of AE layers & $\operatorname{selu}$, $\operatorname{tanh}$ \\
Latent dimension & $\{1\}$ \\
Input/Output data scaling & $[0,1]$ \\
MDN-AE kernels & $\{2,3\}$ \\
MDN-AE hidden units & $50$ \\
MDN-AE multivariate & $1$ \\
MDN-AE covariance scaling factor & $\{0.4,0.6,0.8\}$ \\
\hline
\end{tabular}
\end{table}
\begin{table}[tbhp]
\caption{Hyperparameter tuning of LSTM for MBP}
\label{app:tab:bmp:hyp:lstm}
\centering
\begin{tabular}{ |c|c|c| } 
\hline
\text{Hyperparameter} & 
\text{Values} \\  \hline \hline
Batch size& $32$ \\
Initial learning rate & $10^{-3} $ \\
BPTT sequence length & $\{200, 400 \}$ \\
Number of LSTM layers & $1$ \\
Size of LSTM layers & $\{10,20,40 \}$ \\
Activation of LSTM Cell & $\operatorname{tanh}$ \\
MDN-LSTM kernels & $\{4, 5, 6\}$ \\
MDN-LSTM hidden units & $\{10, 20 \}$ \\
MDN-LSTM multivariate & $0$ \\
MDN-LSTM covariance scaling factor & $\{0.1,0.2,0.3, 0.4\}$ \\
\hline
\end{tabular}
\end{table}
\begin{table}[tbhp]
\caption{Hyperparameters of LED model with lowest validation error on MBP}
\label{app:tab:bmp:hyp:led}
\centering
\begin{tabular}{ |c|c|c| } 
\hline
\text{Hyperparameter} & 
\text{Values} \\  \hline \hline
Number of AE layers & $3$ \\
Size of AE layers & $40$ \\
Activation of AE layers & $\operatorname{tanh}$ \\
Latent dimension & $1$ \\
MDN-AE kernels & $3$ \\
MDN-AE hidden units & $50$ \\
MDN-AE multivariate & $1$ \\
MDN-AE covariance scaling factor & $0.6$ \\
Weight decay rate & $0.0$  \\
BPTT sequence length & $400 $ \\
Number of LSTM layers & $1$ \\
Size of LSTM layers & $20$ \\
Activation of LSTM Cell & $\operatorname{tanh}$ \\
MDN-LSTM kernels & $4$ \\
MDN-LSTM hidden units & $20$ \\
MDN-LSTM multivariate & $0$ \\
MDN-LSTM covariance scaling factor & $0.4$ \\
\hline
\end{tabular}
\end{table}

\clearpage
\newpage
\section{SI: Trp Cage}
\label{sec:appendix:trp}
\begin{figure*}[h!]
\centering
\includegraphics[width=0.9\textwidth,clip]{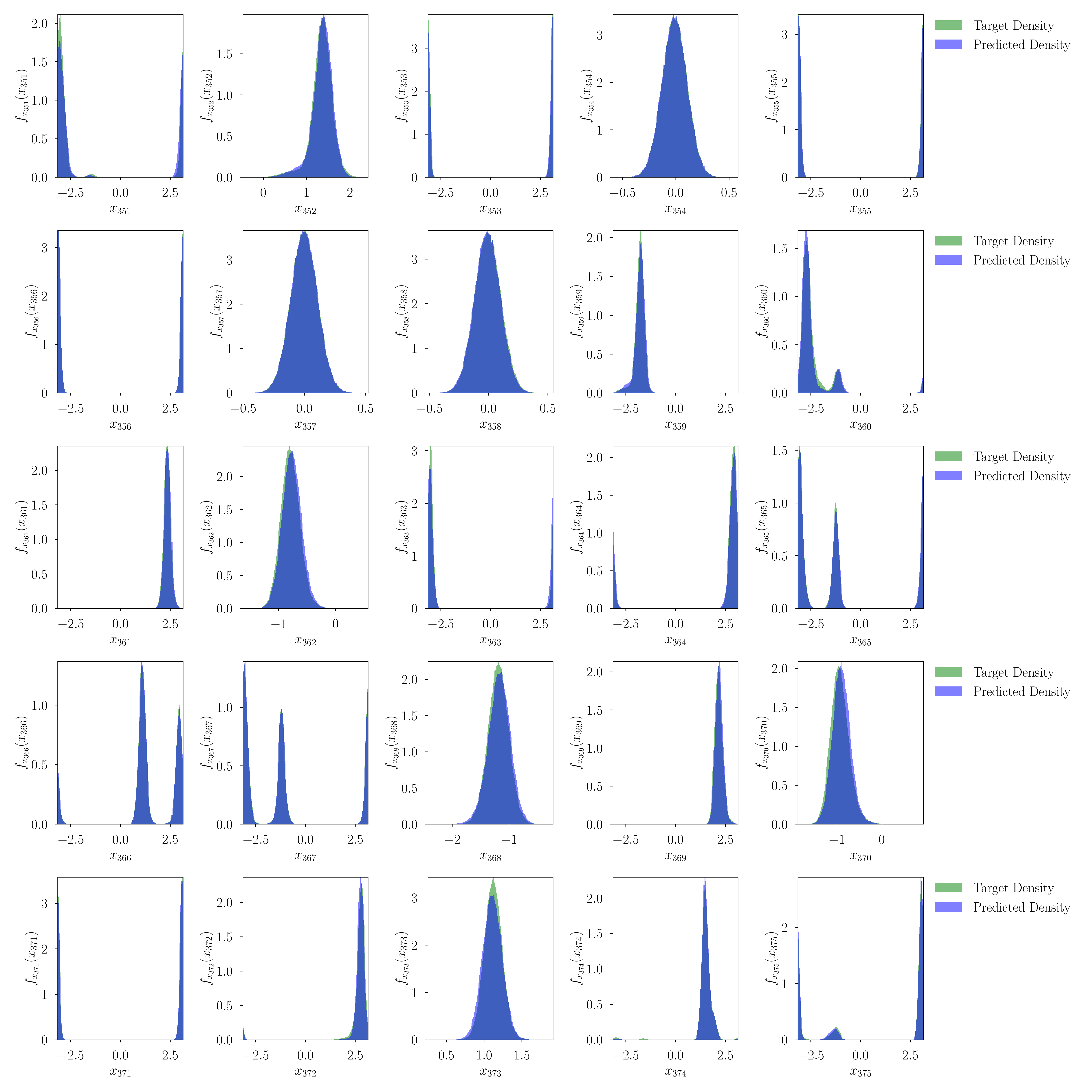}
\caption{
Plot of the marginal state distributions $\boldsymbol{s}_{351}-\boldsymbol{s}_{375}$ in the Trp Cage miniprotein.
Comparison of the state distributions estimated from the MD data (test dataset) and from trajectories sampled from LED.
}
\label{fig:trp:iterative_latent_forecasting_state_dist_bar}
\end{figure*}

\begin{figure*}[h!]
\centering
\includegraphics[width=0.8\textwidth,clip]{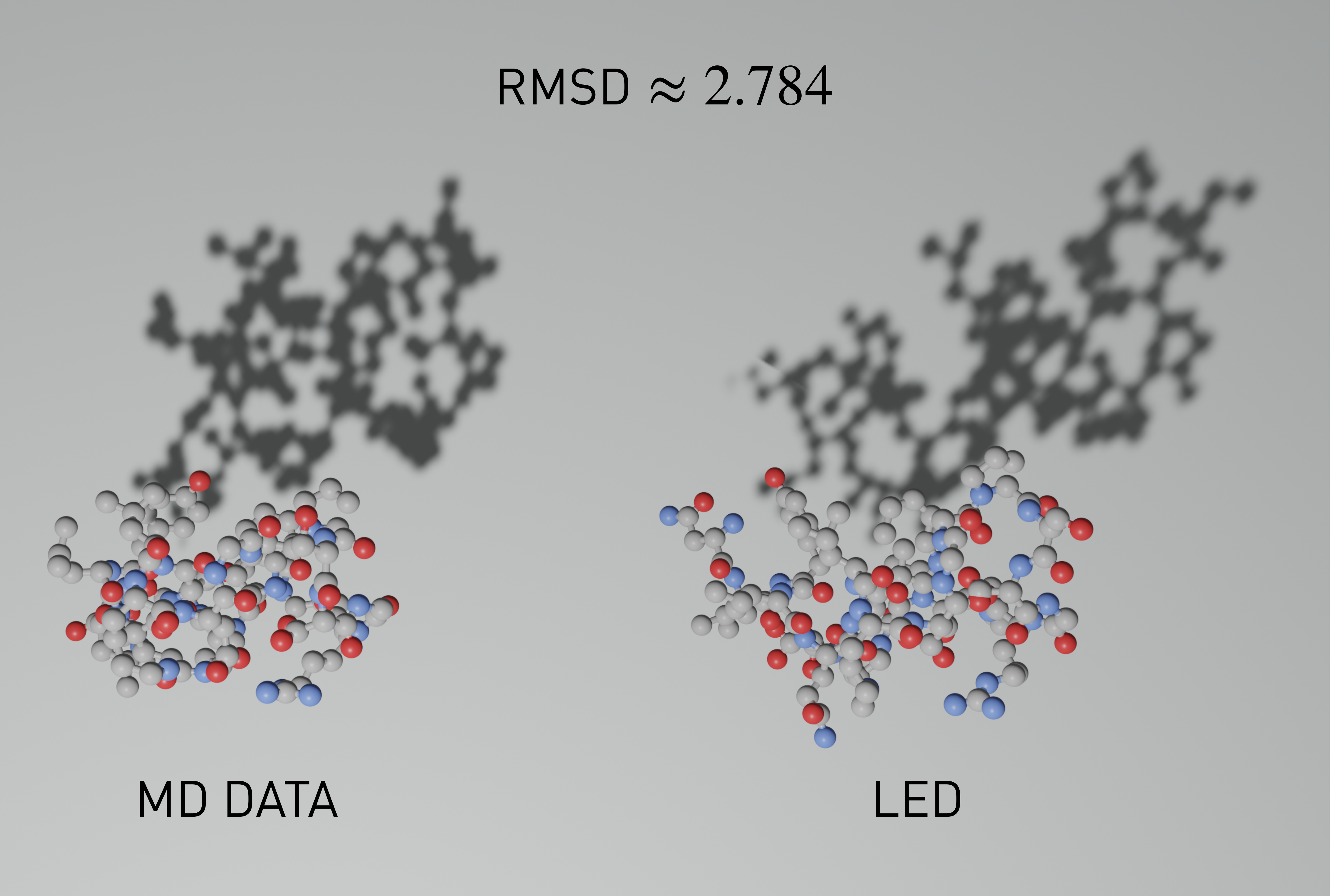}
\caption{
Trp Cage protein configurations found in the MD data compared to a sample of LED that is is in close proximity in the latent space.
The RMSD error between the two configurations is $2.784 \mathring{A}$.
}
\label{fig:trp:TRP}
\end{figure*}

\paragraph*{Marginal State Distributions}
\label{sec:appendix:trp:marginals}
The marginal distributions of the trajectories generated by LED match the ground-truth ones (MD data) closely, as depicted in Figure~\ref{fig:trp:iterative_latent_forecasting_state_dist_bar}.
In Figure~\ref{fig:trp:TRP}, a sample from MD data of the TRP cage is compared with a close sample (in terms of the latent space) of LED.
The RMSD is $2.784 \mathring{A}$.

\paragraph*{LED Hyperparameters}
\label{sec:appendix:trp:hyp}

In the  LED architecture, the number and size of hidden layers are the same for the encoder $\mathcal{E}$, the decoder $\mathcal{D}$, and the latent MDN $\mathcal{Z}$.
The MDN-AE is trained, tuning its hyperparameters based on the grid search reported in Table~\ref{app:tab:trp:hyp:auto}.
The latent space of the MDN-AE is $\boldsymbol{z} \in \mathbb{R}^2$, i.e., $d_{\boldsymbol{z}}=2$.
The MDN-AE model with the lowest error on the state statistics in the validation dataset is picked.
Then, the MDN-AE is coupled with the MDN-LSTM as LED.
The MDN-LSTM is trained to minimize the latent data likelihood.
The hyperparameters of the MDN-LSTM are tuned according to the grid search reported in Table~\ref{app:tab:trp:hyp:lstm}.
The LED model with the lowest error on the state statistics in the validation dataset is selected.
Its hyperparameters are reported in Table~\ref{app:tab:trp:hyp:led}.
The LED is tested in $248$ initial conditions randomly sampled from the testing data.
Starting from these initial conditions, we utilize the iterative propagation in the latent space of the LED to forecast $T=400 \text{ps}$.

\begin{table}[tbhp]
\caption{Hyperparameter tuning of AE for Trp Cage}
\label{app:tab:trp:hyp:auto}
\centering
\begin{tabular}{ |c|c|c| } 
\hline
\text{Hyperparameter} & 
\text{Values} \\  \hline \hline
Batch size& $32$ \\
Initial learning rate & $10^{-3} $ \\
Weight decay rate & $\{0, 10^{-4} , 10^{-5}, 10^{-6}  \}$ \\
Number of AE layers & $\{4,6 \}$ \\
Size of AE layers & $\{100,200,500\}$ \\
Activation of AE layers & $\operatorname{selu}$, $\operatorname{tanh}$ \\
Latent dimension & $2$ \\
Input/Output data scaling & $[0,1]$ \\
MDN-AE kernels & $\{3, 4, 5\}$ \\
MDN-AE hidden units & $\{20,50\}$ \\
MDN-AE covariance scaling factor & $0.8$ \\
\hline
\end{tabular}
\end{table}

\begin{table}[tbhp]
\caption{Hyperparameter tuning of LSTM for Trp Cage}
\label{app:tab:trp:hyp:lstm}
\centering
\begin{tabular}{ |c|c|c| } 
\hline
\text{Hyperparameter} & 
\text{Values} \\  \hline \hline
Batch size& $32$ \\
Initial learning rate & $10^{-3} $ \\
BPTT sequence length & $\{200, 400 \}$ \\
Number of LSTM layers & $1$ \\
Size of LSTM layers & $\{10, 20, 40 \}$ \\
Activation of LSTM Cell & $\operatorname{tanh}$ \\
MDN-LSTM kernels & $\{4,8,12,24\}$ \\
MDN-LSTM hidden units & $\{10, 20, 40, 80\}$ \\
MDN-LSTM multivariate & $\{0, 1\}$ \\
MDN-LSTM covariance scaling factor & $\{ 0.1, 0.2, 0.3, 0.4 \}$ \\
\hline
\end{tabular}
\end{table}
\begin{table}[tbhp]
\caption{Hyperparameters of LED model with lowest validation error on Trp Cage}
\label{app:tab:trp:hyp:led}
\centering
\begin{tabular}{ |c|c|c| } 
\hline
\text{Hyperparameter} & 
\text{Values} \\  \hline \hline
Number of AE layers & $6$ \\
Size of AE layers & $500$ \\
Activation of AE layers & $\operatorname{tanh}$ \\
Latent dimension & $2$ \\
MDN-AE kernels & $5$ \\
MDN-AE hidden units & $50$ \\
MDN-AE multivariate & $0$ \\
MDN-AE covariance scaling factor & $0.8$ \\
Weight decay rate & $0 $ \\
BPTT sequence length & $400 $ \\
Number of LSTM layers & $1$ \\
Size of LSTM layers & $40$ \\
Activation of LSTM Cell & $\operatorname{tanh}$ \\
MDN-LSTM kernels & $4$ \\
MDN-LSTM hidden units & $20$ \\
MDN-LSTM multivariate & $0$ \\
MDN-LSTM covariance scaling factor & $0.2$ \\
\hline
\end{tabular}
\end{table}

\clearpage
\section{SI: Alanine Dipeptide}
\label{sec:appendix:alanine}

A molecule of alanine dipeptide in water is simulated with MD~\cite{guzman2019espressopp}.
The peptide is modeled with the AMBER03 force field~\cite{duan2003point}, while the water is modeled with TIP3P/Fs~\cite{schmitt1999computer}. 
The Velocity Verlet algorithm is employed for the integration.
The simulation domain is a cubic box (edge length $2.7$~nm) with periodic boundary conditions and minimum image convention. 
The temperature is maintained at $298$ K with a local Langevin thermostat~\cite{grest1986molecular}, with the value of the friction constant equal to $1.0/\text{ps}$. 
The cutoff distance for the nonbonded interactions is $r_c=0.9$~nm. 
The reaction field method~\cite{neumann1985dielectric} is used for the electrostatic interaction beyond the cutoff, with the dielectric permittivity of inner and outer regions equal to $1$ and $80$, respectively. 

A timestep of $\delta t=1\text{fs}$ is considered, and the dynamics are integrated up to a total time of $T=100\text{ns}$, creating a dataset with a total of $10^8$ data samples.
The data are subsampled, keeping every $100$\textsuperscript{th} datapoint, creating a trajectory with $N=10^6$ samples.
The coarse time-step of LED is thus $\Delta t=0.1\text{ps}$.
The protein positions are transformed into rototranslational invariant features (internal coordinates), composed of bonds, angles, and dihedral angles.
The data are split to $248$ trajectories of $4000$ samples (each trajectory corresponds to $T=400 \text{ps}$ of MD data), discarding the remaining data.
The first $96$ trajectories (corresponding to a total of $38.4 \text{ns}$ of MD data) are used for training and the next $96$ trajectories for validation.
All $248$ initial conditions are used for testing.

\paragraph*{Marginal State Distributions}

\begin{figure*}[t]
\centering
\includegraphics[width=0.2\textwidth,clip]{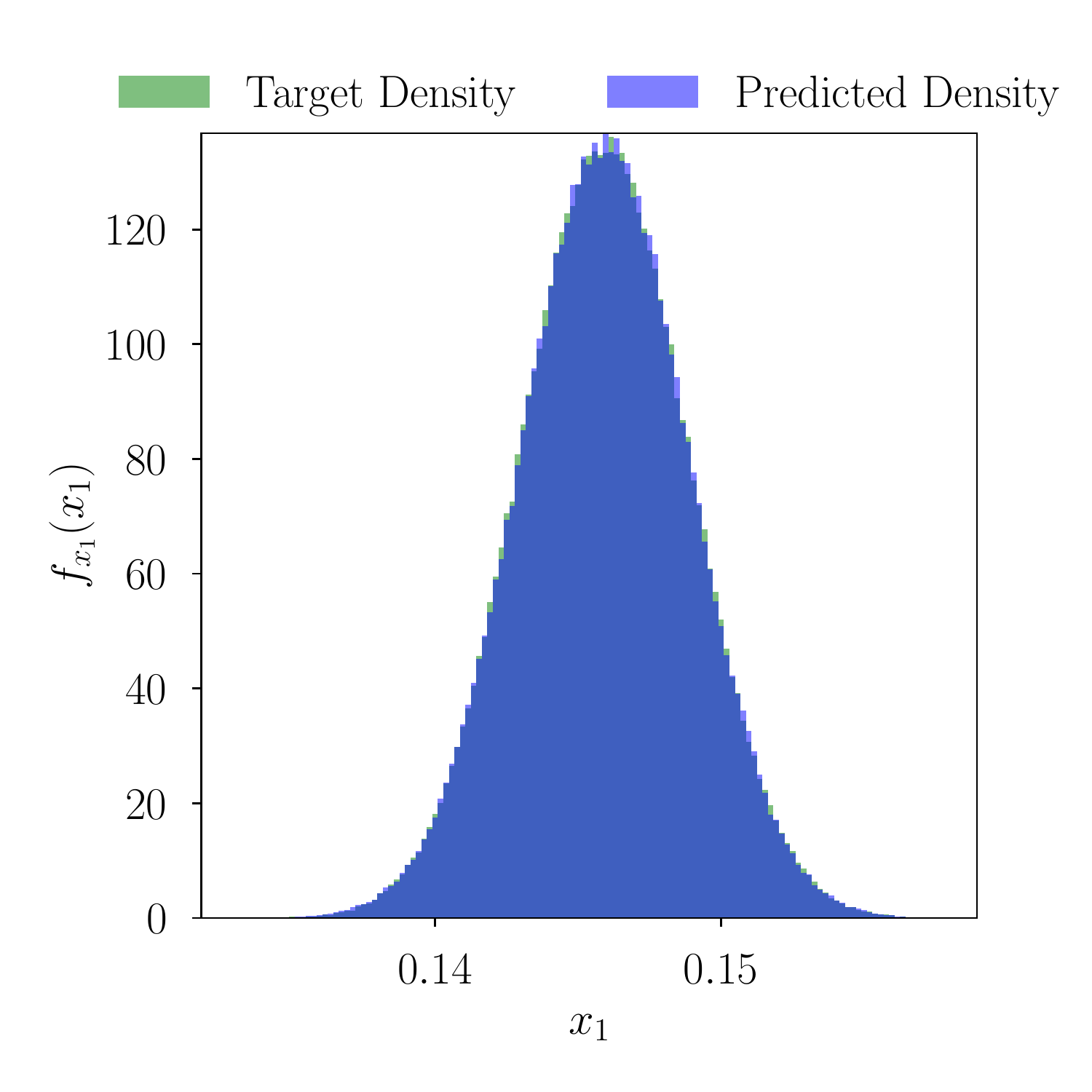}
\includegraphics[width=0.2\textwidth,clip]{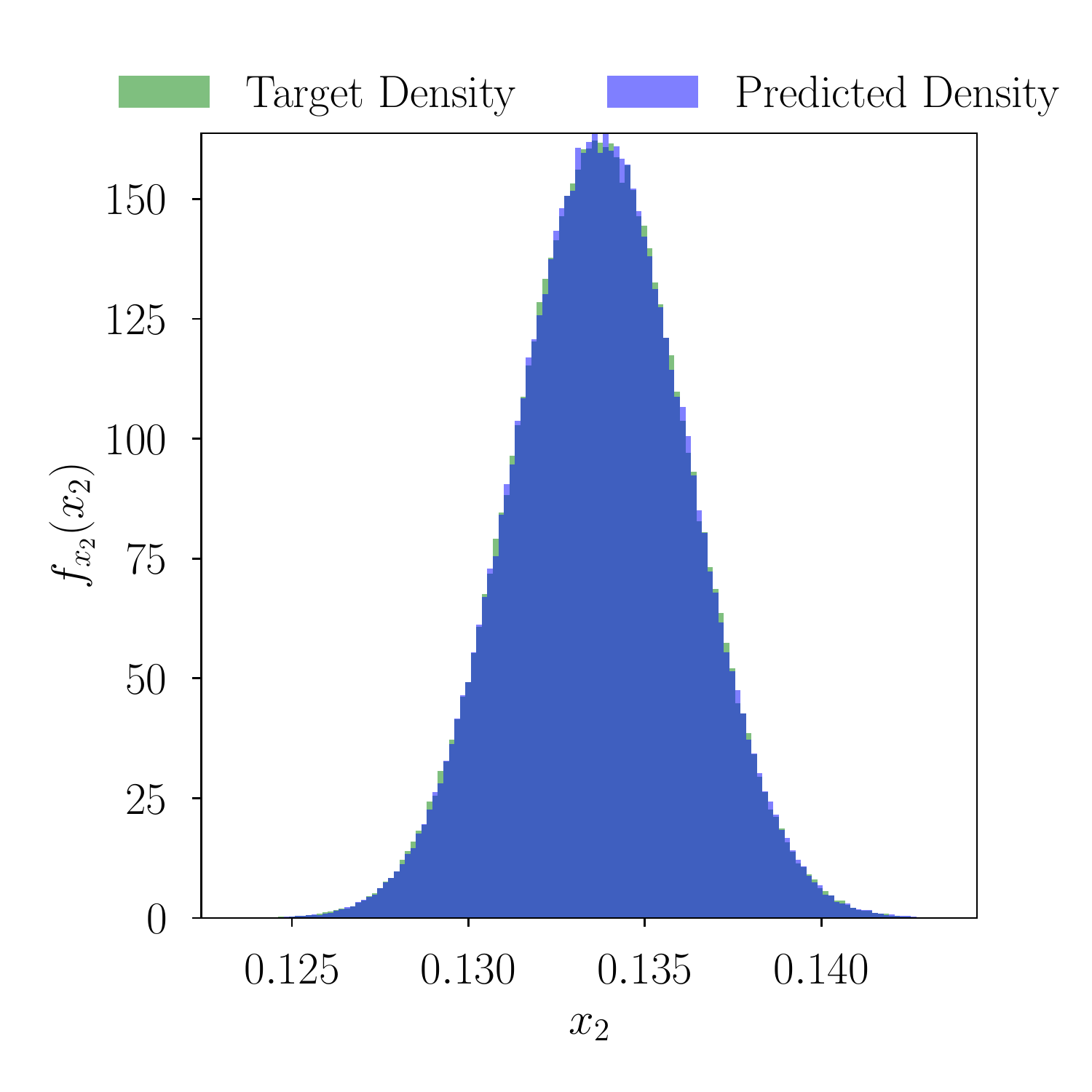}
\includegraphics[width=0.2\textwidth,clip]{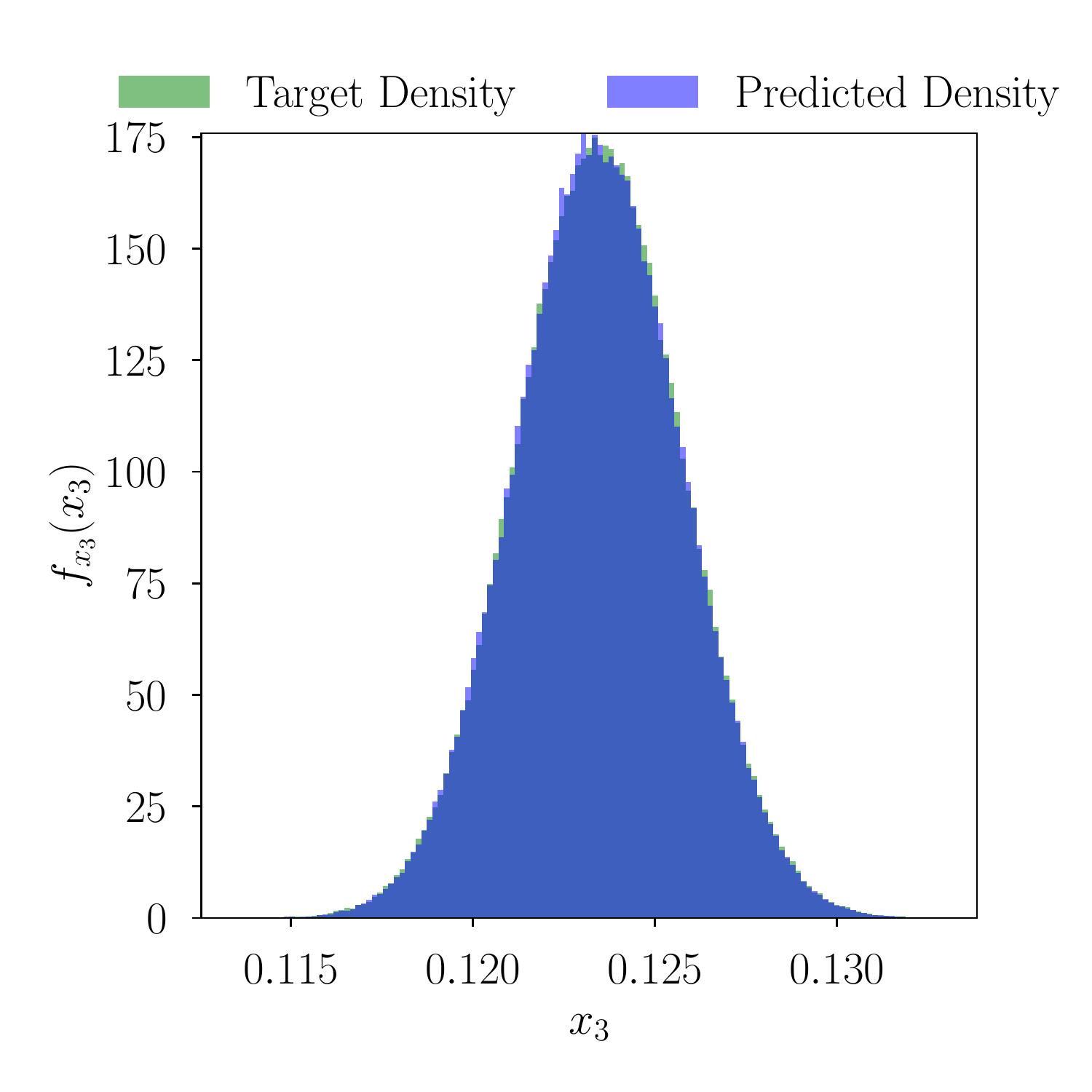}
\includegraphics[width=0.2\textwidth,clip]{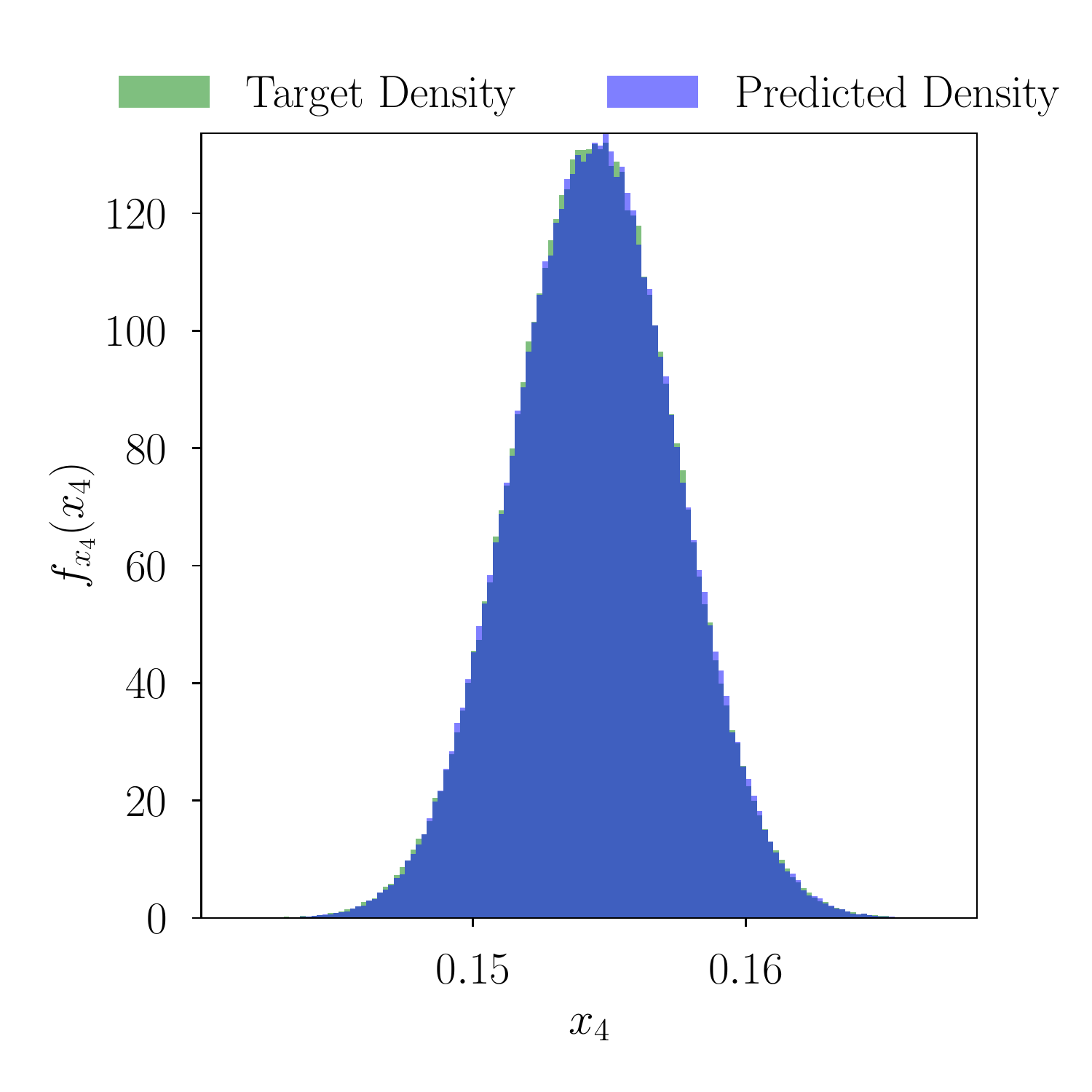}
\includegraphics[width=0.2\textwidth,clip]{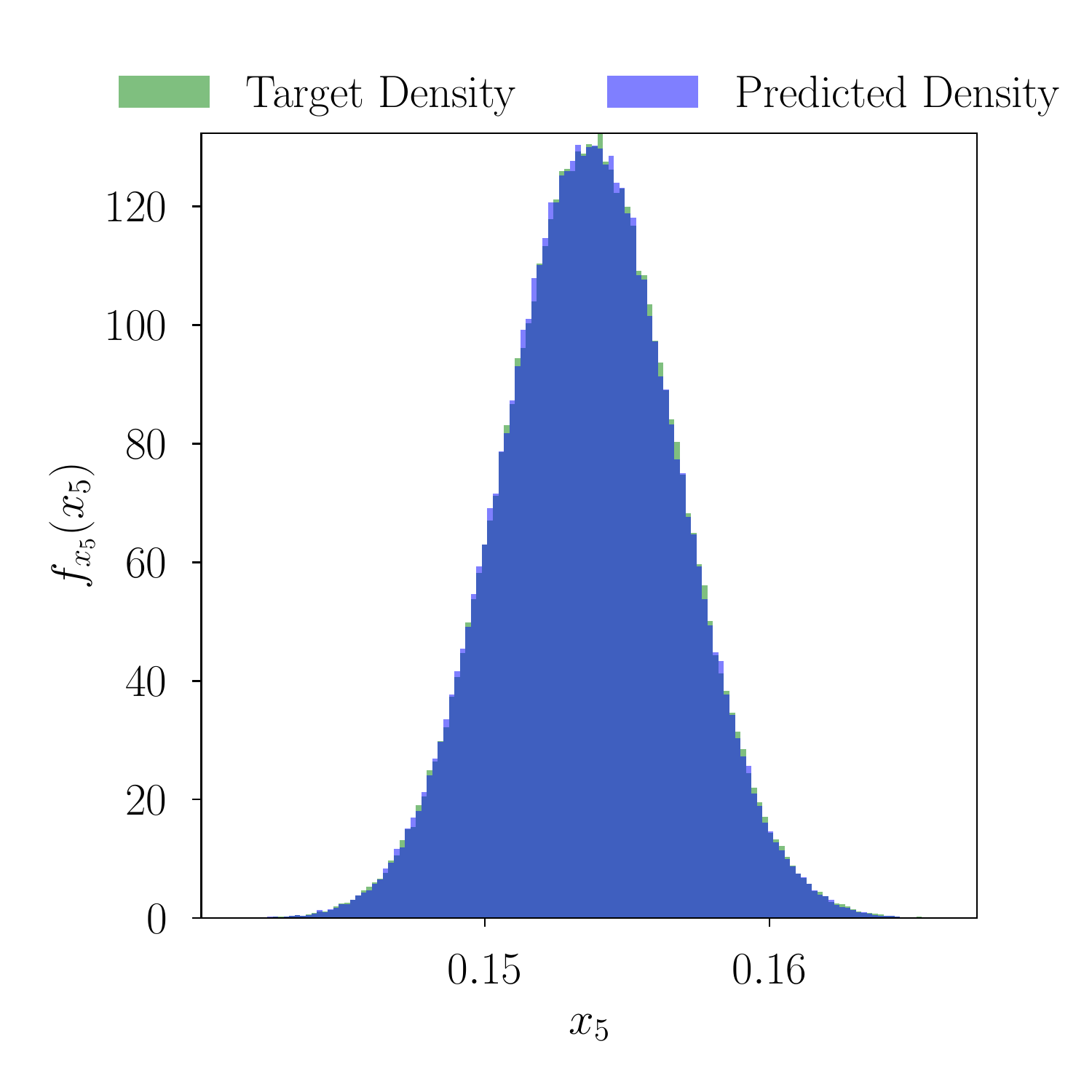}
\includegraphics[width=0.2\textwidth,clip]{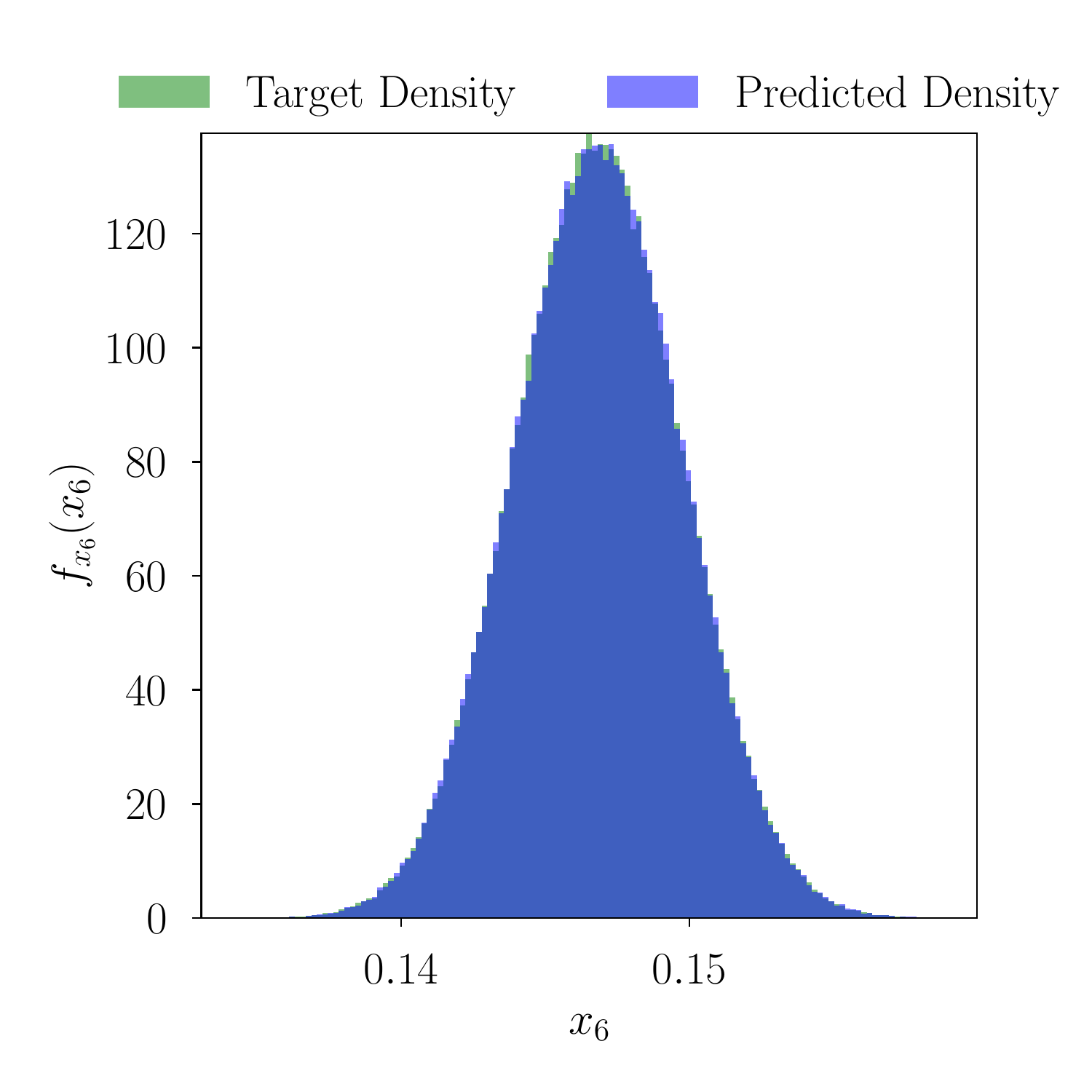}
\includegraphics[width=0.2\textwidth,clip]{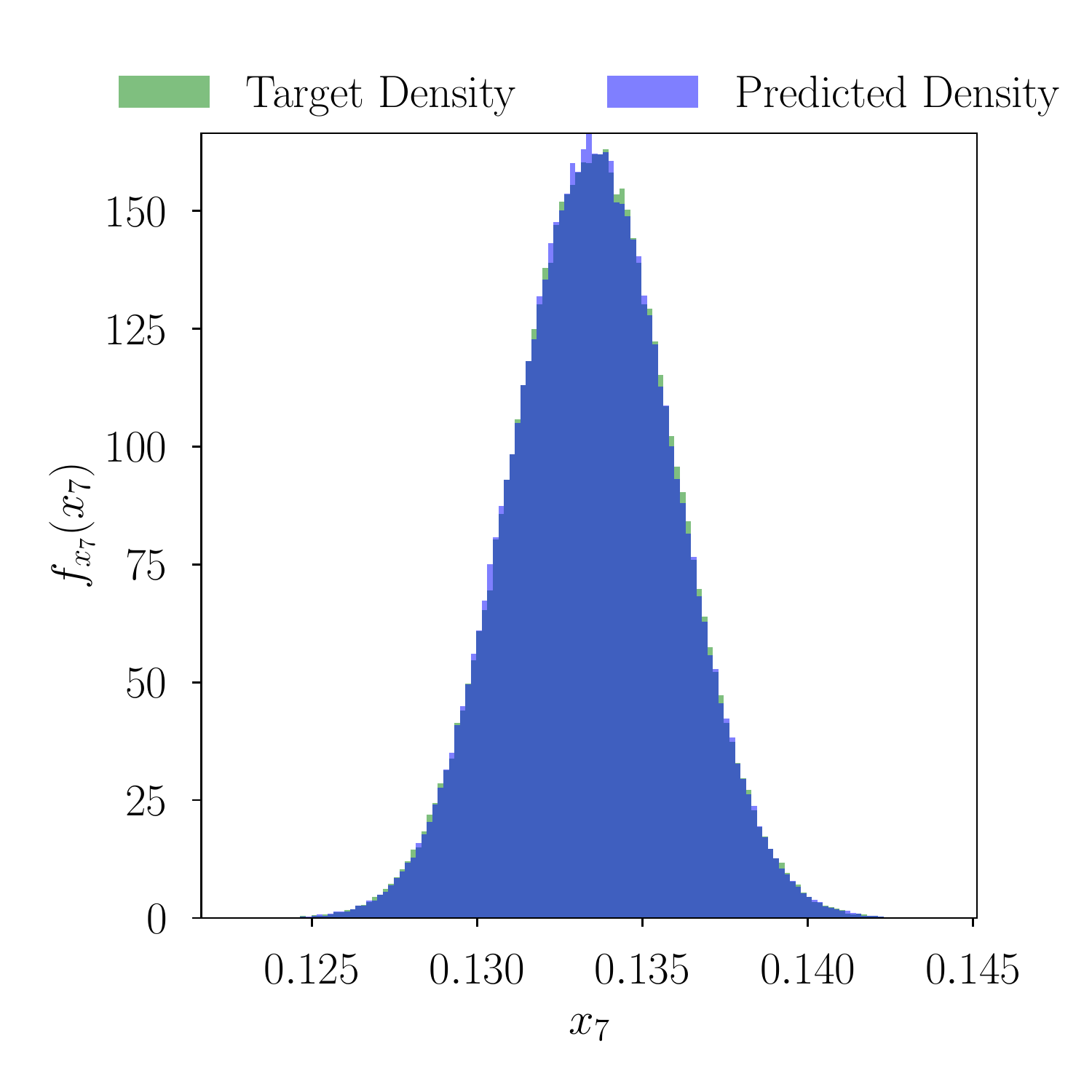}
\includegraphics[width=0.2\textwidth,clip]{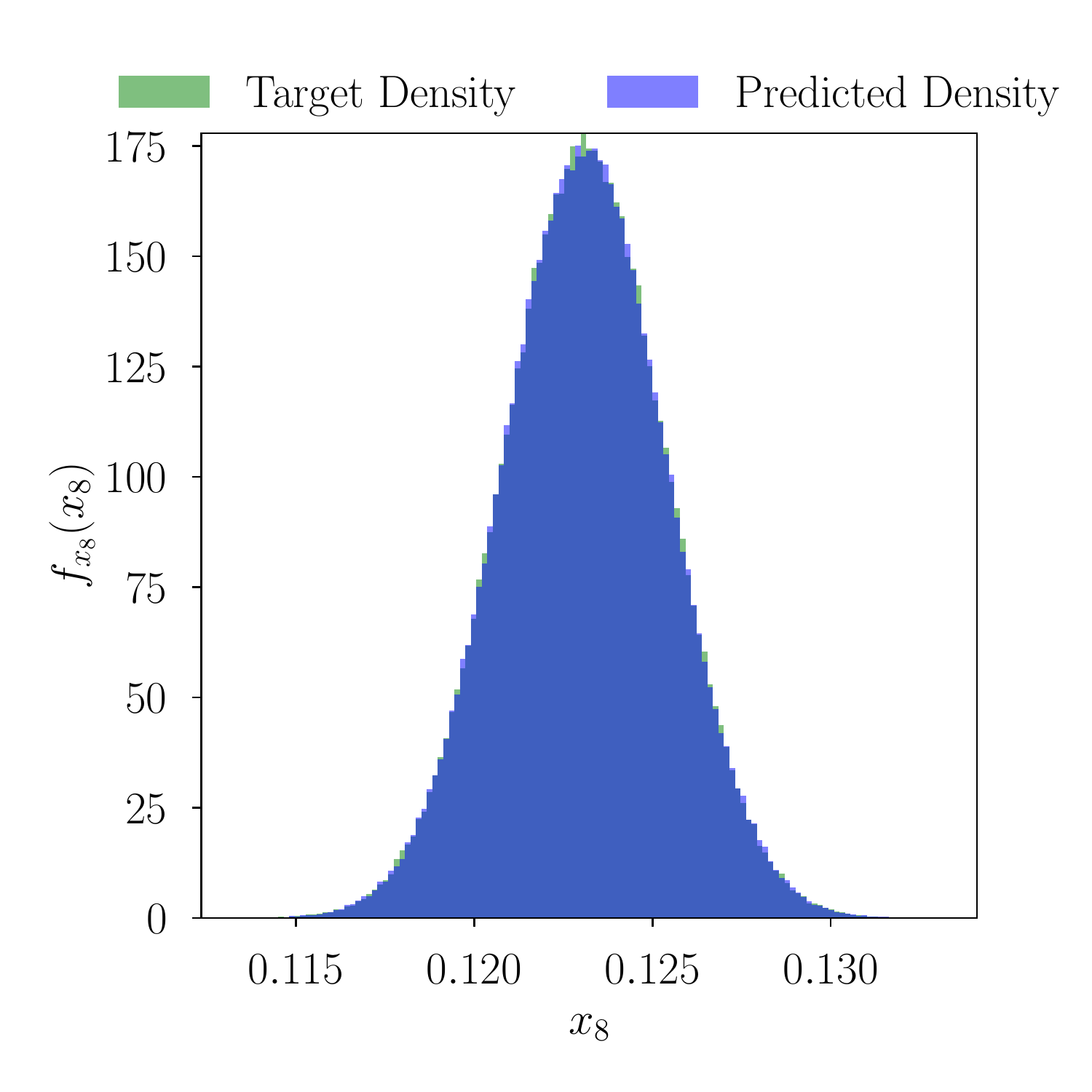}
\includegraphics[width=0.2\textwidth,clip]{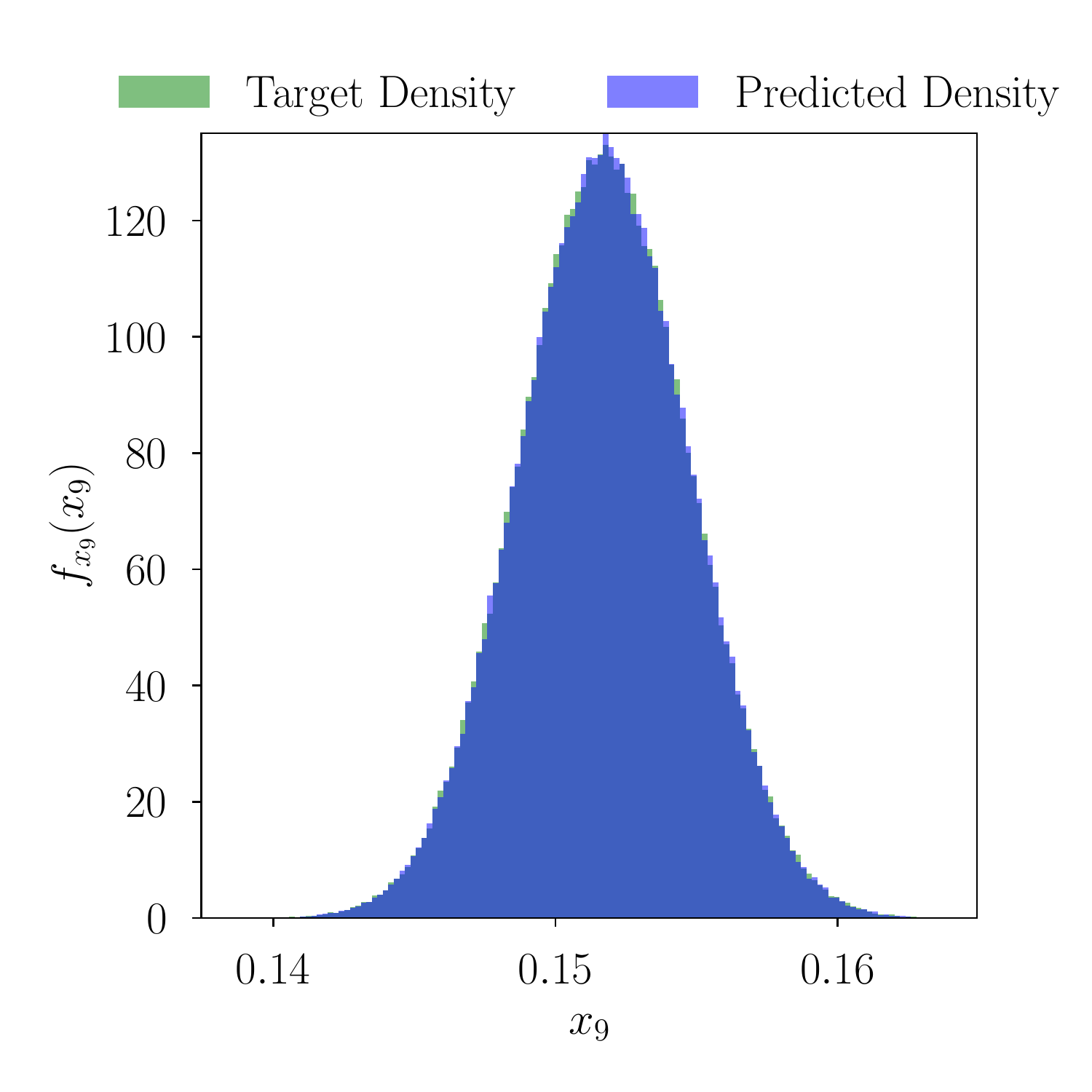}
\includegraphics[width=0.2\textwidth,clip]{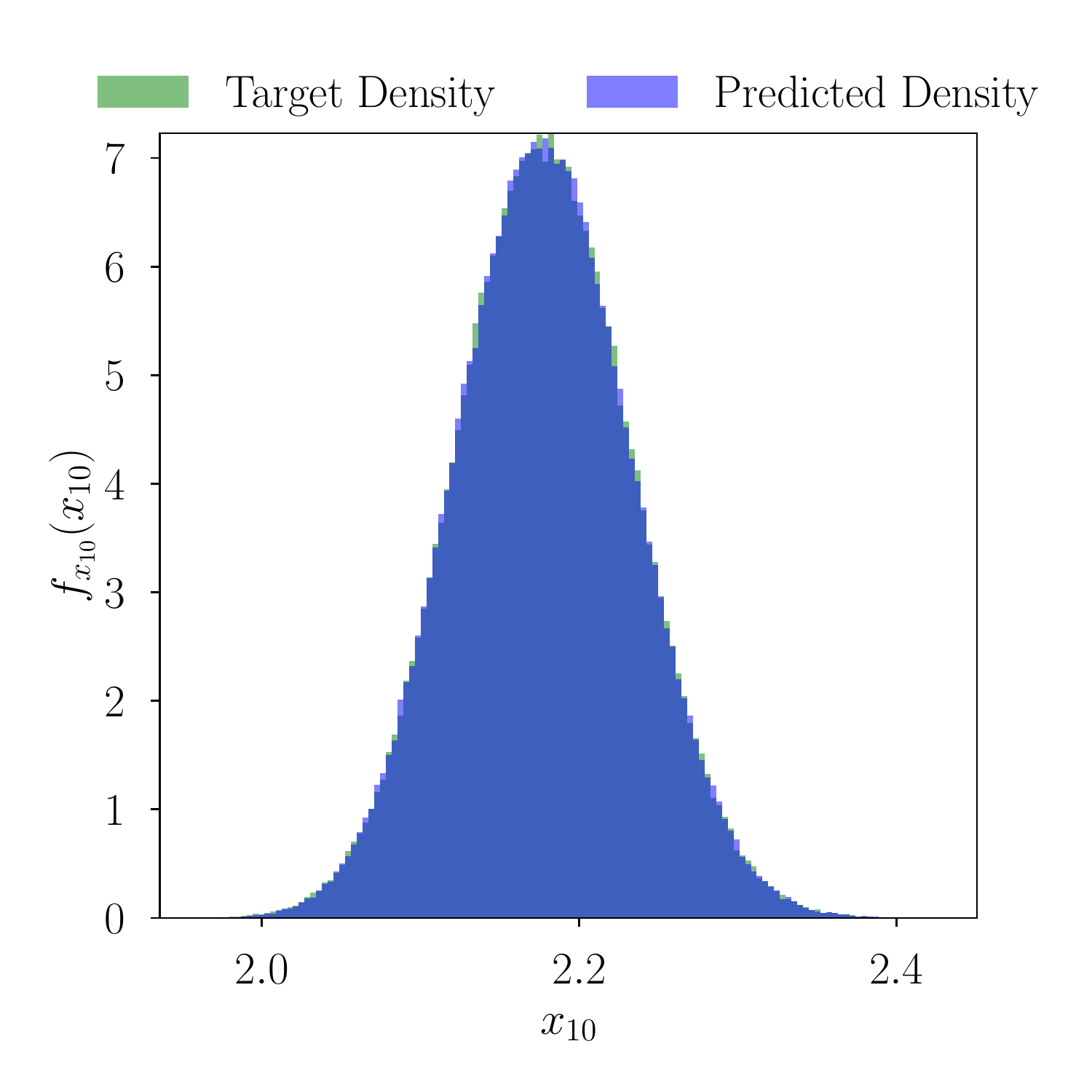}
\includegraphics[width=0.2\textwidth,clip]{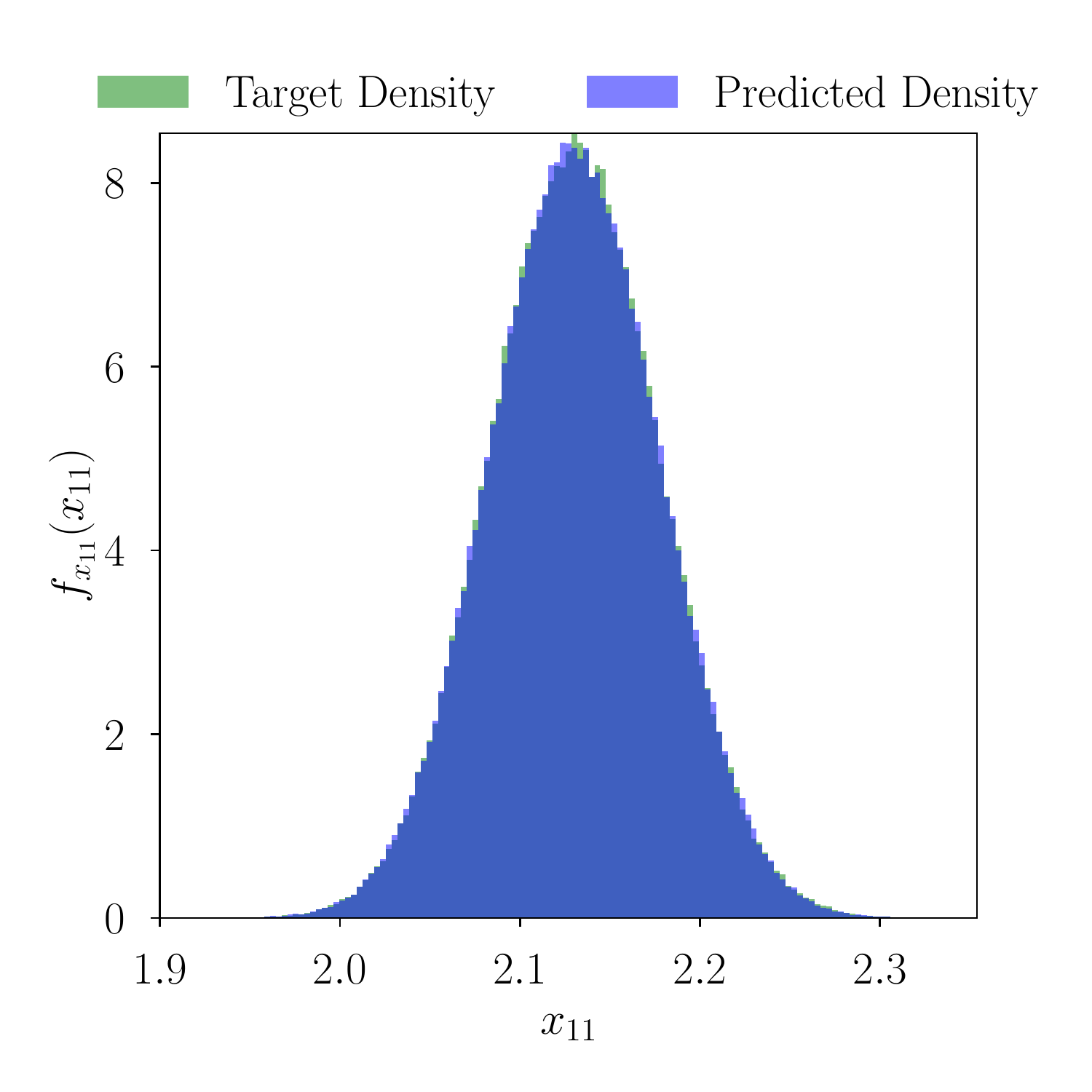}
\includegraphics[width=0.2\textwidth,clip]{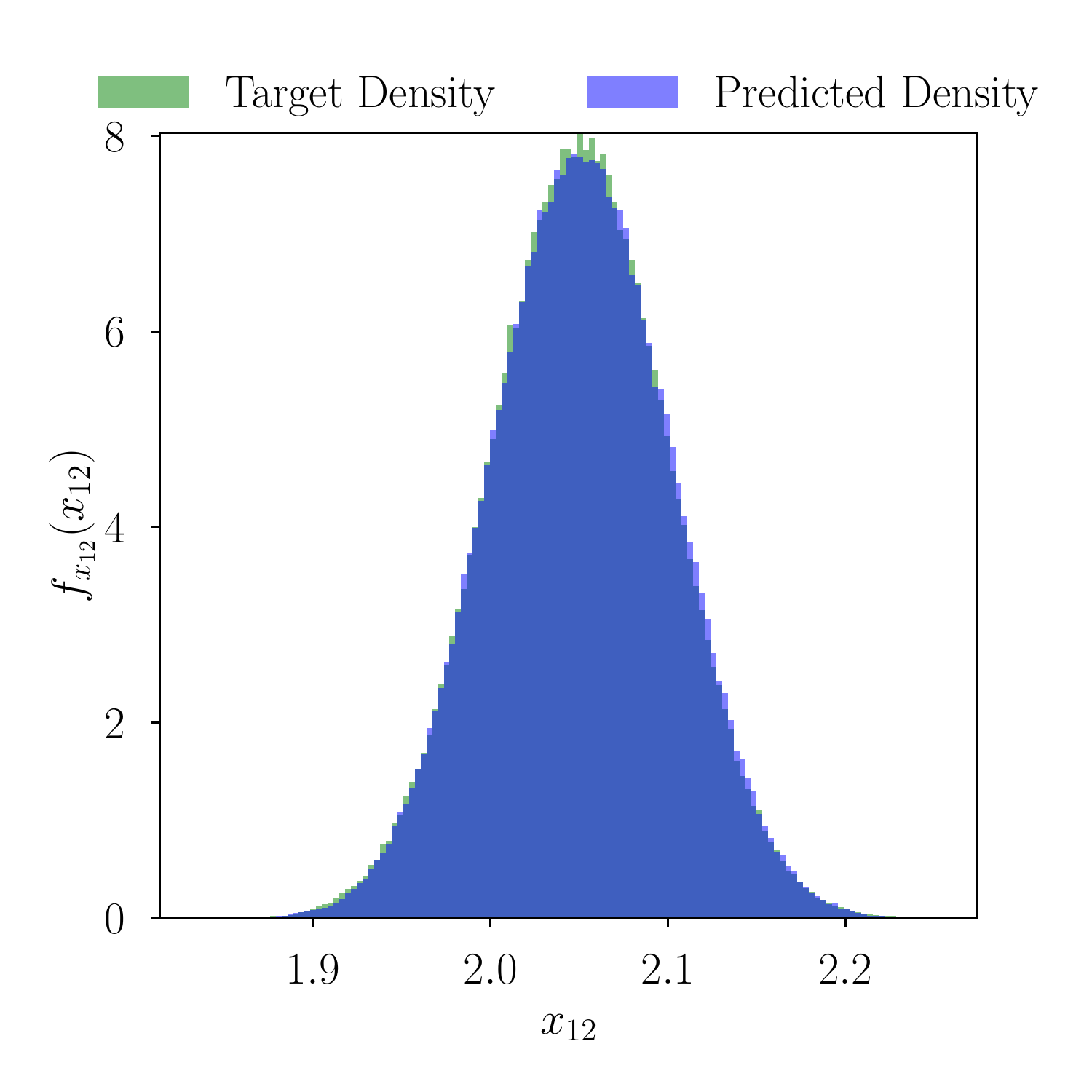}
\includegraphics[width=0.2\textwidth,clip]{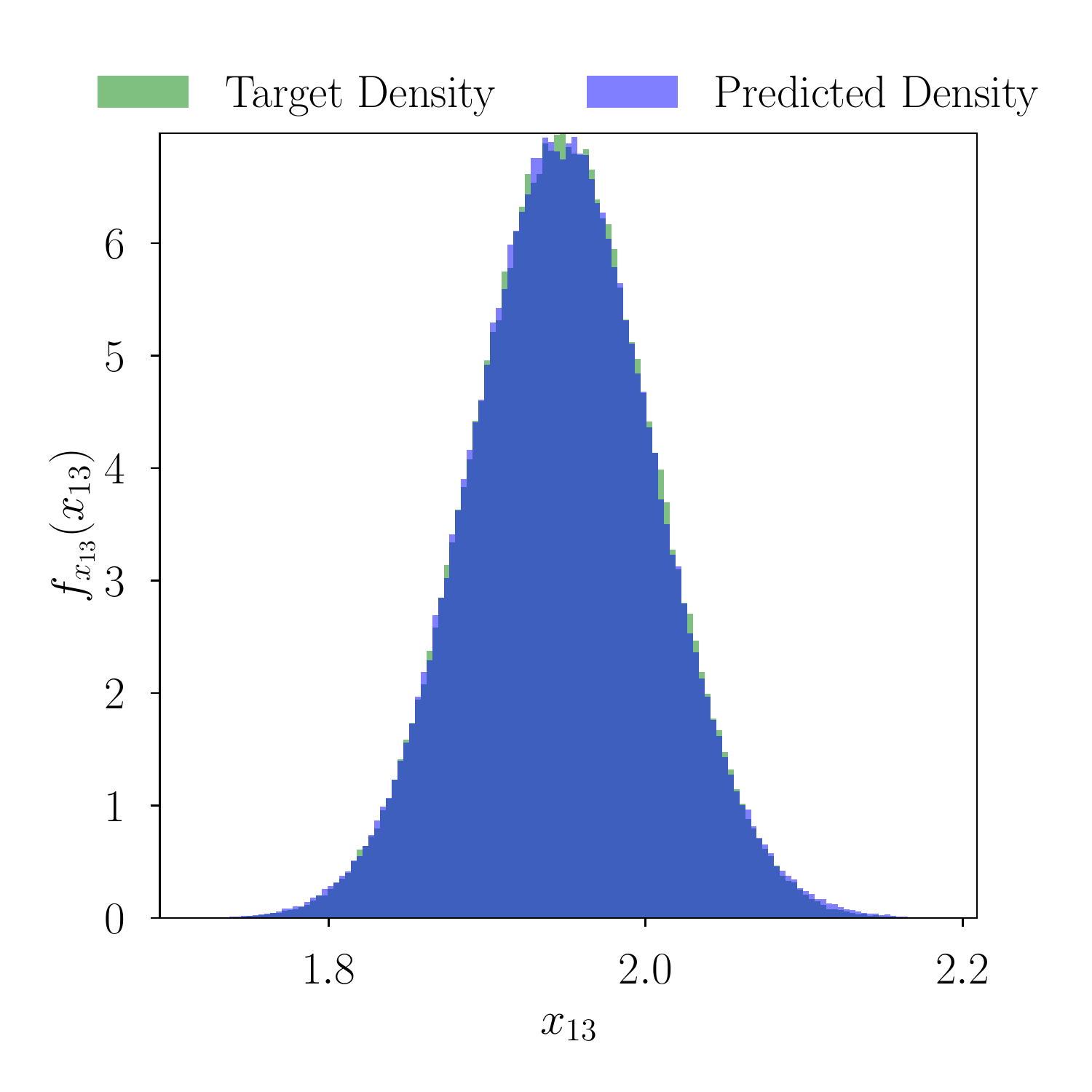}
\includegraphics[width=0.2\textwidth,clip]{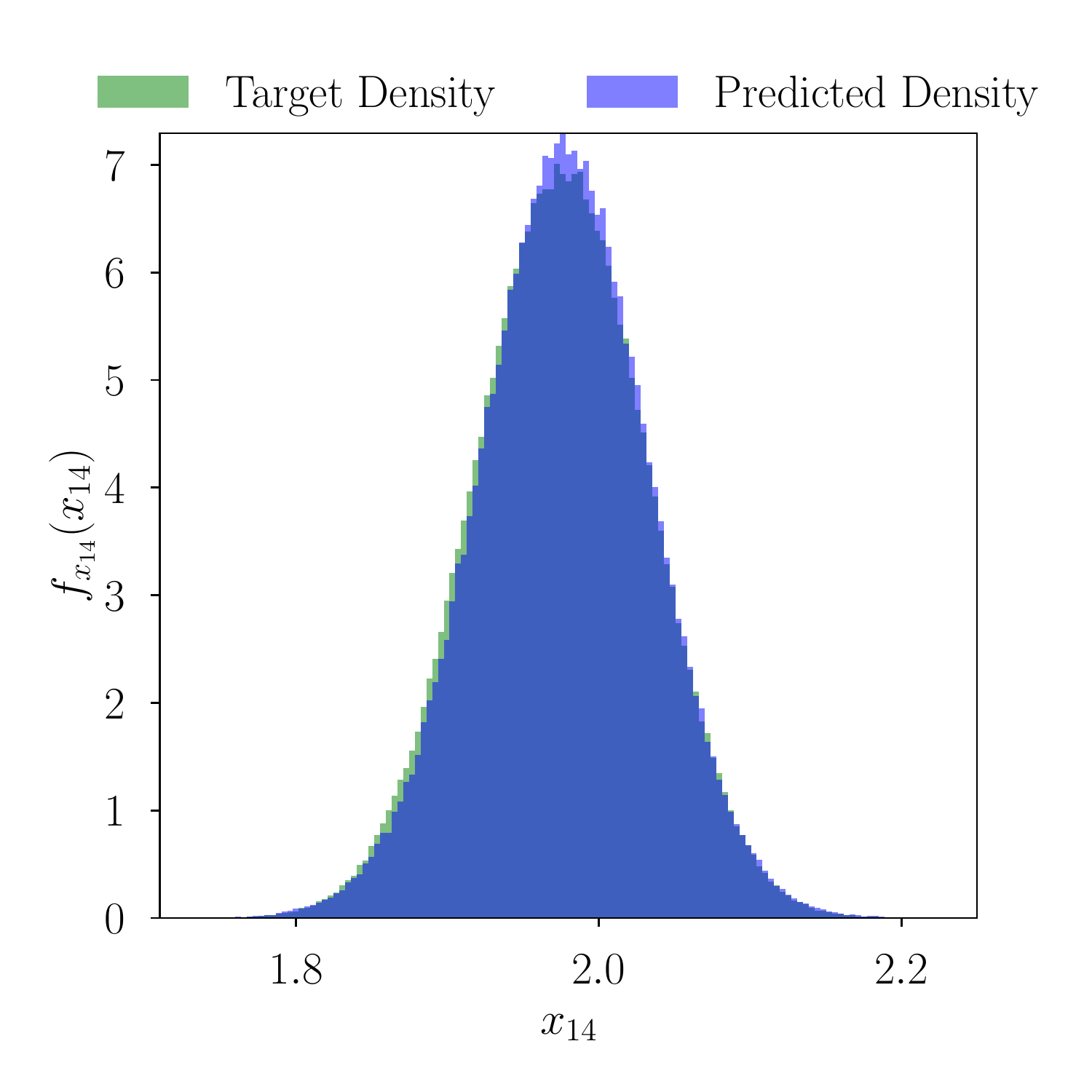}
\includegraphics[width=0.2\textwidth,clip]{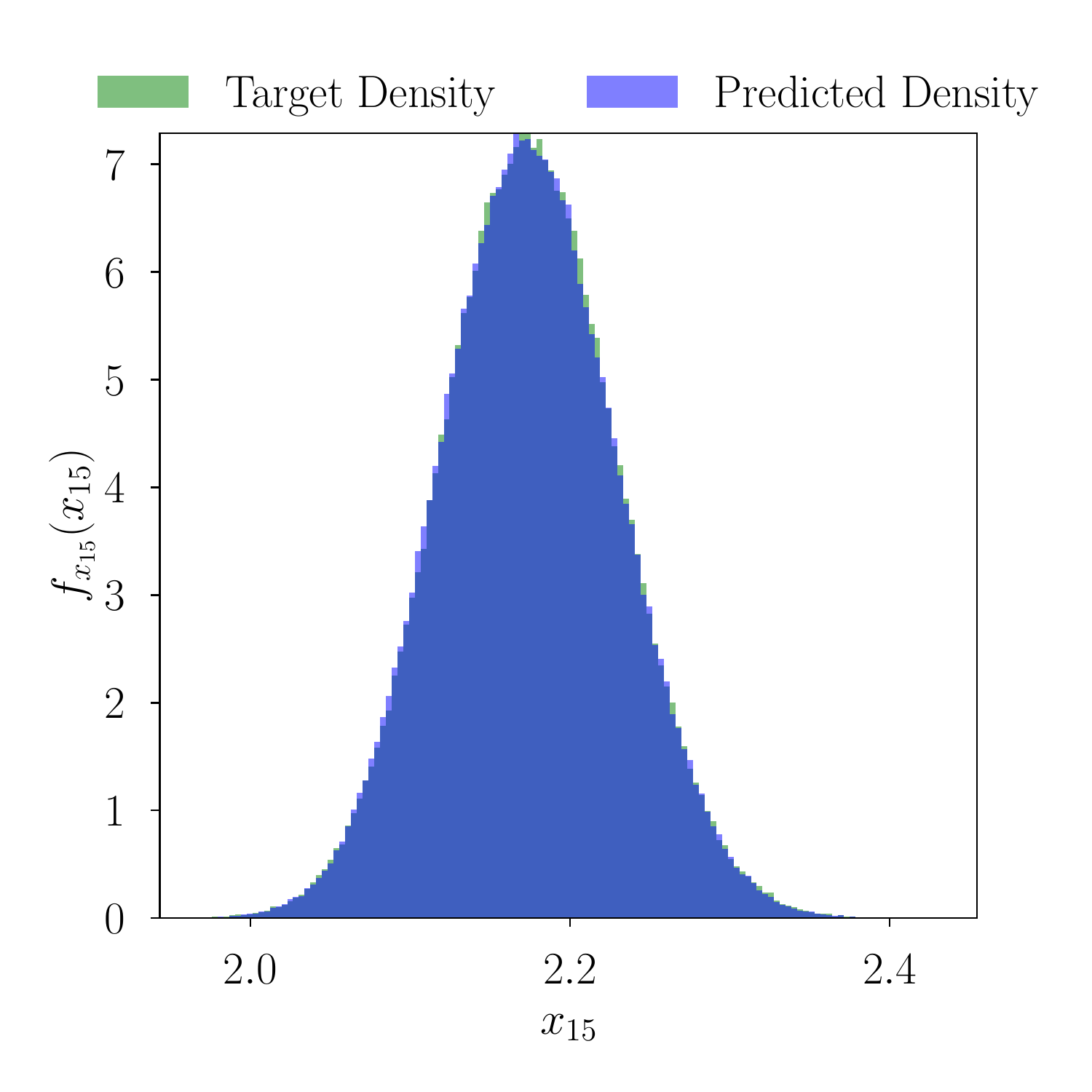}
\includegraphics[width=0.2\textwidth,clip]{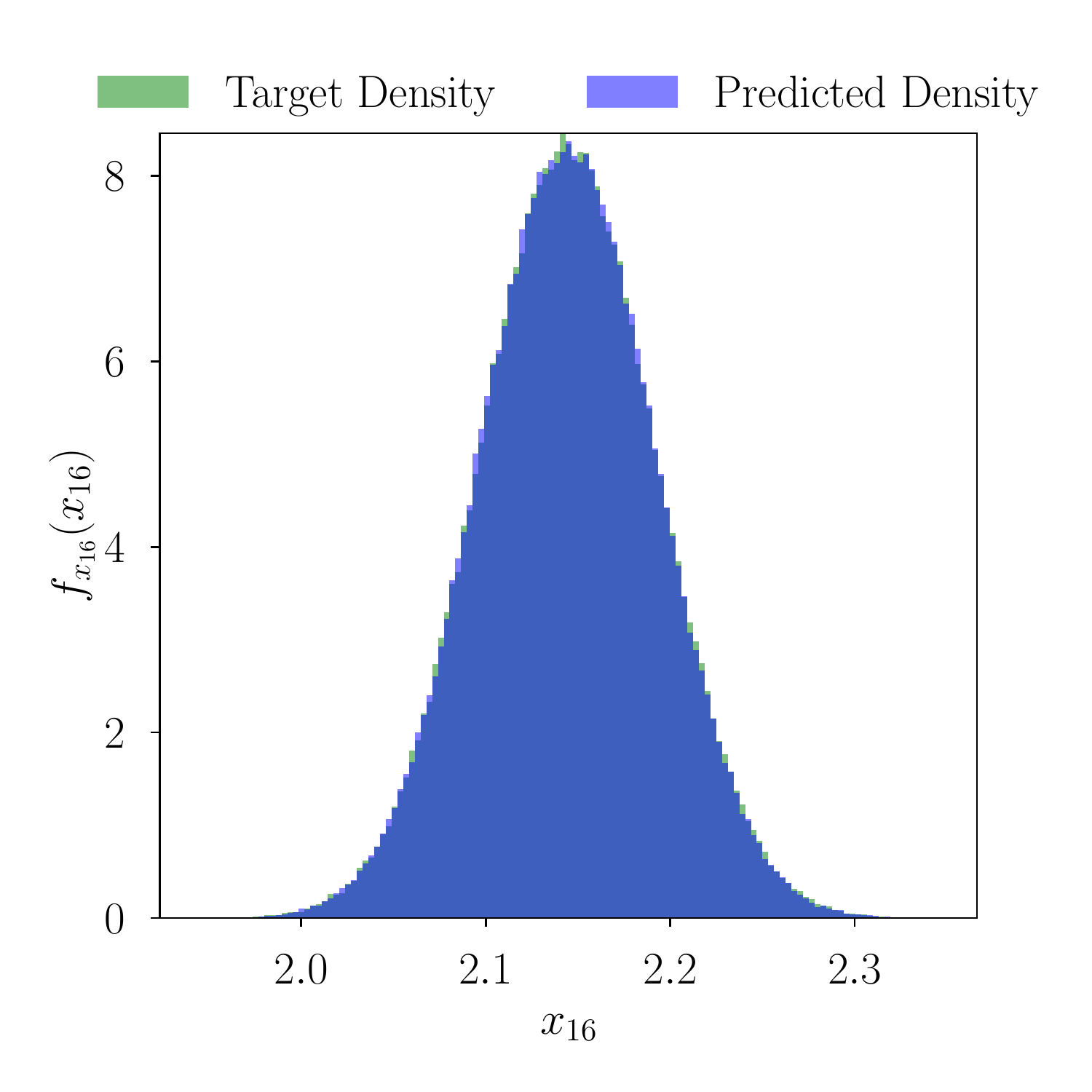}
\includegraphics[width=0.2\textwidth,clip]{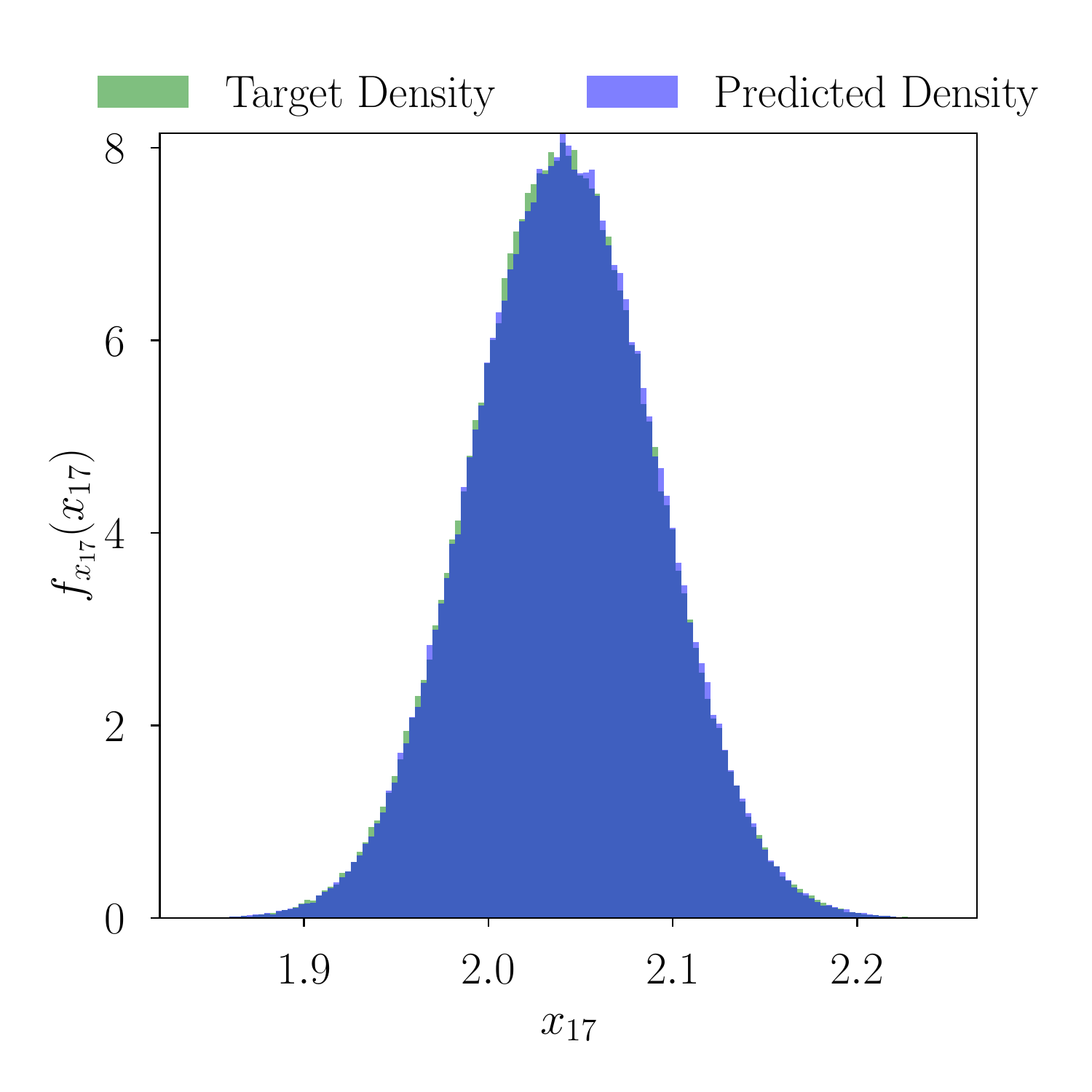}
\includegraphics[width=0.2\textwidth,clip]{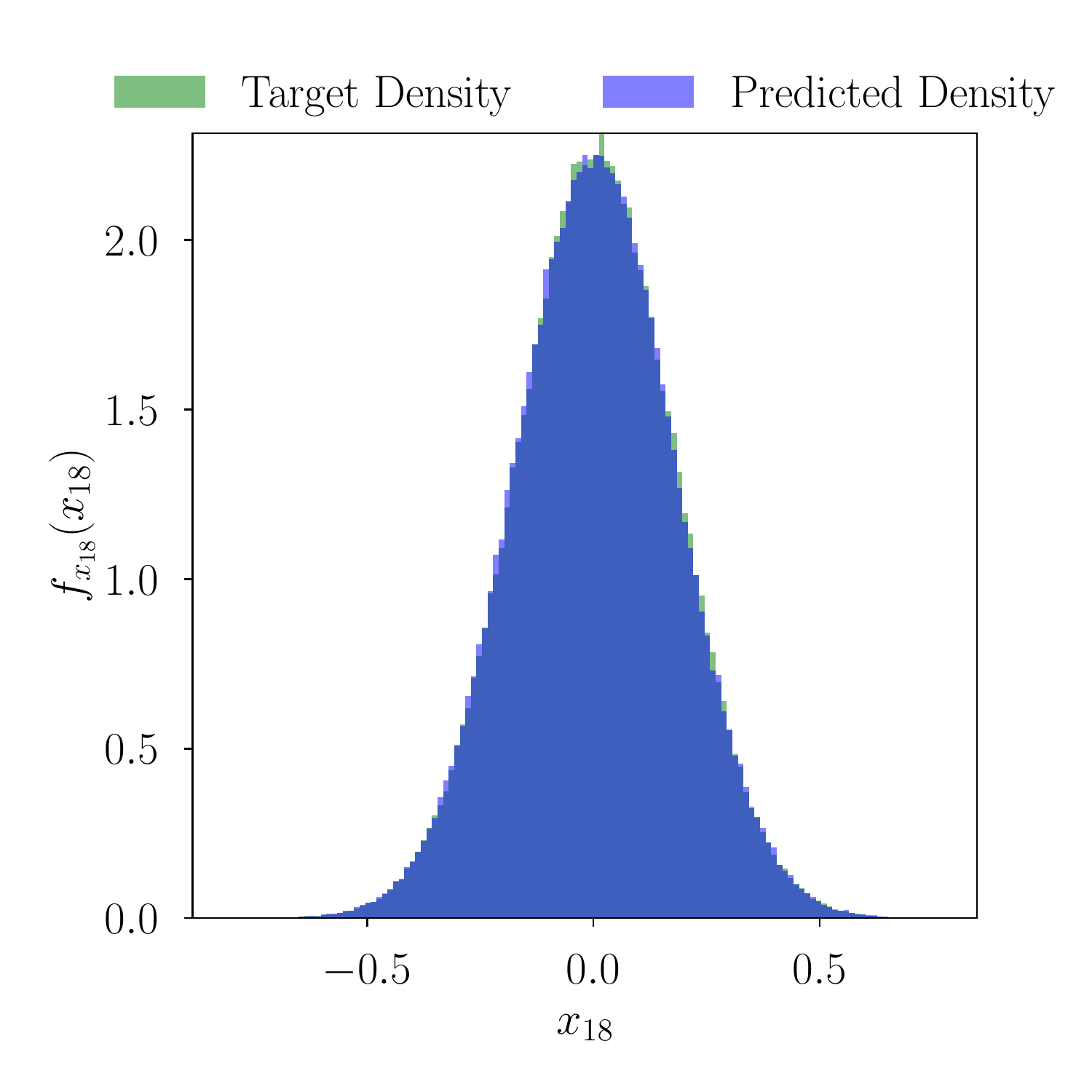}
\includegraphics[width=0.2\textwidth,clip]{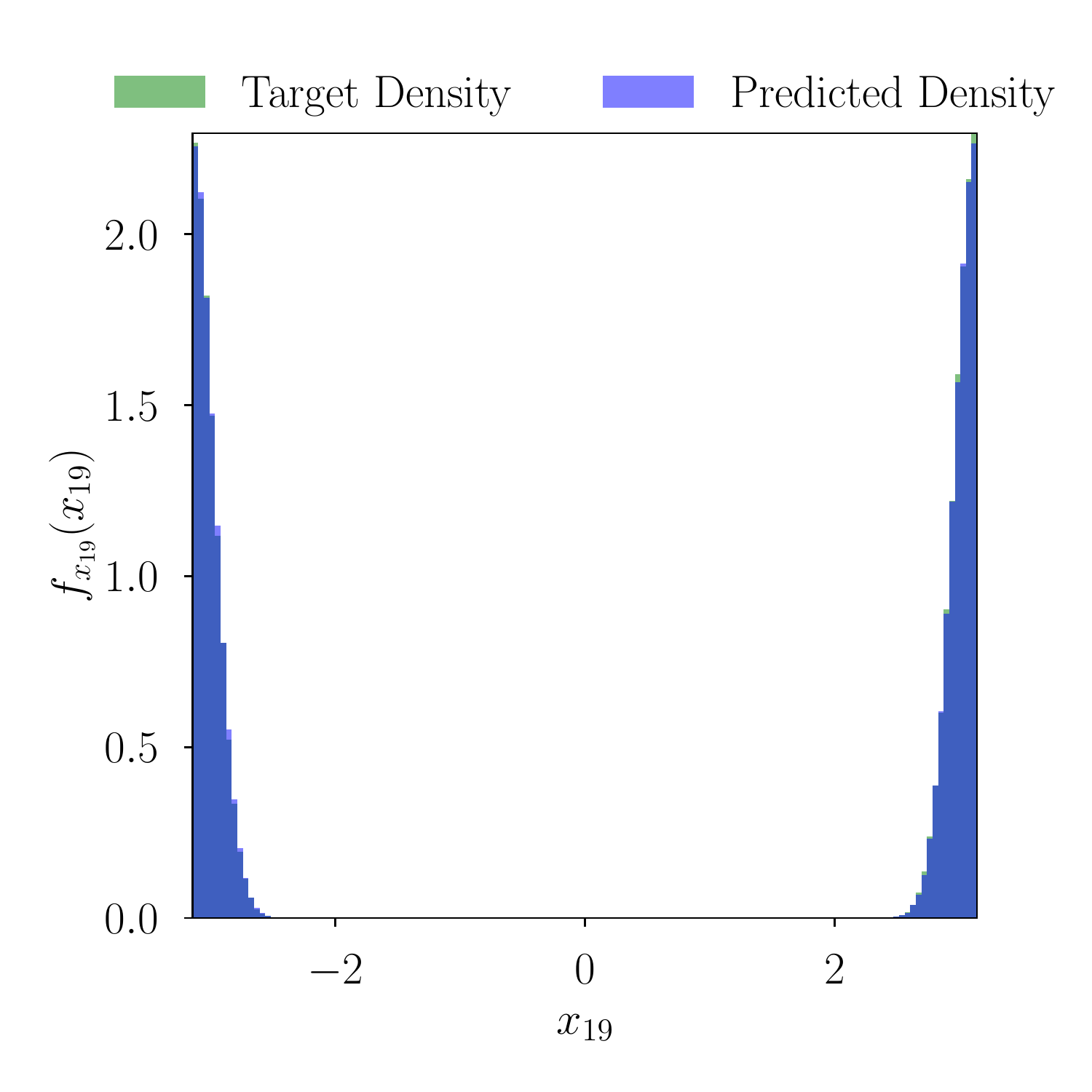}
\includegraphics[width=0.2\textwidth,clip]{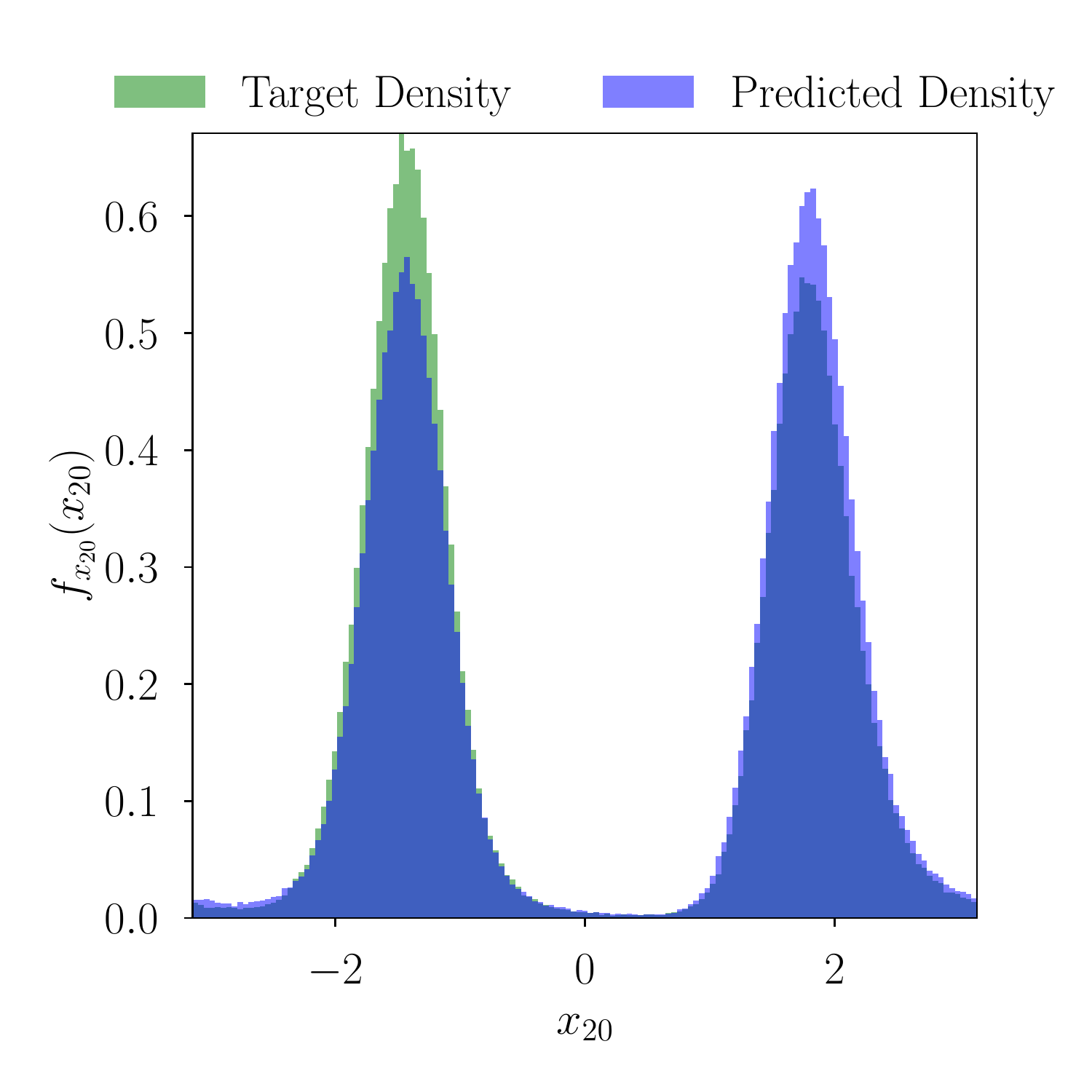}
\includegraphics[width=0.2\textwidth,clip]{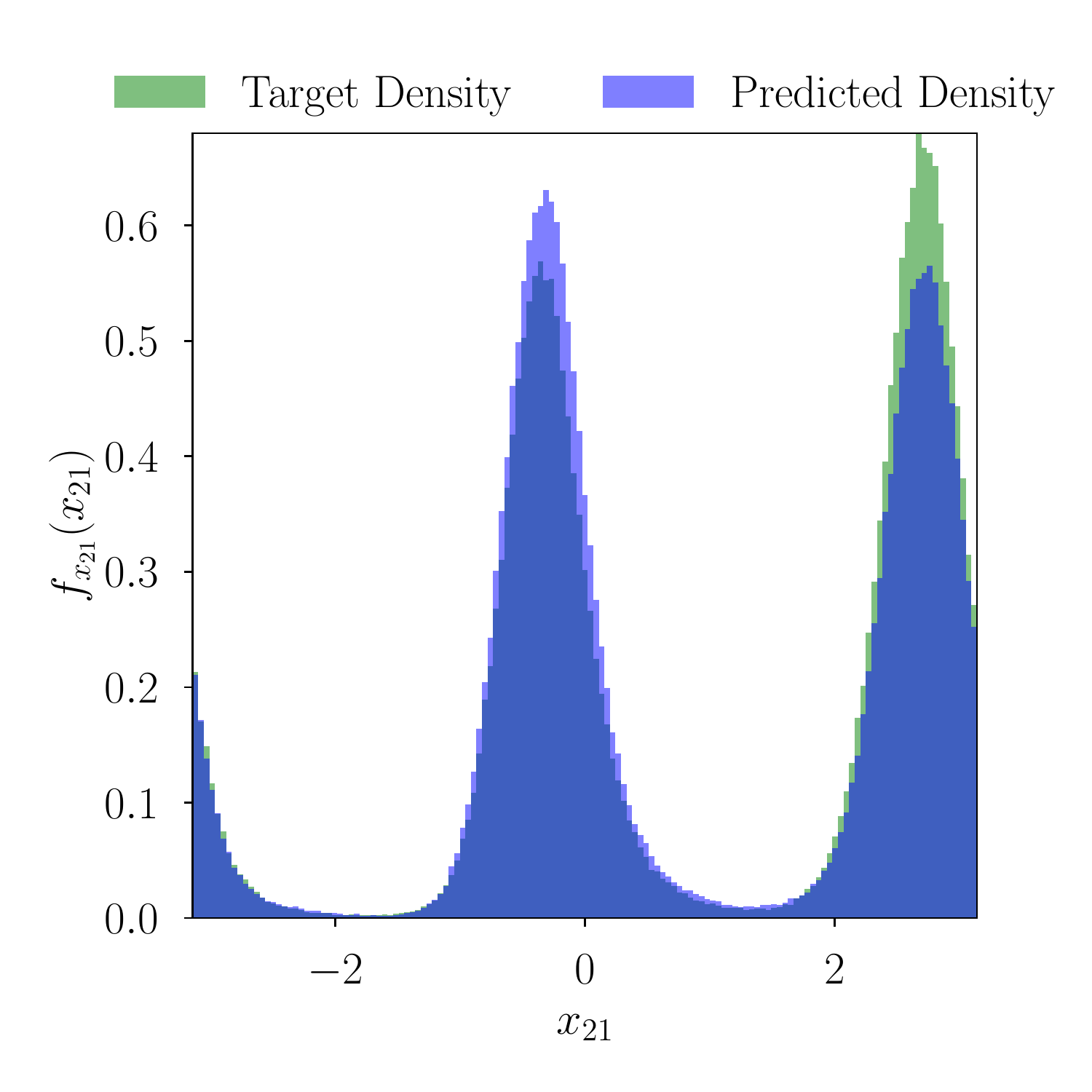}
\includegraphics[width=0.2\textwidth,clip]{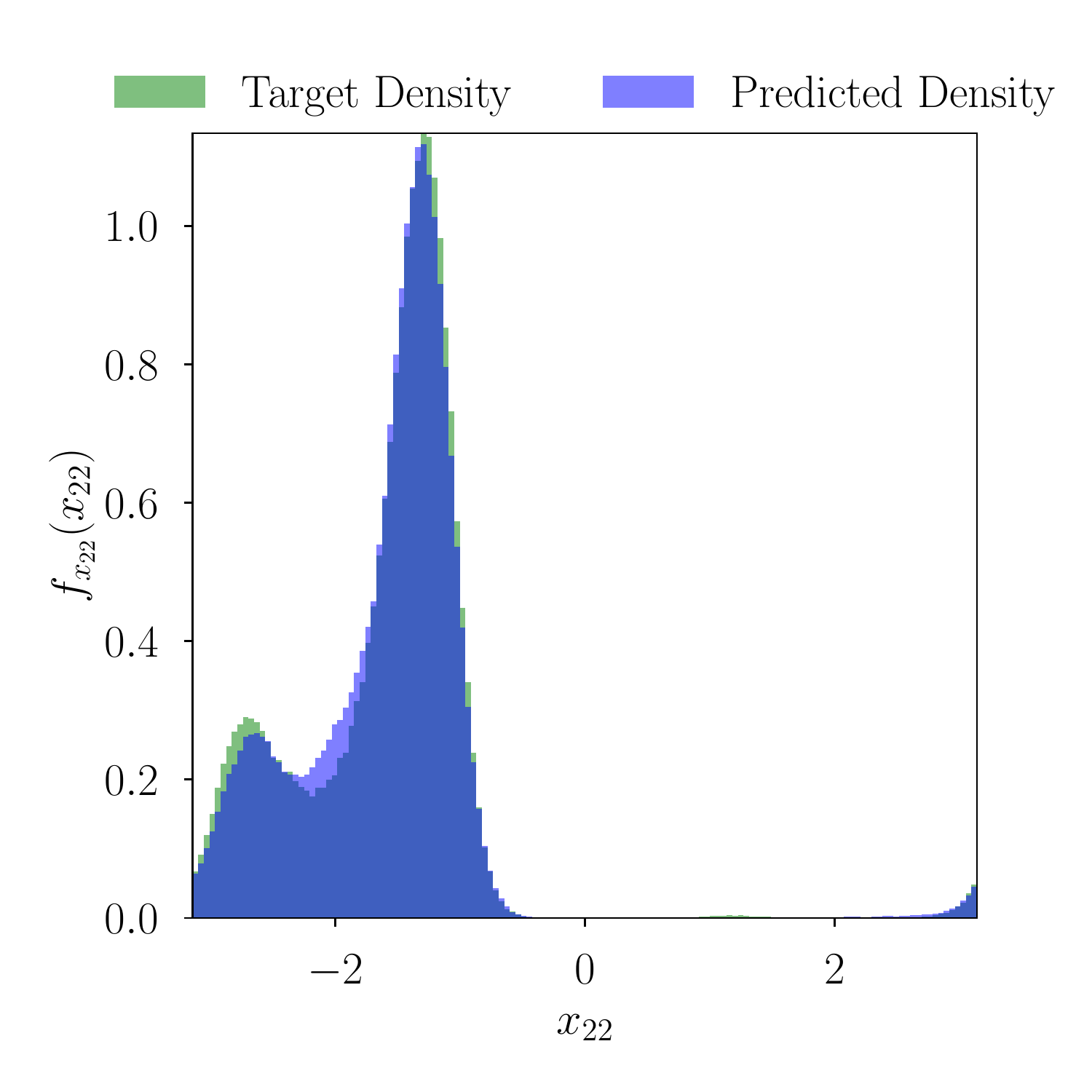}
\includegraphics[width=0.2\textwidth,clip]{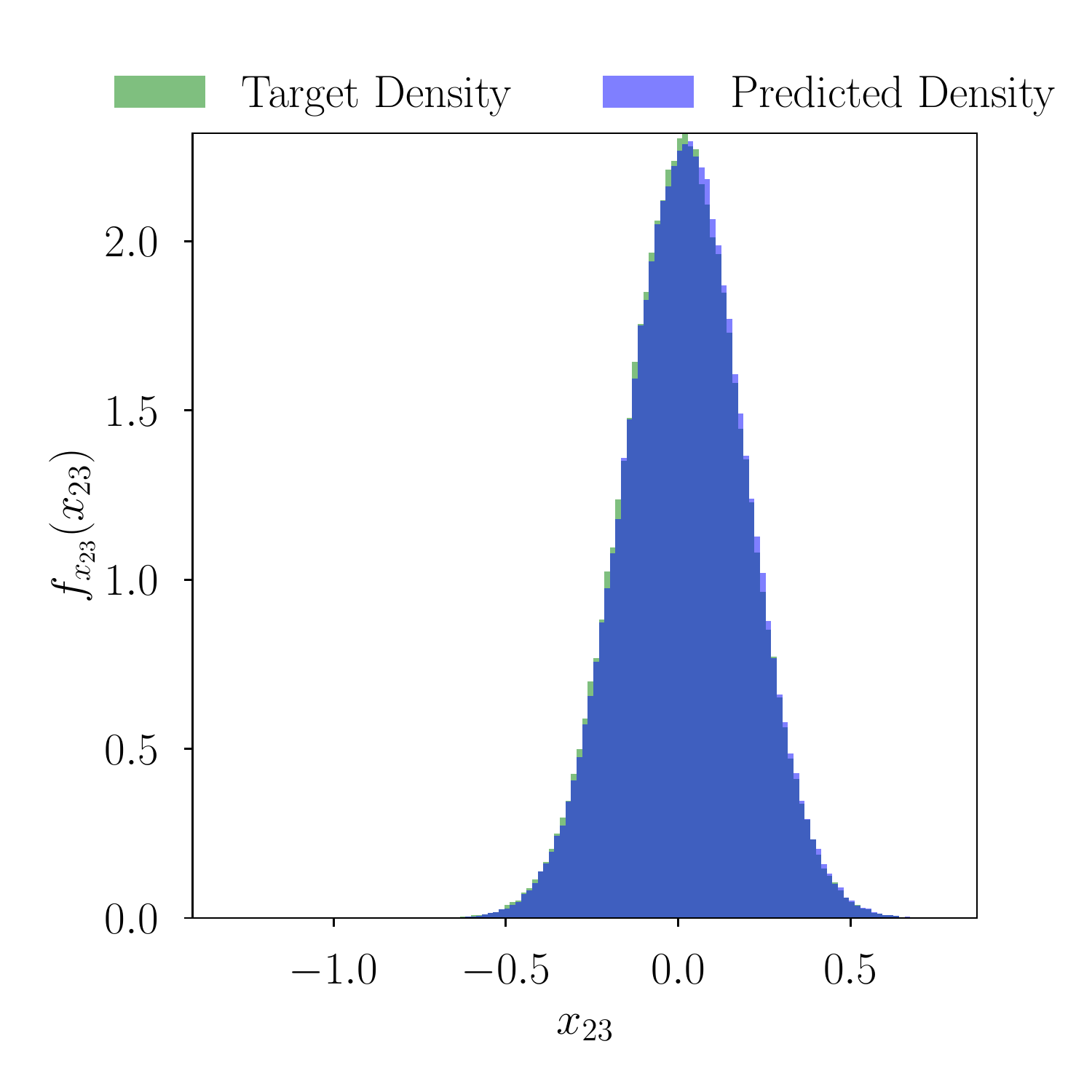}
\includegraphics[width=0.2\textwidth,clip]{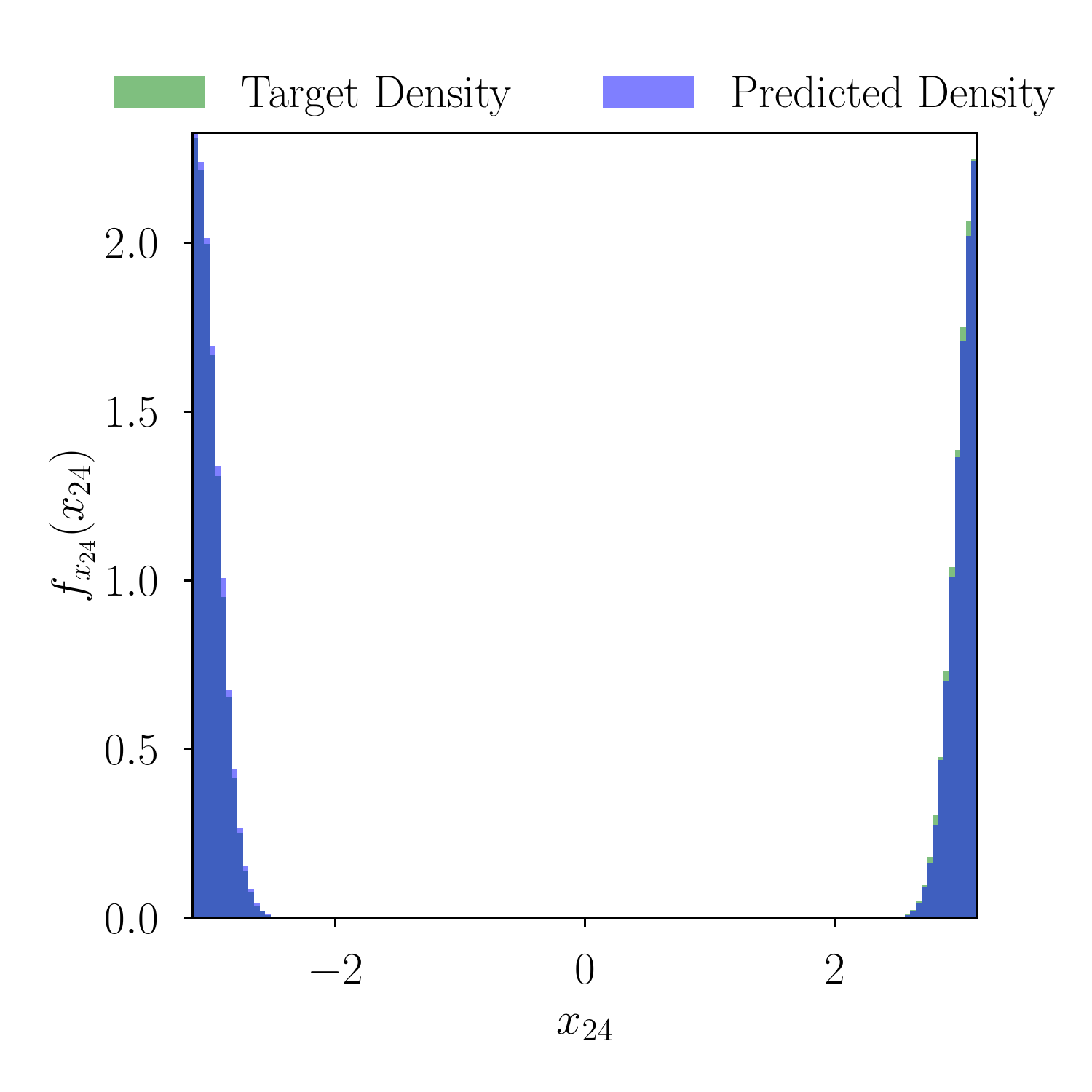}
\caption{
Plot of the marginal state distributions.
Comparison of the state distributions estimated from the MD data (test dataset) and from trajectories sampled from LED.
}
\label{fig:alanine:iterative_latent_forecasting_state_dist_bar}
\end{figure*}

The marginal distributions of the trajectories generated by LED match the ground-truth ones (MD data) closely, as depicted in Figure~\ref{fig:alanine:iterative_latent_forecasting_state_dist_bar}.

\paragraph*{Metastable State Definition}
\label{sec:appendix:alanine:metastablestates}

The protein is considered to lie in each of the five metastable states $\{C5, P_{II},\alpha_R, \alpha_L, C_7^{ax} \}$ if the distance in the Ramachandran plot between the protein state and the metastable state center is smaller than $10$ degrees.
The metastable state centers are defined in Table~\ref{tbl:metastablestatescenters}.
\begin{table}
\caption{
Centers of the metastable states in the Ramachandran plot.
}
\label{tbl:metastablestatescenters}
\centering
\begin{tabular}{|l|c|}
\hline
Metastable state & Center $(\phi, \psi)$ \\
\hline \hline
$ P_{II}$ & $(-75, 150)$\\ \hline
$C5$ & $(-155, 155)$\\ \hline
$\alpha_R$ & $(-75, -20)$\\ \hline
$\alpha_L$ &$(67, 5)$ \\ \hline
$ C_7^{ax}$ &  $(70, 160)$\\ \hline
\end{tabular}
\end{table}

\paragraph*{Metastable States on the Latent Space and Mean First Passage Times}
\label{sec:appendix:alanine:metastablestateslatent}

The metastable states can be defined on the latent space of LED, by projecting the free energy on the latent space, and identifying the local minima.
This alleviates the need for expert knowledge (definition of the metastable states).
The MFPTs between the metastable states on the latent space of the LED are compared with the MFPTs between the corresponding metastable states on the Ramachadran space in Table~\ref{tbl:alanine:mfpt:latent}.
Note that the results depend on how the latent metastable states are defined.
However, in order to capture the order of the timescales without the need of prior expert knowledge, a rough approximation (small region around the minima in the latent space) is adequate.
The LED is able to capture the order of the timescales, alleviating the need for expert knowledge on the definition of the metastable states.

\begin{table}
\caption{
Mean first-passage times (MFPT) between the metastable states of alanine dipeptide in water in [ns].
MFPTs are estimated by fitting MSMs with a time-lag of $10 \text{ps}$ on MD trajectories.
In LED, the metastable states are considered as regions around the local minima of the free energy projection on the latent space.
The average relative error is given for reference.
}
\label{tbl:alanine:mfpt:latent}
\centering
\includegraphics[width=0.8\textwidth]{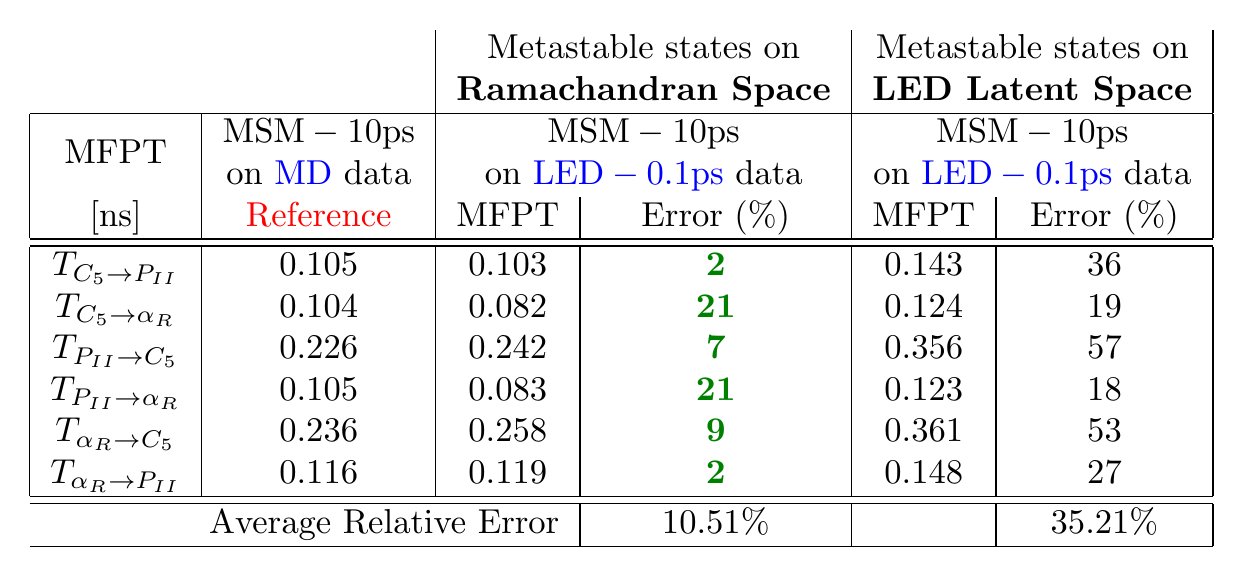}
\end{table}

\paragraph*{Unraveling Novel Configurations with LED}
\label{sec:appendix:alanine:novelconf}

\begin{figure*}[t]
\centering
\includegraphics[width=0.9\textwidth,clip]{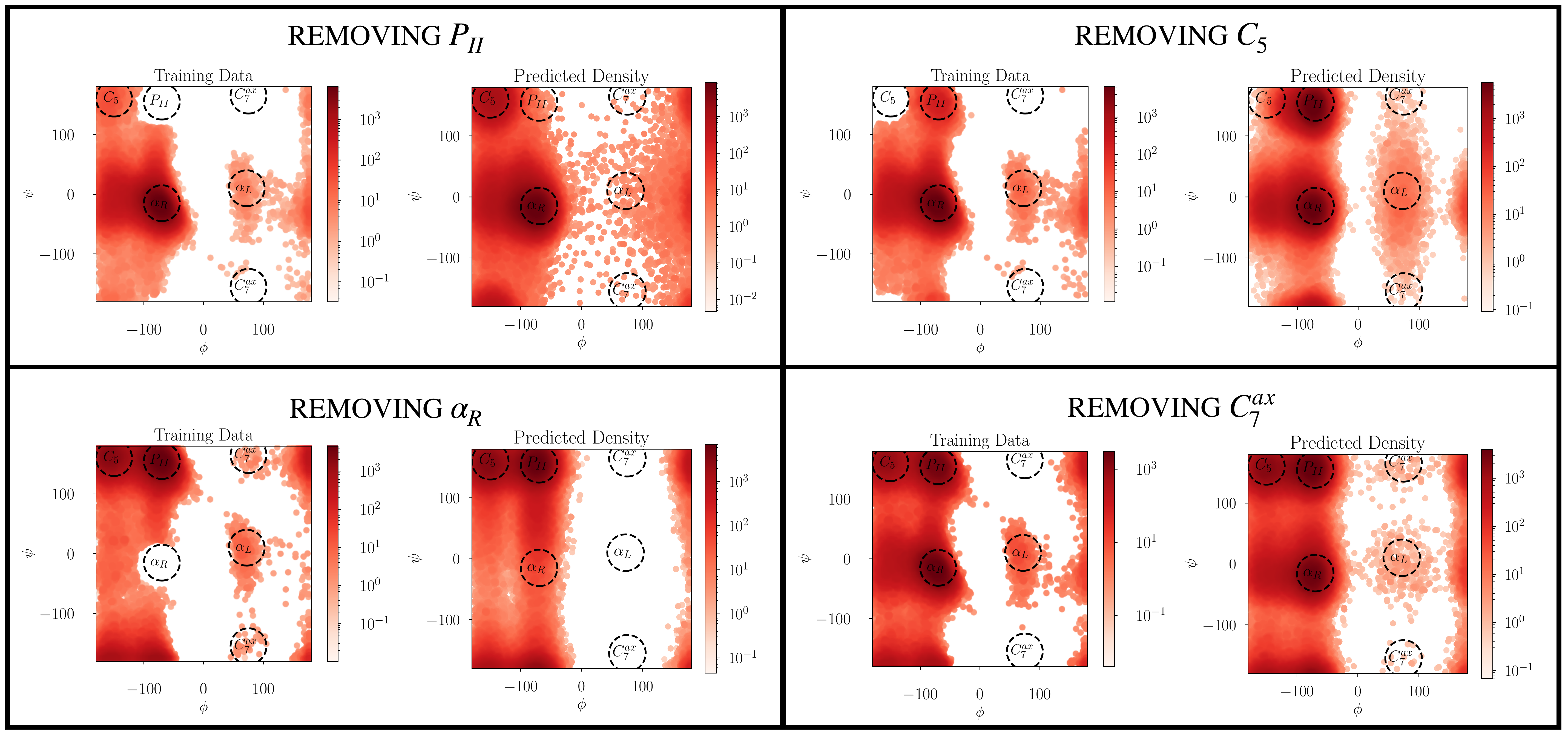}
\centering
\caption{
LED is trained in four scenarios hiding data that lie closer than 40 degrees to one of the metastable states $\{ P_{II}, C_5, \alpha_R, C_7^{ax} \} $ each time.
LED can successfully generate novel probable configurations close to the metastable states $\{ P_{II}, C_5 , \alpha_R, C_7^{ax} \}$.
Due to the limited training data, however, capturing the state density in the Ramachandran plot is challenging.
}
\label{fig:alanine:discovery-alanine}
\end{figure*}

We evaluate LED's effectiveness in unraveling novel configurations of the protein (state-space) absent from the training data.
For this purpose, we create four different small datasets composed of trajectories of the protein, each one not including one of the metastable states $\{C5, P_{II}, \alpha_R, C_7^{ax} \}$.
This is done by removing any state that lies closer than 40 degrees to the metastable states' centers.
In this way, we guarantee that LED has not seen any state close to the metastable state missing from the data. 
Note that in this case, the LED is not trained on a single large MD trajectory but on small trajectories are not temporally adjacent.
We end up with four datasets, each one consisting of approximately $800$ trajectories of length $T=50 \text{ps}$ (500 steps of 0.1 ps).
Each dataset covers approximately $40 \text{ns}$ protein simulation time.
These datasets are created to evaluate the effectiveness of LED in generating truly novel configurations for faster exploration of the state space.
We do not care at this point for accurate reproduction of the statistics due to the minimal data used for training.
In Figure~\ref{fig:alanine:discovery-alanine}, we plot the Ramachandran plots of the training data along with the ones obtained by analyzing the trajectories of the trained LED models in each of the four cases.
We observe that the LED can unravel the metastable states $P_{II}$, $C_5$,  $C_7^{ax}$, and $\alpha_R$, even though they were not part of the training data.
However, by removing states that lie close to the metastable state $\alpha_R$, the LED cannot capture the $\alpha_R$ and $C_7^{ax}$ metastable states.
This is because the LED is trained on only a small subset of the training dataset and the transitions to these metastable states are rare.

\clearpage
\newpage
\paragraph*{LED Hyperparameters}
\label{sec:appendix:alanine:hyp}

Regarding the LED architecture, the number and size of hidden layers are the same for the encoder $\mathcal{E}$, the decoder $\mathcal{D}$, and the latent MDN $\mathcal{Z}$.
The MDN-AE is trained, tuning its hyperparameters based on the grid search reported in Table~\ref{app:tab:alanine:hyp:auto}.
The latent space of the MDN-AE is $\boldsymbol{z} \in \mathbb{R}^2$, i.e., $d_{\boldsymbol{z}}=1$.
The MDN-AE model with the lowest validation error on the state statistics is picked.
Then, the MDN-AE is coupled with the MDN-LSTM in LED.
The MDN-LSTM is trained to minimize the latent data likelihood.
The hyperparameters of the MDN-LSTM are tuned according to the grid search reported in Table~\ref{app:tab:alanine:hyp:lstm}.
The LED model with the lowest error on the state statistics in the validation dataset is selected.
Its hyperparameters are reported in Table~\ref{app:tab:alanine:hyp:led}.
The LED is tested in $248$ initial conditions randomly sampled from the testing data.
Starting from these initial conditions, we utilize the iterative propagation in the latent space of the LED to forecast $T=400 \text{ps}$.

\begin{table}[tbhp]
\caption{Hyperparameter tuning of AE for alanine dipeptide}
\label{app:tab:alanine:hyp:auto}
\centering
\begin{tabular}{ |c|c|c| } 
\hline
\text{Hyperparameter} & 
\text{Values} \\  \hline \hline
Batch size& $32$ \\
Initial learning rate & $10^{-3} $ \\
Weight decay rate & $\{0, 10^{-5}  \}$ \\
Number of AE layers & $\{4,6 \}$ \\
Size of AE layers & $\{50,100\}$ \\
Activation of AE layers & $\operatorname{selu}$, $\operatorname{tanh}$ \\
Latent dimension & $2$ \\
Input/Output data scaling & $[0,1]$ \\
MDN-AE kernels & $5$ \\
MDN-AE hidden units & $\{20,50\}$ \\
MDN-AE multivariate & $0$ \\
MDN-AE covariance scaling factor & $0.8$ \\
\hline
\end{tabular}
\end{table}
\begin{table}[tbhp]
\caption{Hyperparameter tuning of LSTM for alanine dipeptide}
\label{app:tab:alanine:hyp:lstm}
\centering
\begin{tabular}{ |c|c|c| } 
\hline
\text{Hyperparameter} & 
\text{Values} \\  \hline \hline
Batch size& $32$ \\
Initial learning rate & $10^{-3} $ \\
BPTT sequence length & $\{200, 400 \}$ \\
Number of LSTM layers & $1$ \\
Size of LSTM layers & $\{10, 20, 40 \}$ \\
Activation of LSTM Cell & $\operatorname{tanh}$ \\
MDN-LSTM kernels & $\{4,5,6\}$ \\
MDN-LSTM hidden units & $\{10, 20\}$ \\
MDN-LSTM multivariate & $\{0,1\}$ \\
MDN-LSTM covariance scaling factor & $\{ 0.1, 0.2, 0.3, 0.4 \}$ \\
\hline
\end{tabular}
\end{table}
\begin{table}[tbhp]
\caption{Hyperparameters of LED model with lowest validation error on alanine dipeptide}
\label{app:tab:alanine:hyp:led}
\centering
\begin{tabular}{ |c|c|c| } 
\hline
\text{Hyperparameter} & 
\text{Values} \\  \hline \hline
Number of AE layers & $4$ \\
Size of AE layers & $50$ \\
Activation of AE layers & $\operatorname{tanh}$ \\
Latent dimension & $2$ \\
MDN-AE kernels & $5$ \\
MDN-AE hidden units & $50$ \\
MDN-AE multivariate & $0$ \\
MDN-AE covariance scaling factor & $0.8$ \\
Weight decay rate & $0 $ \\
BPTT sequence length & $400 $ \\
Number of LSTM layers & $1$ \\
Size of LSTM layers & $20$ \\
Activation of LSTM Cell & $\operatorname{tanh}$ \\
MDN-LSTM kernels & $5$ \\
MDN-LSTM hidden units & $20$ \\
MDN-LSTM multivariate & $0$ \\
MDN-LSTM covariance scaling factor & $0.4$ \\
\hline
\end{tabular}
\end{table}

\end{document}